\documentclass[twocolumn]{aastex63}
\usepackage{amsmath,amssymb}
\usepackage[overload]{textcase}
\usepackage{seqsplit}

\def\arcsec{{^{\prime\prime}}}

\def\farcs{\hbox{$.\!\!^{\prime\prime}$}}

\def\gtrsim{\mathrel{\hbox{\rlap{\hbox{\lower4pt\hbox{$\sim$}}}\hbox{$>$}}}}
\def\lessim{\mathrel{\hbox{\rlap{\hbox{\lower4pt\hbox{$\sim$}}}\hbox{$<$}}}}

\def\Ls{L_{\odot}}

\accepted{April 29, 2023}

\shorttitle{\MakeTextLowercase{e}Disk I: Overview and first results}
\shortauthors{Ohashi et al.}

\begin{document}

\title{Early Planet Formation in Embedded Disks (eDisk). I. Overview of the Program and First Results}

\correspondingauthor{Nagayoshi Ohashi}
\email{ohashi@asiaa.sinica.edu.tw}

\author[0000-0003-0998-5064]{Nagayoshi Ohashi}
\affiliation{Academia Sinica Institute of Astronomy \& Astrophysics,
11F of Astronomy-Mathematics Building, AS/NTU, No.1, Sec. 4, Roosevelt Rd,
Taipei 10617, Taiwan, R.O.C.}
\author[0000-0002-6195-0152]{John J. Tobin}
\affiliation{National Radio Astronomy Observatory, 
520 Edgemont Rd., Charlottesville, VA 22903 USA}
\author[0000-0001-9133-8047]{Jes K. J{\o}rgensen}
\affiliation{Niels Bohr Institute, University of Copenhagen,
{\O}ster Voldgade 5-7, 1350, Copenhagen K, Denmark}
\author[0000-0003-0845-128X]{Shigehisa Takakuwa}
\affiliation{Department of Physics and Astronomy, Graduate School of Science and Engineering, Kagoshima University, 1-21-35 Korimoto, Kagoshima,Kagoshima 890-0065, Japan}
\affiliation{Academia Sinica Institute of Astronomy \& Astrophysics,
11F of Astronomy-Mathematics Building, AS/NTU, No.1, Sec. 4, Roosevelt Rd,
Taipei 10617, Taiwan, R.O.C.}
\author[0000-0002-9209-8708]{Patrick Sheehan}
\affiliation{National Radio Astronomy Observatory, 520 Edgemont Rd., Charlottesville, VA 22903 USA}

\author[0000-0003-3283-6884]{Yuri Aikawa}
\affiliation{Department of Astronomy, Graduate School of Science, The University of Tokyo, 7-3-1 Hongo, Bunkyo-ku, Tokyo 113-0033, Japan}
\author[0000-0002-7402-6487]{Zhi-Yun Li}
\affiliation{University of Virginia, 530 McCormick Rd., Charlottesville, Virginia 22904, USA}
\author[0000-0002-4540-6587]{Leslie W. Looney}
\affiliation{Department of Astronomy, University of Illinois, 1002 West Green St, Urbana, IL 61801, USA}
\author[0000-0001-5058-695X]{Jonathan P. Williams}
\affiliation{Institute for Astronomy, University of Hawai‘i at Mānoa, 2680 Woodlawn Dr., Honolulu, HI 96822, USA}

\author[0000-0002-8238-7709]{Yusuke Aso}
\affiliation{Korea Astronomy and Space Science Institute, 776 Daedeok-daero, Yuseong-gu, Daejeon 34055, Republic of Korea}
\author[0000-0002-0549-544X]{Rajeeb Sharma}
\affiliation{Niels Bohr Institute, University of Copenhagen, {\O}ster Voldgade 5-7, 1350, Copenhagen K, Denmark}
\author[0000-0003-4361-5577]{Jinshi Sai (Insa Choi)}
\affiliation{Academia Sinica Institute of Astronomy \& Astrophysics,
11F of Astronomy-Mathematics Building, AS/NTU, No.1, Sec. 4, Roosevelt Rd, Taipei 10617, Taiwan, R.O.C.}
\author[0000-0003-4099-6941]{Yoshihide Yamato}
\affiliation{Department of Astronomy, Graduate School of Science, The University of Tokyo, 7-3-1 Hongo, Bunkyo-ku, Tokyo 113-0033, Japan}
\author[0000-0003-3119-2087]{Jeong-Eun Lee}
\affiliation{Department of Physics and Astronomy, Seoul National University, 1 Gwanak-ro, Gwanak-gu, Seoul 08826, Korea}
\author[0000-0001-8105-8113]{Kengo Tomida}
\affiliation{Astronomical Institute, Graduate School of Science, Tohoku University, Sendai 980-8578, Japan}
\author[0000-0003-1412-893X]{Hsi-Wei Yen}
\affiliation{Academia Sinica Institute of Astronomy \& Astrophysics,
11F of Astronomy-Mathematics Building, AS/NTU, No.1, Sec. 4, Roosevelt Rd, Taipei 10617, Taiwan, R.O.C.}

\author[0000-0002-3566-6270]{Frankie J. Encalada} \affiliation{Department of Astronomy, University of Illinois, 1002 West Green St, Urbana, IL 61801, USA}
\author[0000-0002-8591-472X]{Christian Flores}
\affiliation{Academia Sinica Institute of Astronomy \& Astrophysics,
11F of Astronomy-Mathematics Building, AS/NTU, No.1, Sec. 4, Roosevelt Rd,
Taipei 10617, Taiwan, R.O.C.}
\author[0000-0001-5782-915X]{Sacha Gavino}
\affiliation{Niels Bohr Institute, University of Copenhagen, {\O}ster Voldgade 5-7, 1350, Copenhagen K, Denmark}
\author[0000-0002-2902-4239]{Miyu Kido}
\affiliation{Department of Physics and Astronomy, Graduate School of Science and Engineering, Kagoshima University, 1-21-35 Korimoto, Kagoshima,Kagoshima 890-0065, Japan}
\author[0000-0002-9143-1433]{Ilseung Han}
\affiliation{Division of Astronomy and Space Science, University of Science and Technology, 217 Gajeong-ro, Yuseong-gu, Daejeon 34113, Republic of Korea}
\affiliation{Korea Astronomy and Space Science Institute, 776 Daedeok-daero, Yuseong-gu, Daejeon 34055, Republic of Korea}
\author[0000-0001-7233-4171]{Zhe-Yu Daniel Lin}
\affiliation{University of Virginia, 530 McCormick Rd., Charlottesville, Virginia 22904, USA}
\author[0000-0002-0244-6650]{Suchitra Narayanan}
\affiliation{Institute for Astronomy, University of Hawai‘i at Mānoa, 2680 Woodlawn Dr., Honolulu, HI 96822, USA}
\author[0000-0002-4372-5509]{Nguyen Thi Phuong}
\affiliation{Korea Astronomy and Space Science Institute, 776 Daedeok-daero, Yuseong-gu, Daejeon 34055, Republic of Korea}
\affiliation{Department of Astrophysics, Vietnam National Space Center, Vietnam Academy of Science and Technology, 18 Hoang Quoc Viet, Cau Giay, Hanoi, Vietnam}
\author[0000-0001-6267-2820]{Alejandro Santamaría-Miranda}
\affiliation{European Southern Observatory, Alonso de Cordova 3107, Casilla 19, Vitacura, Santiago, Chile}
\author[0000-0003-0334-1583]{Travis J. Thieme}
\affiliation{Institute of Astronomy, National Tsing Hua University, No. 101, Section 2, Kuang-Fu Road, Hsinchu 30013, Taiwan}
\affiliation{Center for Informatics and Computation in Astronomy, National Tsing Hua University, No. 101, Section 2, Kuang-Fu Road, Hsinchu 30013, Taiwan}
\affiliation{Department of Physics, National Tsing Hua University, No. 101, Section 2, Kuang-Fu Road, Hsinchu 30013, Taiwan}
\author[0000-0002-2555-9869]{Merel L.R. van 't Hoff}
\affil{Department of Astronomy, University of Michigan, 1085 S. University Ave., Ann Arbor, MI 48109-1107, USA}

\author[0000-0003-4518-407X]{Itziar de Gregorio-Monsalvo}
\affiliation{European Southern Observatory, Alonso de Cordova 3107, Casilla 19, Vitacura, Santiago, Chile}
\author[0000-0003-2777-5861]{Patrick M. Koch}
\affiliation{Academia Sinica Institute of Astronomy \& Astrophysics,
11F of Astronomy-Mathematics Building, AS/NTU, No.1, Sec. 4, Roosevelt Rd,
Taipei 10617, Taiwan, R.O.C.}
\author[0000-0003-4022-4132]{Woojin Kwon}
\affiliation{Department of Earth Science Education, Seoul National University, 1 Gwanak-ro, Gwanak-gu, Seoul 08826, Republic of Korea}
\affiliation{SNU Astronomy Research Center, Seoul National University, 1 Gwanak-ro, Gwanak-gu, Seoul 08826, Republic of Korea}
\author[0000-0001-5522-486X]{Shih-Ping Lai}
\affiliation{Institute of Astronomy, National Tsing Hua University, No. 101, Section 2, Kuang-Fu Road, Hsinchu 30013, Taiwan}
\affiliation{Center for Informatics and Computation in Astronomy, National Tsing Hua University, No. 101, Section 2, Kuang-Fu Road, Hsinchu 30013, Taiwan}
\affiliation{Department of Physics, National Tsing Hua University, No. 101, Section 2, Kuang-Fu Road, Hsinchu 30013, Taiwan}
\affiliation{Academia Sinica Institute of Astronomy \& Astrophysics,
11F of Astronomy-Mathematics Building, AS/NTU, No.1, Sec. 4, Roosevelt Rd,
Taipei 10617, Taiwan, R.O.C.}
\author[0000-0002-3179-6334]{Chang Won Lee}
\affiliation{Division of Astronomy and Space Science, University of Science and Technology, 217 Gajeong-ro, Yuseong-gu, Daejeon 34113, Republic of Korea}
\affiliation{Korea Astronomy and Space Science Institute, 776 Daedeok-daero, Yuseong-gu, Daejeon 34055, Republic of Korea}
\author[0000-0002-9912-5705]{Adele Plunkett}
\affiliation{National Radio Astronomy Observatory, 520 Edgemont Rd., Charlottesville, VA 22903 USA}
\author[0000-0003-1549-6435]{Kazuya Saigo}
\affiliation{Department of Physics and Astronomy, Graduate School of Science and Engineering, Kagoshima University, 1-21-35 Korimoto, Kagoshima,Kagoshima 890-0065, Japan}

\author[0000-0002-4317-767X]{Shingo Hirano}
\affiliation{Department of Astronomy, Graduate School of Science, The University of Tokyo, 7-3-1 Hongo, Bunkyo-ku, Tokyo 113-0033, Japan}
\author[0000-0003-3581-1834]{Ka Ho Lam}
\affiliation{University of Virginia, 530 McCormick Rd., Charlottesville, Virginia 22903, USA}
\author[0000-0002-7002-939X]{Shoji Mori}
\affiliation{Astronomical Institute, Graduate School of Science, Tohoku University, Sendai 980-8578, Japan}











\begin{abstract}

We present an overview of the Large Program, ``Early Planet Formation in Embedded Disks (eDisk)'', conducted with the Atacama Large Millimeter/submillimeter Array (ALMA). 
The ubiquitous detections of substructures, particularly rings and gaps, in
protoplanetary disks around T Tauri stars raise the possibility that
at least some planet formation may have
already started during the embedded stages of star formation.
In order to address exactly how and when planet formation is initiated,
the program focuses on searching for substructures in disks around 12 Class 0 and 7 Class I protostars in nearby ($< $200 pc) star-forming regions
through 1.3~mm continuum observations at a resolution of $\sim7$~au (0\farcs{04}).
The initial results show that the continuum emission, mostly arising from dust disks around the sample protostars, has relatively few distinctive substructures, such as rings and spirals, in marked contrast to Class II disks. The dramatic difference may suggest that substructures quickly develop in disks when the systems evolve from protostars to Class II sources or alternatively that high optical depth of the continuum emission could obscure internal structures.
Kinematic information obtained through
CO isotopologue lines and other lines reveals the presence of Keplerian disks around protostars,
providing us with crucial physical parameters, in particular, the dynamical mass of the central protostars. 
We describe the background of the eDisk program, the sample selection and their ALMA observations, the data reduction, and also highlight representative first-look results.\\

\end{abstract}


\section{Introduction} \label{sec:intro}

Protoplanetary disks are ubiquitously found around young stellar objects (YSOs), particularly around Class II YSOs \citep[e.g.,][]{Mann_2014,2016ApJ...828...46A,2016ApJ...831..125P,2019MNRAS.482..698C,2019A&A...626A..11C,2019A&A...628A..85V,2021ApJ...913..123G}. Recent direct imaging of protoplanets in the disk around the T Tauri star PDS~70 \citep{2018A&A...617A..44K,2018A&A...617L...2M,2019NatAs...3..749H,2019Isella,2021Benisty,2022Casassus}  lends strong support to the notion that these disks are the sites of planet formation. 
The detection of rings and gaps in the HL Tau disk \citep{2015ApJ...808L...3A,2016Yen,2019Yen}
revolutionized the field of planet formation and made the use of disk substructures as signposts for
young planets a reality \citep[e.g.,][]{2015ApJ...809...93D,2018ApJ...869L..47Z}. Although it is still debated
whether such substructures are in fact always created by planets \citep[e.g.,][]{2015ApJ...806L...7Z,2015A&A...574A..68F}, they are important in their own right. For example, regions of density enhancement, such
as rings \citep[either complete or partial, e.g.,][]{2013Sci...340.1199V} and spirals \citep[e.g.,][]{2016Sci...353.1519P}, have a local pressure maximum that is conducive to the trapping of relatively large grains,
which, in turn, facilitates the formation of planetesimals and ultimately planets \citep{Youdin_2005,2007Natur.448.1022J,Johansen_2009}.

Given their fundamental importance, there have been a number of ALMA programs aiming at
detecting and characterizing disk substructures \citep[e.g.,][]{2016ApJ...818..158A, 2016ApJ...820L..40A,2018ApJ...869...17L}. The Cycle 4 Large Program, Disk Substructures at High Angular Resolution Project \citep[DSHARP; see][]{2018ApJ...869L..41A}, surveyed 20 nearby Class
II disks in dust emission at 1.25 mm at an angular resolution of 0\farcs{04} and showed that substructures
are ubiquitous, mostly in the form of concentric rings and gaps. 
Similar high-resolution observations of Class II disks were also performed for 7 sources in Ophiuchus at resolutions of 3-5 au as a part of the Ophiuchus DIsc Survey Employing ALMA (ODISEA) project \citep{2021MNRAS.501.2934C}, finding ring/gap structures in all of them. These surveys strongly suggest that the important disk substructures are well developed in the Class II stage of
star formation. 
This raises the obvious question: \textit{when do substructures start to develop in disks?}


If planets are indeed responsible for producing the rings and gaps in Class II disks, planet formation
could have started earlier, i.e., during the Class I or even Class 0 stages. Here, we consider that the dust growth forming millimeter- or centimeter-sized dust grains at least should occur prior to the Class II stage.  The general lack of sufficient
solids in Class II disks to form giant planets also points to an earlier epoch of planet formation \citep{2018A&A...618L...3M,2018ApJS..238...19T,2018AJ....155...54W}. Theoretical studies have suggested that Class 0/I
disks are more plausible sites for planetesimal formation than Class II disks because of their large
amount of gas and dust \citep{2017ApJ...838..151T}. A systematic survey focused on protostellar disks
can directly test this possibility by searching for structural signatures of early planet formation. For
example, if such early planets are formed through core accretion \citep[e.g.,][]{2014ApJ...789...69H}, dust traps
that facilitate the formation of planetesimals and planetary cores should be seen. Alternatively, if
planets formed through gravitational instabilities \citep[e.g.,][]{1997Sci...276.1836B}, prominent spiral arms are expected \citep[e.g.,][]{2017ApJ...835L..11T}
.

In order to investigate the possibility of early planet formation, it is obviously necessary for us to investigate disks around Class 0/I protostars. However, observations of Class 0/I disks are more difficult than those of Class II disks because they are still deeply embedded in envelopes. In particular, the inner parts of envelopes within a few 1000~au often show flattened structures and velocity gradients and it is not straightforward to distinguish embedded disks from these structures. Careful analysis, including their kinematics, is required to identify rotationally-supported embedded disks.

In this paper, we present the ALMA Large Program ``Early Planet Formation in Embedded Disks (eDisk)'', in which 19 Class 0/I embedded protostars in nearby star-forming regions are systematically surveyed at resolutions of $\sim5$~au (0\farcs{04}) for the first time. The main scientific objective of the eDisk program is to investigate whether or not substructures exist in disks around embedded, Class 0/I, protostars that might be indicative of early planet formation. This, of course, also involves the important task of securely identifying the nature of the embedded disks, e.g., by searching for the kinematical signatures of rotationally supported (Keplerian) disks through molecular line observations. In turn, this will provide fundamental (global) parameters of the young stellar objects, such as the disk radii and dynamical stellar masses based on the Keplerian motions.
The first homogeneous samples of embedded disks obtained in the eDisk program can be directly compared with the corresponding samples of Class II disks, enabling us to shed light on the disk evolution from the Class 0/I to Class II phases. 

The paper is organized as follows: an overview of
previous studies of embedded disks around protostars is provided in Sect.~\ref{sec:intro-sub}.
Section~\ref{sec:obs} describes the selection of the targeted sample of sources and the overall observing strategy, while the data reduction procedures are presented in Section \ref{sec:reduction}. In Sect.~\ref{sec:result}, the first results of the program are presented with emphasis on observations of the 1.3~mm continuum of the sample of sources as well as an example of the C$^{18}$O 2-1 emission toward one, R~CrA~IRS7B. These parts of the continuum and line observations are key to achieving the main objectives mentioned above. In Sect.~\ref{sec:dis}, we briefly discuss implications of the first results from the survey, and brief descriptions of the data release from the eDisk program are provided in Sect.~\ref{sec:data release} Finally, a summary of the paper is provided in Sect.~\ref{sec:summary}. 

\section{Previous studies of embedded disks around protostars} \label{sec:intro-sub}

Even though disks around embedded protostars are expected theoretically to be a natural consequence of the collapse of a rotating dense molecular core \citep[e.g.,][]{1984ApJ...286..529T} their firm observational detections are challenging as such disks are still deeply embedded in the envelopes surrounding them.
Envelopes often show flattened structures with typical sizes of a few thousand au, which is roughly 10-50 times larger than typical disks, and also have rotating as well as infalling motions \citep[e.g.,][]{1997ApJ...475..211O,1998ApJ...504..314M}. High resolution and sensitivity are therefore required to carefully disentangle emission from disks and envelopes.

The confusion between the envelope and disk emission also makes it difficult to unambiguously address the disk sizes from the continuum emission by themselves and thus whether significant disks may develop early in protostellar evolution. However, some attempts suggest that very large ($>$~100~au) disks are rare around deeply embedded protostars \citep{2019A&A...621A..76M}. 

\subsection{Kinematic identification of embedded disks}
One approach to distinguish embedded disks and their envelopes is to utilize their kinematic difference; embedded disks are expected to have Keplerian rotation similarly to Class II disks, while envelopes are expected to have rotation conserving their specific angular momenta. As a result, rotational velocities of embedded disks are proportional to $r^{-0.5}$, while those of envelopes are proportional to $r^{-1}$, where $r$ is the cylindrical radius from the rotational axis \citep[e.g.,][]{1997ApJ...488..317O,2000ApJ...545.1034S, 2013EAS....62...25B}

This approach was taken to distinguish the embedded disk and its surrounding envelope around the protostar HH111 \citep{2011ApJ...741...62L} and also those around 6 other protostars \citep{2013ApJ...772...22Y} in C$^{18}$O 2-1 with the Submillimeter Array (SMA) at angular resolution of $\sim$1$\arcsec$-4$\arcsec$. 
In some cases (L1527 IRS and L1448-mm), only rotation proportional to $r^{-1}$ was found because of insufficient angular resolution. However, L1527 IRS was also observed with the Combined Array for Millimeter-wave Astronomy (CARMA) at an angular resolution of $\sim1\arcsec$ in $^{13}$CO 2-1, displaying Keplerian rotation instead of that proportional to $r^{-1}$ \citep{2012Natur.492...83T}. 

Further observations following up these pioneering works were enabled by ALMA with its high angular resolution and sensitivity.  L1527 IRS was observed again as a benchmark at sub-arcsecond resolutions in C$^{18}$O 2-1, clearly demonstrating that its infalling envelope transforms into a Keperian disk at a radius of $\sim$70~au \citep{2014ApJ...796..131O,2017ApJ...849...56A}. Similar findings have been found in other low-mass Class 0/I protostars \citep[e.g.,][]{2014ApJ...793....1Y,2017ApJ...834..178Y,2015ApJ...812...27A,2016Natur.538..483T,2018NatAs...2..646H,2020ApJ...893...51S,2020A&A...635A..15M}. C$^{18}$O has often been adopted to investigate kinematics of envelopes and disks, as well as other more rare isotopes, such as C$^{17}$O or C$^{34}$S \citep[e.g.,][]{2019A&A...626A..71A}. Important products from detected Keplerian motions are dynamical masses of protostars and mass accretion rates. Notably, direct measurements of masses for optically invisible protostars are usually difficult because their stellar photospheres are not observable. These physical parameters could shed light on the formation and evolution of protostars \citep[e.g.,][]{2017ApJ...834..178Y,2019A&A...626A..71A} based on more systematic observations of embedded disks.

\subsection{Chemical identification of embedded disks}
The presence of disks around deeply embedded protostars may also affect their chemical structures and consequently, the molecular line emission observed toward them. For example, such embedded disks have been suggested to be associated with distinct features such as slow accretion shocks that in turn may reveal the presence of the disks \citep[e.g.,][]{2014Natur.507...78S,2016ApJ...820L..34S,2016ApJ...824...88O,2018ApJ...864L..25O}. Also, grain growth to millimeter- or centimeter-sized grains in disks may cause the continuum emission to become optically thick affecting the line emission signatures observed from the embedded disks \citep[e.g.,][]{2018NatAs...2..646H}. 

We should, however, note that consistent chemical indicators across larger samples of disks are still lacking. For example, it has been suggested that SO is enhanced at the disk-envelope surface due to slow accretion shocks in L1527 IRS \citep{2014Natur.507...78S}, while in L483 and Elias 29 SO appears to trace
the Keplerian rotating disks themselves \citep{Oya_2017,Oya_2019}.
\citet{Oya_2017} also suggested that the emission from CS and complex organic molecules are arising from the disk around L483, while \citet{2019A&A...629A..29J} demonstrated that those species show velocity structures that can be explained better by infalling and rotating material rather than a Keplerian disk.
These observations indicate that the molecules tracing the disks may vary among objects, depending on their physical properties (e.g. stellar luminosities) and chemical history in prestellar stage \citep[e.g.,][]{Oya_2019}. 

\subsection{Substructures of embedded disks}
Although embedded disks have been successfully observed around a number of protostars in the past decade, only a few have shown detectable substructures suggestive of ongoing planet formation to date because higher angular resolutions of 0$\farcs{1}$ or less are required to look for substructures in embedded disks.
\citet{2017Sheehan} have found a central cavity and ring toward the Class I protostar WL 17, in 3 mm continuum at an angular resolution of 0$\farcs$06. \citet{2018Sheehan} have also found concentric rings/gaps within the disk of the Class~I protostar GY~91, also known as ISO-Oph 54 \citep[see][]{2021MNRAS.501.2934C} in the 0.9 mm continuum emission at an angular resolution of 0\farcs{13}. More recently, \citet{2020ApJ...902..141S} presented seven protostars including two Class 0 sources from the Orion Very Large Array/ALMA Nascent Disk and Multiplicity (VANDAM) survey that exhibit ring/gap substructures in their disks, although detailed modeling showed an offset in the inner disk for a few sources, which may suggest that the substructures could be due to close binary formation.
In addition, \citet{2020Natur.586..228S} have found four annular structures in the Class I protostar Oph IRS~63 in 1.3~mm continuum at an angular resolution of $\sim$0$\farcs$04 although their contrast is relatively low. 

In addition to ring structures, spiral structures have also been observed in embedded disks.
\citet{2020NatAs...4..142L} have found a pair of symmetric spiral structures in the disk around the Class I protostar HH111 VLA1 at a resolution of 0\farcs{04} (16~au). The disk is relatively massive (0.33-0.5~$M_{\sun}$), with Toomre's Q parameter near unity in the outer part of the disk, where the spiral structures are detected, supporting the idea that mass accretion from the envelope is driving the outer disk to be gravitationally unstable. Obviously, it is definitely crucial to observe more embedded sources at angular resolutions higher than 0$\farcs$1 to search for substructures, particularly around Class 0 protostars, which show less substructures in their disks to date.  


\section{Description of the eDisk program} \label{sec:obs}
The foundation of eDisk is the ALMA large program (2019.1.00261.L: PI N. Ohashi) with supplemental data obtained through ALMA DDT observations (2019.A.00034.S: PI J. Tobin). We also used some ALMA archival data. In this section we present the sample (\S\ref{subsec:sample}) and the details of the observations (\S\ref{subsec:obs}). 

\subsection{Sample} \label{subsec:sample}

For the eDisk program we focus on sources located in nearby ($d<$ 200 pc) star forming regions that are relatively bright ($L_{\rm bol} > 0.1~\Ls$). The first criterion is required to spatially resolve observed disks at a given angular resolution ($\sim$0\farcs{04}; see below). The second criterion is required to detect disks around sample protostars within a reasonable observation time. We checked how many such protostars there are based on available catalogs of young stellar objects (YSOs): the largest catalogs of YSOs in nearby star-forming regions were provided through large survey observations at mid- and far-infrared wavelengths using the Spitzer Space Telescope. \citet{2010ApJS..186..259R} counted the number of Class 0/I protostars in Taurus based on the near- to mid-IR slopes, $\alpha$ obtained in the Spitzer observations, finding that there are 21 Class 0/I protostars with $L_{\rm bol} > 0.1~\Ls$ \footnote{Since $T_{\rm bol}$ and $L_{\rm bol}$ were not provided for these sources in \citet{2010ApJS..186..259R}, we calculated them based on 2MASS, Spitzer IRAC and MIPS, Herschel data, except for one source, for which IR photometry data from \citet{2012AJ....144...31K} was used instead of Spitzer data.}, excluding potential Taurus members newly identified. \citet{2015ApJS..220...11D} also counted the number of Class 0/I protostars based on Spitzer observations and found 37 Class 0/I protostars in the regions of Chamaeleon, Corona Australis, Lupus, Musca, and Ophiuchus. Note that two of the 37 sources show unreasonably high $T_{\rm bol}$ ($>$ 3000~K) for embedded protostars. Based on these previous works, there would be $\sim$60 protostars within a distance of 200~pc and with luminosities more than 0.1~$\Ls$. Note that protostars in isolated star-forming regions are not included here. 

Within a reasonable observation time, it is impossible for us to observe all these protostars from a data volume point of view as well as the restrictions of the maximum time allocated to large program in a given LST range. In the Cycle 7 semester 645 hours were available in total for Large Programs, of which we aimed for a required observation time for the eDisk program of $\sim$100~hours. According to our previous ALMA observations of embedded disks around protostars and also our simulations of observations, we set a target sensitivity for molecular line observations at $\sim$2.7~mJy~beam$^{-1}$ with a velocity resolution of $\sim$0.17~km~s$^{-1}$. This sensitivity was able to be achieved with an on-source integration time of $\sim$2.4~hours, requiring $\sim$6~hours observation time, including overhead. Note that this observation time can provide us with a sensitivity of $\sim$13 $\mu$Jy~beam$^{-1}$ for the continuum emission, which is sufficiently high to detect continuum emission around protostars. With these estimations of the required observation time, we found that $\sim$17 sources could be observed within $\sim$100 hours of observation time.

When 17 sources were selected we tried to avoid any bias as much as possible; we selected 17 ``representative'' protostars, which have been relatively well studied before, are located in differnet star-forming regions (i.e., Chamaeleon, Corona Australis, Lupus, Ophiuchus, Taurus, and also isolated Bok globules), and present a wide distribution of the bolometric luminosities ($L_{\rm bol}$) and temperatures ($T_{\rm bol}$). Selecting sources spreading out over various star-forming regions can also help to carry out our observations more efficiently within a limited period of the observations. In addition to these 17 sources, two sources, B335 and TMC-1A, for which necessary data has been already obtained with ALMA before, were added to the sample. The 19 sample protostellar systems selected for the eDisk program are listed in Table \ref{tab:sample}.

\begin{figure*}[ht!]
\epsscale{1.15}
\plotone{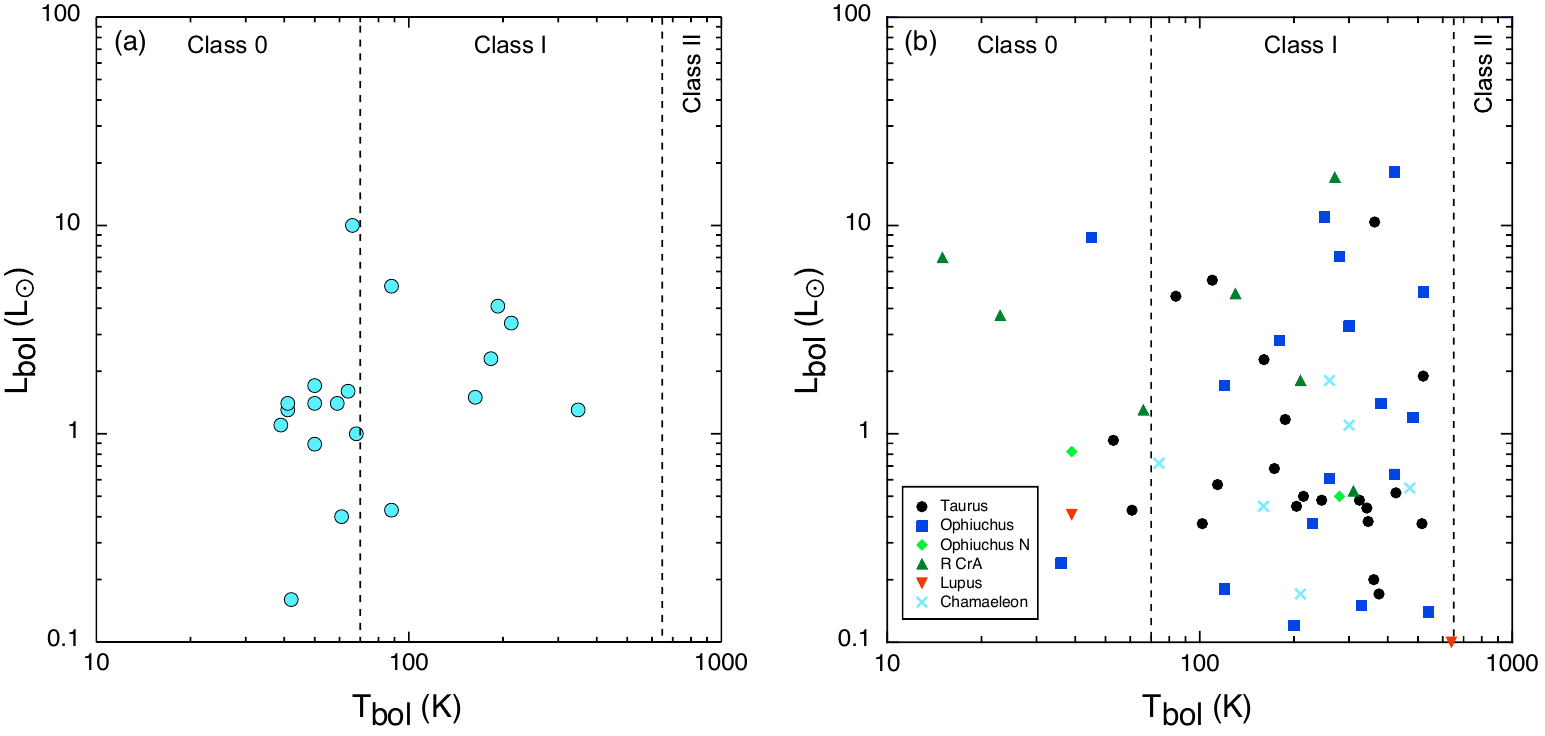}
\caption{(a) $T_{\rm bol}$-$L_{\rm bol}$ diagram of the eDisk sample sources. (b) $T_{\rm bol}$-$L_{\rm bol}$ diagram of the protostars with $d<200~pc$ and $L_{\rm bol}>0.1~\Ls$. Two dashed vertical lines in each diagram indicate the boundaries between Class 0 and Class I \citep[70~K; see][]{2009ApJS..181..321E} and Class I and Class II \citep[650~K; see][]{2009ApJS..181..321E}.  \label{fig:sample}}
\end{figure*}

\begin{deluxetable*}{llllcclDl}
\tablecaption{eDisk sample} \label{tab:sample}
\tablewidth{0pt}
\tablehead{
\colhead{Source name} & \colhead{Alternative name} & \colhead{ICRS R.A.} & \colhead{ICRS Dec.} & \colhead{Class} &
\colhead{Distance} & \colhead{$T_{\rm bol}$} & \multicolumn2c{$L_{\rm bol}$} & \colhead{Ref.} \\
\colhead{} & \colhead{} & \colhead{(h m s)} & \colhead{(d m s)} & \colhead{} &
\colhead{(pc)} & \colhead{(K)} & \multicolumn2c{($L_\odot$)} & \colhead{}
}
\decimalcolnumbers
\startdata
L1489 IRS & IRAS 04016+2610 & 04:04:42.85 & +26:18:56.3 & I & 146 & 213 & 3.4 & 1 \\
IRAS 04166+2706 & & 04:19:42.50 & +27:13:36.0 & 0 & 156 & 61 & 0.40 & 2 \\
IRAS 04169+2702 & & 04:19:58.45 & +27:09:57.1 & I & 156 & 163 & 1.5 & 2 \\
IRAS 04302+2247 & Butterfly star & 04:33:16.50 & +22:53:20.4 & I & 160 & 88 & 0.43 & 1,3,4\\
L1527 IRS & IRAS 04368+2557 & 04:39:53.91 & +26:03:09.8 & 0 & 140 & 41 & 1.3 & 5,6 \\
Ced110 IRS4 & IRAS 11051-7706 & 11:06:46.44 & --77:22:32.2 & 0 & 189 & 68 & 1.0 & 7 \\
BHR71 IRS2 & & 12:01:34.09 & --65:08:47.36 & 0 & 176 & 39 & 1.1 & 8 \\
BHR71 IRS1 & IRAS 11590–6452 & 12:01:36.81 & --65:08:49.22 & 0 & 176 & 66 & 10 & 8 \\
IRAS 15398-3359 & B228 & 15:43:02.24 & --34:09:06.8 & 0 & 155 & 50 & 1.4 & 9,10,11 \\
GSS30 IRS3 & SMM J162622-24225 & 16:26:21.72 & --24:22:50.7 & 0 & 138 & 50 & 1.7 & 12 \\
Oph IRS43 & YLW 15 & 16:27:26.7 & --24:40:52.3 & I & 137 & 193 & 4.1 & 13 \\
IRAS 16253-2429 & & 16:28:21.6 & --24:36:23.4 & 0 & 139 & 42 & 0.16 & 10,14 \\
Oph IRS63 & IRAS 16285-2355 & 16:31:35.7 & --24:01:29.6 & I & 132 & 348 & 1.3 & 10,12 \\
IRAS 16544--1604 & CB 68 & 16:57:19.64 & --16:09:23.9 & 0 & 151 & 50 & 0.89 & 10 \\
R CrA IRS5N & CrA-20 & 19:01:48.46 & --36:57:14.7 & 0 & 147 & 59 & 1.4 & 10 \\
R CrA IRS7B & SMM 1B & 19:01:56.42 & --36:57:28.4 & I & 152 & 88 & 5.1 & 9 \\
R CrA IRAS 32 & IRAS 18595-3712 & 19:02:58.67 & --37:07:35.9 & 0 & 150 & 64 & 1.6 & 9 \\
\hline
TMC-1A$^{\dagger}$ & IRAS 04365+2535 & 04:39:35.2 & +25:41:44.4 & I & 137 & 183 & 2.3 & 6 \\
B335$^{\dagger}$ & IRAS 19347+0727 & 19:37:0.89 & +07:34:10.0 & 0 & 165 & 41 & 1.4 & 15 \\
\enddata
\tablecomments{$^{\dagger}$ For these sources, we used only ALMA archival data without new observations. Some of the archival data we used have slightly different phase centers from those shown in this table (see Table \ref{tab:archive-logs}). Column 1: Target name. Column 2: Alternative common name also used in previous works. Column 3: Right ascension of the observation phase center in the International Celestial Reference System (ICRS). Column 4. Declination of the observation phase center in ICRS. Column 5. Evolutionary stage classification based on $T_{\rm bol}$. Column 6. Distance to the target based on recent estimations from Gaia measurements. Column 7: Bolometric temperature re-derived from the newly compiled SED. See Appendix~\ref{sec:app-SEDs}. Column 8: Bolometric luminosity re-derived from the newly compiled SED. See Appendix~\ref{sec:app-SEDs}. Column 9: Reference for the distance estimations.}
\tablerefs{1. \citet{2020AA...638A..85R}; 2. \citet{2021AJ....162..110K}; 3. \citet{2018ApJ...859...33G}; 4. \citet{2018AJ....156..271L}; 
5. \citet{1994AJ....108.1872K};
6. \citet{2019ApJ...879..125Z};
7. \citet{2021AA...646A..46G};
8. \citet{2018AA...610A..64V}; 
9. \citet{2020AA...643A.148G}; 
10. \citet{2020AA...633A..51Z}; 
11. \citet{2018ApJ...867..151D}; 
12. \citet{2018ApJ...869L..33O}; 
13. \citet{2017ApJ...834..141O};
14. \citet{1989AA...216...44D};
15. \citet{2020RNAAS...4...88W}
}
\end{deluxetable*}


All of the selected sources have complete spectral energy distributions (SEDs) measured from the near-infrared to submillimeter that are compiled in Appendix~\ref{sec:app-SEDs}. Based on those SEDs we re-derived their bolometric temperatures ($T_{\rm bol}$) and luminosities ($L_{\rm bol}$) also utilising the most recent distances from Gaia measurements listed in Table~\ref{tab:sample}. Note that none of the 19 targets were detected with Gaia directly due to their embedded natures, but distances to nearby sources, presumably related to the same star-forming complexes, were typically estimated based on Gaia observations and provided the distance estimates quoted in literature. The final sample comprises 12 Class 0 and 7 Class I protostars and represents various evolutionary stages of protostars with a roughly uniform distribution of $L_{\rm bol}$ and $T_{\rm bol}$ (Figure \ref{fig:sample}a). For comparison purposes, Figure \ref{fig:sample}b shows the same $T_{\rm bol}$-$L_{\rm bol}$ diagram, but for the 56 protostars with $d<$200~pc and $L_{\rm bol}>0.1~\Ls$ discussed above, except for two sources with unreasonably high $T_{\rm bol}$, indicating that the final 19 protostars reasonably represent the nearby and relatively bright protostars.

\subsection{Observations} \label{subsec:obs}
Observations of the eDisk program using the ALMA 12 m array in band 6 were conducted from April 2021 to July 2022. 
We aimed to achieve high angular resolutions of $\sim$0\farcs{04}, corresponding to $\sim$6~au
at a distance of 140~pc. Such high angular resolution is necessary for us to investigate substructures of embedded disks and make direct comparisons with previous maps of Class II disks obtained at similar angular resolutions, such as those obtained by DSHARP or ODISEA. 

The correlator was configured to observe both 1.3~mm continuum emission as well as CO ($J=2-1$) isotopologues and other lines providing additional information about the physical and chemical structures of the gas on the scales of the circumstellar disks surrounding the embedded protostars. 
Baseband 1 of the correlator was configured in the lower sideband with 4 spectral windows (spw) to include our primary molecular lines, C$^{18}$O (2--1), $^{13}$CO (2--1), and SO ($6_5-5_4$), at a velocity resolution of $\sim$0.17~km~s$^{-1}$. The $^{12}$CO (2--1) emission was included in baseband 4 in the opposite sideband to study protostellar outflows. The bandwidth of baseband 4 was set to be 937.5 MHz to secure the continuum bandwidth, and as a result, the velocity resolution of the $^{12}$CO data is 4 times worse than that of the C$^{18}$O data. The remaining two basebands are located in the upper and lower sidebands, respectively, with the maximum bandwidth of 1875 MHz. While these two basebands were adopted for the 1.3~mm continuum emission, the central frequency of baseband 2 was adjusted to incorporate as many important molecular lines as possible, such as CH$_3$OH, SiO, DCN, H$_2$CO, and $c$-C$_3$H$_2$. Even though the velocity resolution of these observations, $\sim$1.3~km~s$^{-1}$, is not ideal to study the kinematics from these molecular transitions, their detection can still provide useful constraints on circumstellar structures based on, e.g. the integrated intensities. 
All the major emission lines observed in our observations are listed in Table \ref{tab:obsline}.

In order to achieve an angular resolution of $\sim$0\farcs{04}, the C43-8 antenna configuration was adopted for our observations, providing long baselines up to $\sim$12.6~km.
In addition to the C43-8 antenna configuration, the C43-5 antenna configuration was also adopted to cover shorter baselines with a minimum of $\sim$15~m.
This additional antenna configuration covering shorter baselines is crucial for us to investigate relatively extended structures up to $\sim 2\arcsec - 3\arcsec$ in the envelopes
surrounding our sampled protostars. We should note that a large fraction of the emission from the envelopes around the sampled protostars is still filtered out in our observations. However, this missing flux has less impact on the main objectives of the eDisk program focusing on dust and gas on the scales of the disks in the vicinity of the protostars. The Field of View (FoV) of the 12 m array at the observed frequency is $\sim 26\arcsec$.
Logs of the observations are provided in Appendix \ref{sec:app-obslog} (see Table~\ref{tab:logs}).

For two sources in our sample, TMC-1A and B335, comparable observations were obtained with ALMA prior to our eDisk program and we collected those data from the ALMA archive.  
In addition, due to scheduling constraints only short baseline data were obtained for Oph IRS~63 as a part of eDisk. Hence we relied on ALMA archival observations of Oph IRS~63 for long baseline data. The information on the ALMA archival data we use in the eDisk program is summarized in Appendix \ref{sec:app-obslog} (see Table~\ref{tab:archive-logs}).

\begin{deluxetable*}{ccccl}
\tablecaption{Main molecular lines covered in the spectral setups for the eDisk program\label{tab:obsline}}
\tablewidth{0pt}
\tablehead{
\colhead{Rest frequency} & \colhead{Molecule} & \colhead{Transition} & \colhead{Velocity resolution} & \colhead{Note} \\
\colhead{(GHz)} & \colhead{} & \colhead{} & \colhead{(km~s$^{-1}$)} & \colhead{}
}
\decimalcolnumbers
\startdata
\multicolumn{5}{c}{Baseband 2 (216.8125--218.6875 GHz)} \\
\tableline
217.10498 & SiO & $5-4$ & 1.35 & outflows, shocks \\ 
217.23860 & DCN & $3-2$ & 1.35 & envelopes (cold gas) \\
217.82204$^{\dagger}$ & $c$-C$_3$H$_2$ & $6_{0,6}-5_{1,5}$ & 1.34 & envelopes (UV irradiated regions) \\
217.82215$^{\dagger}$ & $c$-C$_3$H$_2$ & $6_{1,6}-5_{0,5}$ & 1.34 & envelopes (UV irradiated regions) \\
217.94005 & $c$-C$_3$H$_2$ & $5_{1,4}-4_{2,3}$ & 1.34 & envelopes (UV irradiated regions) \\
218.16044 & $c$-C$_3$H$_2$ & $5_{2,4}-4_{1,3}$ & 1.34 & envelopes (UV irradiated regions) \\
218.22219 & H$_2$CO & $3_{0,3}-2_{0,2}$ & 1.34 & high density gas, temperature tracer\\
218.44006 & CH$_3$OH & $4_2-3_1$ $E$ & 1.34 & envelopes, outflows \\
218.47563 & H$_2$CO & $3_{2,2}-2_{2,1}$ & 1.34 & high density gas, temperature tracer\\
\tableline
\multicolumn{5}{c}{Baseband 1, spw 0 (218.730773--218.789367 GHz)} \rule[0pt]{0pt}{20pt}\\
\tableline
218.76007 & H$_2$CO & $3_{2,1}-2_{2,0}$ & 0.17 & high density gas, temperature tracer\\
\tableline
\multicolumn{5}{c}{Baseband 1, spw 3 (219.531063--219.589657 GHz)$^{\ddagger}$} \rule[0pt]{0pt}{20pt} \\
\tableline
219.56035 & C$^{18}$O & 2--1 & 0.17 & envelopes, disks \\
\tableline
\multicolumn{5}{c}{Baseband 1, spw 2 (219.920143--219.978737 GHz)} \rule[0pt]{0pt}{20pt} \\
\tableline
219.94944 & SO & $6_5-5_4$ & 0.17 & high density gas, shocks \\
\tableline
\multicolumn{5}{c}{Baseband 1, spw 1 (220.350703--220.409297 GHz)$^{\ddagger}$} \rule[0pt]{0pt}{20pt} \\
\tableline
220.39868 & $^{13}$CO & 2--1 & 0.17 & disks, envelopes, outflows \\
\tableline
\multicolumn{5}{c}{Baseband 4 (230.09125--231.02875 GHz)} \rule[0pt]{0pt}{20pt} \\
\tableline
230.53800 & $^{12}$CO & 2--1 & 0.63 & outflows, envelopes, disks \\
\tableline
\multicolumn{5}{c}{Baseband 3 (233.3125--235.1875 GHz)} \rule[0pt]{0pt}{20pt} \\
\tableline
\multicolumn{5}{c}{Mostly for the 1.3~mm continuum emission} \\
\enddata
\tablecomments{$\dagger$ These two lines are blended. $\ddagger$ In the DDT observations, spw 1 and spw 3 in the baseband 1 were swapped, and the frequency range of the spw 3 in the DDT observations was set at 220.369383--220.427977~GHz.}
\end{deluxetable*}

\section{Data reduction} \label{sec:reduction}
\subsection{Detailed procedure} 
The eDisk data were all calibrated using the ALMA calibration pipeline packaged with the Common Astronomy Software Application (CASA) \citep{2007ASPC..376..127M} version 6.2.1 and pipeline
version 2021.2.0.128. After delivery, the data were restored to a calibrated state running \textit{scriptForPI.py} in the same CASA version that the calibration pipeline ran with. For
four sources (BHR71 IRS1/IRS2, IRAS 04166+2706, and IRAS 04169+2702), there were Execution Blocks (EBs) that were given a QA0 status of `semi-pass' (see Table \ref{tab:logs} in Appendix \ref{sec:app-obslog}), which in the case of these datasets was because of excessive phase decorrelation during the observations. For those we downloaded the raw data and ran the ALMA calibration pipeline manually to inspect the datasets and determine if they could be salvaged using self-calibration. The QA0 semi-pass data that are used in the final images are denoted in the Observing Logs included in Appendix~\ref{sec:app-obslog}.

Outside of standard calibration, we needed to perform additional processing to improve the S/N of the continuum data through self-calibration, to remove the continuum emission from the line data, and to combine data taken in two configurations in order to 
achieve the balance between flux recovery and angular resolution needed to achieve the science goals of the project. Nearly all of the eDisk targets were sufficiently bright that
images created using the standard ALMA calibration were dynamic range limited, meaning that the images were not reaching the thermal noise limit, and were instead
limited by phase and/or amplitude variations during the science target observations. To both combine the two configurations, self-calibrate, continuum subtract, and image
the data, we constructed a set of semi-interactive scripts for each target. Our methodology was based on the DSHARP data reduction procedure \citep{2018ApJ...869L..41A} with 
substantial modification. One particular choice we made was to operate on each measurement set individually for calculation and application of phase and amplitude gain solutions and not concatenate our data to a single measurement set.
This avoids issues that can be encountered when operating on concatenated measurement sets within various CASA tasks; for example, interpolation of gain solutions between distinct observations that might be far apart in time, spectral window numbers being offset, and possible issues with combined metadata to name a few. Many of the issues with concatenated measurement sets can be worked around, but we avoid the known issues and the possible unknown issues by operating on individual measurement sets. The scripts used for each source are provided at http://github.com/jjtobin/edisk.

The reduction procedure starts with the construction of a spectrally-averaged continuum dataset. We identified line-free regions of the individual spectral windows to include in the continuum dataset in two ways: first, we made use of the \textit{cont.dat} file produced by the ALMA imaging pipeline, and edits to the continuum ranges in that file were made if necessary, based on inspection of the \textit{hif\_findcont} weblog. Second, we manually flagged the spectral lines that are expected to be present in all datasets regardless of whether they were already included in the \textit{cont.dat} file. We then translated these frequency ranges defined in the kinematic local standard of rest frame (LSRK) to the channel ranges for each EB and flagged those channels. Then the separate spws from each EB were spectrally averaged individually. The four high-resolution spws were averaged over 480 channels and the three low-resolution spws were averaged 60 channels, resulting in $\sim$30 MHz channels, approximating the spectral resolution of the low-spectral resolution (time domain mode; TDM) of the ALMA correlator.

We next checked the relative flux density scaling of each individual EB using their azimuthally-averaged visibility amplitude data. To do this, we first imaged each EB individually and shifted
the brightest source in the image to the phase center using the CASA task \textit{fixvis}. We then computed the azimuthally-averaged visibility amplitudes for each EB, plotted them together, and computed the average scale factors between each EB. The scale factors were estimated between the minimum overlapping \textit{uv}-distance out to 800~k$\lambda$; this outer \textit{uv}-distance was chosen to avoid a possibility that decorrelation at longer baselines could significantly impact the scale factors. The offsets were typically at the 10--15\% level between EBs and we scaled each EB to the `best' EB which was judged by having the least amount of scatter in its data. Thus, our goal was a relative flux normalization of all the EBs, given that we were not improving on the overall flux density scale accuracy. We also note that there is possibility that some scale factors include intrinsic source variability in addition to calibration differences.
While the ALMA Technical Handbook indicates that the accuracy of the flux density scale is expected to be $\sim$5\%, the scaling that we have had to apply to datasets indicates that the flux density scale can have uncertainties up to the rescaling factors that we apply to the data.\footnote{There is the Jorsater \& van Moorsel effect in clean (see Section \ref{sec:JvM}) that may also add uncertainties to the measured flux densities in the cleaned images}. The scale factor applied to each EB is provided in Appendix~\ref{sec:app-scale-factor}.

There were some cases where this flux normalization could not be performed at this stage due to decorrelation, which caused the visibility amplitude profile of one or more EBs to have a different behavior as a function of \textit{uv}-distance than others. In those cases, we did not determine scale factors at this stage, but we completed the self-calibration steps that followed and determined the scale factors after self-calibration. We then re-ran the whole procedure from the start, applying these scale factors \textit{a priori}; we referred to this as the `two-pass' method. The `two-pass' method was applied to Oph IRS43, IRAS 04166+2706, IRAS 04169+2702, and BHR71 IRS1 and IRS2. The relative scaling of each EB was then applied to the continuum datasets that did not have the brightest source shifted to the phase center because the unshifted data were what we ultimately used in the self-calibration process.

Following flux normalization of each EB we began the self-calibration process.
For the combined processing of the compact and extended configuration data, we first self-calibrated the compact configuration data and then we self-calibrated the pre-self-calibrated compact configuration data with the extended configuration data. Self-calibration was performed using the data that did not have the phase center shifted. It was not necessary to align our data prior to self-calibration because the time difference between observations was less than 1 year and all the long baseline data, where proper motion would have the greatest effect, were all taken within a 3 month time period. We create our model for self-calibration using the data from all the measurement sets; however, it is only the calculation and application of the phase and amplitude gain solutions that happen on a per-measurement set basis.

To prepare for self-calibration, we first calculated a ranked list of reference antennas for each EB, equally weighing the amount of flagging on a particular antenna and its geometric position within the array. Following this step, we made an initial image using all the continuum datasets to determine the peak S/N of the image from standard calibration; the creation of all images during the self-calibration process made use of the \textit{tclean} \textit{auto-multithresh} algorithm \citep{2020PASP..132b4505K} to automatically create clean masks for deconvolution. Following the generation of this initial image, we analyzed the scan lengths within an observation and determined calibration solution intervals that most evenly divide a scan by the number of integrations per scan. This approach will avoid the uneven division of scans which can result in some solutions being calculated from significantly less data. The list of solution intervals to attempt always includes a solution that spans the length of an EB, referred to as `inf\_EB' where
\textit{gaincal} is run with the parameter \textit{combine=`scan,spw'}. The `inf\_EB' solution interval corrects systematic effects, in particular, antenna position offsets that translate to systematic phase offsets between the calibrator and science target. Then for shorter solution intervals, we are correcting the atmospheric phase variations. We include an `inf' solution, where the `scan' setting is removed from the \textit{gaincal} parameters and solutions break at scan boundaries, then for shorter solutions, we divide the median scan length by 2 and all subsequent solution intervals divide the previous interval by 2 until the final solution interval of `int' (one integration) is reached.

Then for each solution that is to be attempted, we also determine the clean depths in terms N$\times$ $\sigma$ to use for each solution interval in order to progressively clean deeper during each round of self-calibration. We begin with the assumption that we would trust features with intensities $>$ (Peak S/N)/15, which means that for a Peak S/N = 150, we would trust features $>$ 10$\sigma$. Note that these numbers are empirical, but small changes do not appreciably affect the outcome of self-calibration. So we divide the S/N of the initial image by 15 to obtain our starting clean threshold, and decreased the thresholds for subsequent solution intervals logarithmically until the clean depth is equivalent to 3$\sigma$.

With a set of solution intervals to attempt and their accompanying clean thresholds, the self-calibration process begins. We collect all the model creation with \textit{tclean}, calibration solution generation with \textit{gaincal}, application of the calibration solutions with \textit{applycal}, and checking the outcome of a self-calibration iteration with \textit{tclean} into a single Python function to simplify to process. We always use the \textit{combine=`spw'} parameter in \textit{gaincal} to use all the available bandwidth for self-calibration and we use the parameter \textit{interp='linearPD'}  which will scale the phase solutions with frequency to compensate for the fact that the optimal phase solutions will be different between the low and high frequency end of the bandwidth.

At the start of each self-calibration iteration, the self-calibrated data from the previous iteration (corrected data column) is split out into a new measurement set such that each self-calibration solution interval is incremental upon the previous iteration. We compare the image S/N before self-calibration and after self-calibration of each iteration is applied to ensure that the S/N is increasing. We do this by running \textit{tclean} again but using the self-calibration model that was created at the start of a self-calibration iteration as the \textit{startmodel} for \textit{tclean}. 
This approach enables us to effectively determine the improvement in S/N just due to the gain corrections derived from that model.
When the image S/N increases within an iteration, we consider that iteration successful, if the S/N decreases, that iteration has failed and we consider the last successful iteration as the final self-calibration iteration. The above steps consider phase-only self-calibration and we do attempt amplitude self-calibration (solving for amplitude and phase simultaneously) following the phase-only self-calibration using the same criteria for success or failure of amplitude self-calibration. Following the completion of self-calibration for the successful solution intervals, the self-calibrated data for each EB are saved to new measurement sets and are considered the final continuum data.

Once self-calibration is completed on the continuum data, we process the line data. First, we apply the flux normalization to each full measurement set using the scaling factors derived from the continuum. Then
we apply the self-calibration solutions to the full measurement sets for each EB. However, the S/N improvement for the line data are typically not significant because the line data are not dynamic range limited like the continuum. For each EB, we apply all the gain tables for each successful self-calibration iteration, which is necessary because each table is incremental, so applying the full correction requires all the tables. Once
self-calibration solutions are applied, we subtract continuum emission from the line data using the \textit{uvcontsub} task. To subtract the continuum data, we select the line-free regions of the data that will have their continuum data fitted and subtracted. To select the frequency ranges to fit, we select the complement of the frequency ranges that were flagged to produce the continuum dataset, that is the line-free regions of the bandwidth. Following \textit{uv} continuum subtraction, we have the final spectral line data set and it is ready for imaging.

We first create the continuum images using \textit{tclean} and the \textit{auto-multithresh} algorithm to create the clean masks. Then we create images using Briggs weighting, for the robust parameters of -2, -1, -0.5, 0, 0.5, 1, and 2 in order to compare the features present with different weighting. We also created images using \textit{uv}-tapering, to reduce the influence of the long baselines and bring out larger-scale structure, and we created tapered images with the tapers starting at 1000, 2000, and 3000~k$\lambda$ and used robust parameters of 1 and 2. The typical synthesized beam size for the robust=0.5 images was $\sim$0\farcs05. The images for most sources use 6000 pixels with a pixel size of 0\farcs003; however, some sources require image sizes of $\sim$14000 in order to include sources toward the edge of the primary beam.

Then we create the spectral line images for each line listed in Table \ref{tab:obsline}. We typically created images using Briggs weighting with robust values of 0.5 and 2, and \textit{uv}-tapering starting at 2000~k$\lambda$. The resulting images have synthesized beams of $\sim$0\farcs1 and $\sim$0\farcs15 for the two robust values and applied tapering, respectively. The tapering was essential for the spectral line images because the surface brightness sensitivity was otherwise too low at the native resolution of the data to detect the spectral lines with high S/N. The spectral line images typically used 4000 pixels and a pixel size of 0\farcs01. The spectral line images were all created with velocity-defined channels around the spectral lines of interest using their associated rest frequencies (Table \ref{tab:obsline}).

We then save the non-primary beam corrected images (flat noise), the primary beam corrected images, the final clean mask, and the primary beam response for each individual image. As such, there are $\sim$52 images created per target when all the combinations of robust parameters, lines, and tapering are considered. The images presented in this overview paper and subsequent papers from the eDisk program are selected based on the balance between resolution and sensitivity, and also the structures discussed.

\subsection{Jorsater \& van Moorsel effect} \label{sec:JvM}
Recently it has been pointed out that data imaged with the clean algorithm may suffer systematic errors of the flux scale due to the so-called Jorsater \& van Moorsel (JvM) effect 
(\citet[][]{1995AJ....110.2037J}; see, in particular, \citet[][]{2021ApJS..257....2C}).
The reason for this is that the dirty beams in the CLEAN processes deviate from the used Gaussian restoring beams. As discussed in \citet{2021ApJS..257....2C}, the JvM effect is more significant when data taken from different antenna configurations are combined like in the case of the eDisk program. The impact due to the JvM effect to the flux scales of our eDisk maps differs from source to source: maps with more extended diffuse emission are more impacted by the JvM effect. However, it has also been pointed out that the correction method to fix the flux scale issue, implemented by \citet{2021ApJS..257....2C}, may artificially manipulate the noise levels of maps after the correction \citep{2022Casassus}. As the main discussions of the first set of papers from eDisk are based primarily on the morphology of maps and velocity structure of molecular emission rather than the flux scale of maps, we did not apply any correction to our maps in this overview paper and the related first set of papers.



\begin{figure*}[ht!]
\epsscale{0.95}
\plotone{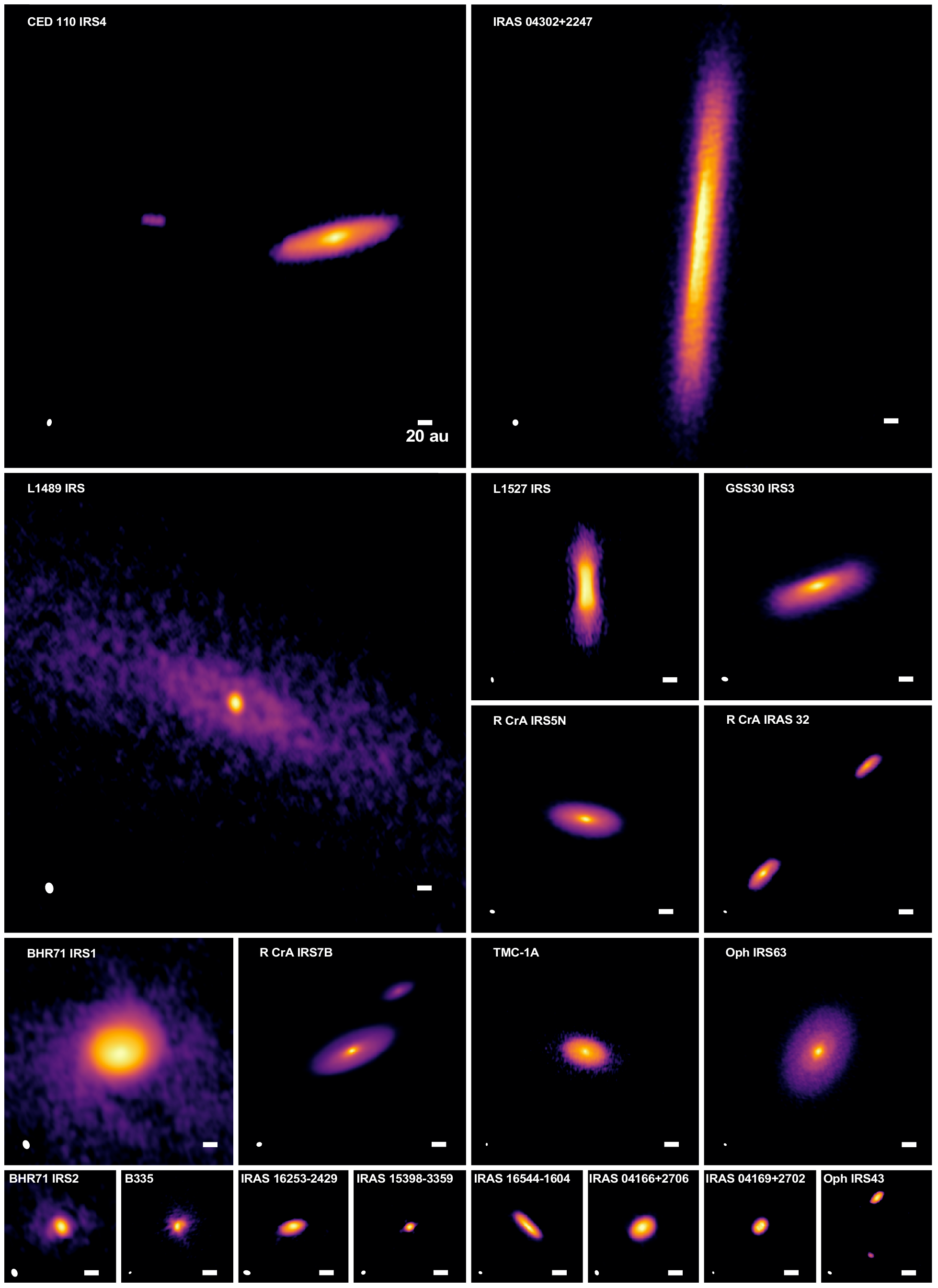}
\caption{Gallery of the continuum maps obtained in the eDisk program. All maps are shown in the same linear scale (large panels: 640 au $\times$ 640 au, medium panels: 320 au $\times$ 320 au, small panels: 160 au $\times$ 160 au). All the maps are shown in individual asinh-stretch intensity scales, except for IRAS 04169+2702 that is shown in a linear intensity scale. All the map centers are source positions listed in Table \ref{tab:cont-tab}, except for Ced110 IRS4, R CrA IRAS~32, and Oph IRS43, where the map centers are shifted from the source positions by (0$\farcs{75}$, 0$\farcs{0}$), ($-0\farcs{5}$, $0\farcs{5}$), and (0$\farcs{0}$, $-0\farcs{3}$), respectively to show the binary. The beam size and the scale of 20~au are shown at the bottom left and right corners in each panel, respectively.  \label{fig:continuum-gallery}}
\end{figure*}
\begin{figure*}[ht!]
\epsscale{1.15}
\plotone{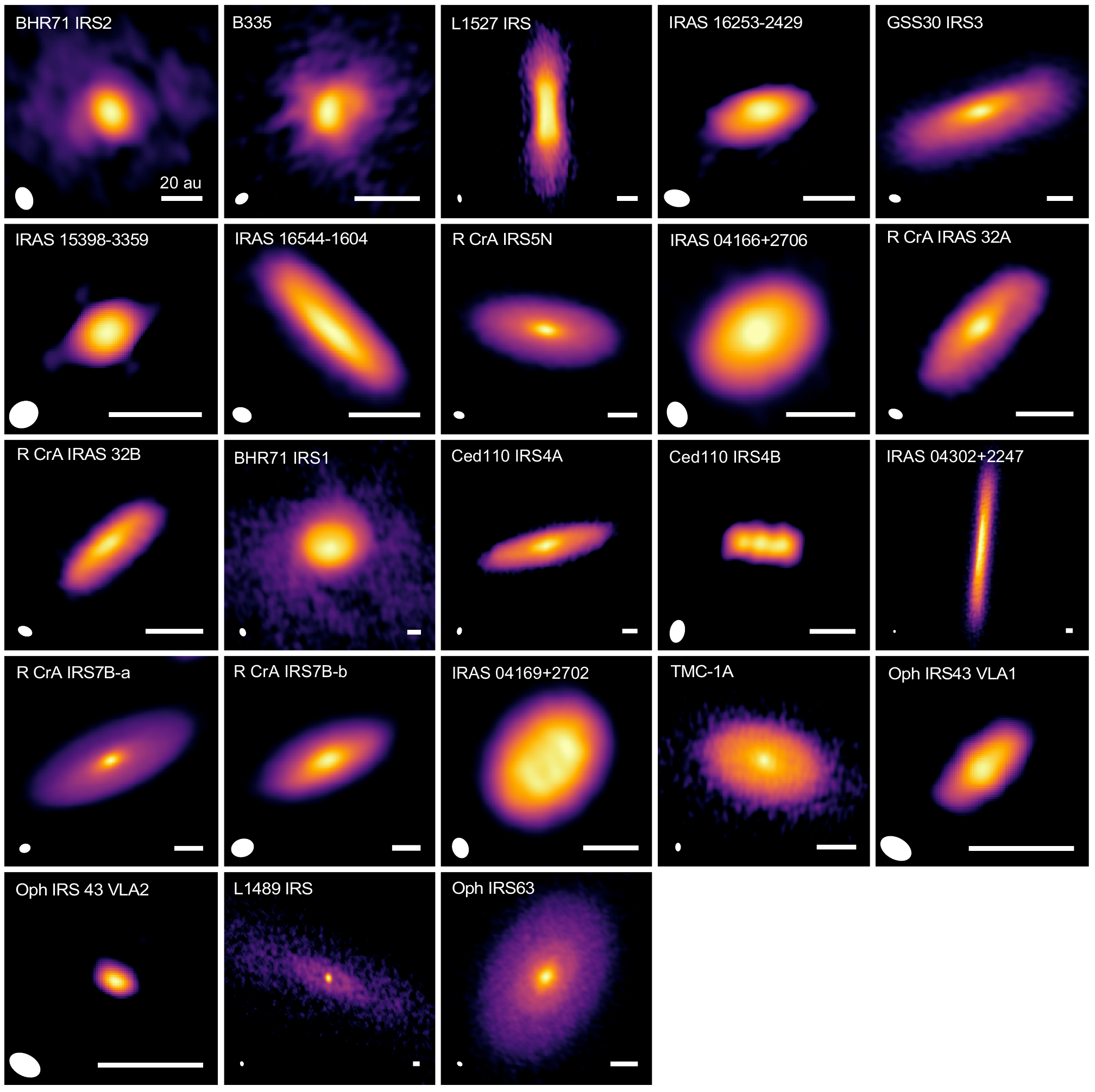}
\caption{Enlarged continuum images. Images of IRAS 04302+2247 and L1489 IRS are not enlarged from those in Figure \ref{fig:continuum-gallery}. The intensity scales are the same as those in Figure \ref{fig:continuum-gallery} except for the 4 companions. All the map centers are source positions listed in Table \ref{tab:cont-tab}. Images are presented in an order of T$_{\rm bol}$ listed in Table \ref{tab:sample}.
The beam size and the scale of 20~au are shown at the bottom left and right corners in each panel, respectively. \label{fig:zoom-cont-maps}}
\end{figure*}

\section{First results} \label{sec:result}
In this section, we present the first results of our eDisk program, an overview of the observations of the 1.3~mm continuum for the sample of sources as well as an example of analysis of the C$^{18}$O 2--1 emission toward R CrA IRS7B. The structures of embedded disks can be traced well in the 1.3~mm dust continuum emission, while C$^{18}$O is the key molecular line to study the kinematics of the gas on small scales, e.g. searches for the presence of Keplerian disks and estimates of the dynamical masses of the central protostars.
In this sense, the presentation here provides a glance at the type of scientific results that can be obtained from the eDisk data. More details of the first results about each targeted source can be found in a series of subsequent papers (see the references in Table \ref{tab:cont-tab}). 

\subsection{Continuum emission} \label{subsec:cont}

We present representative maps of the continuum images obtained in the eDisk program on the same spatial scale in Figure \ref{fig:continuum-gallery}. The same continuum images but enlarged are also presented in Figure \ref{fig:zoom-cont-maps}. Robust parameters used to create these continuum images and their angular resolutions are shown in Table \ref{tab:cont-tab}. The rms noise level of these continuum maps is $\sim$25~$\mu$Jy~beam$^{-1}$ on average.
Continuum emission is clearly detected at all source positions, and four out of the 19 sources also show continuum emission arising from companions; three of them, i.e., Ced110 IRS4, R~CrA~IRS7B, and R~CrA~IRAS~32, are newly found to be binaries in our observations. These newly found binaries are named based on the IAU convention; Ced110 IRS4A and Ced110 IRS4B, R~CrA~IRS7B-a and R~CrA~IRS7B-b, and R~CrA~IRAS~32A and R~CrA~IRAS~32B.

\begin{figure}[ht!]
\epsscale{1.15}
\plotone{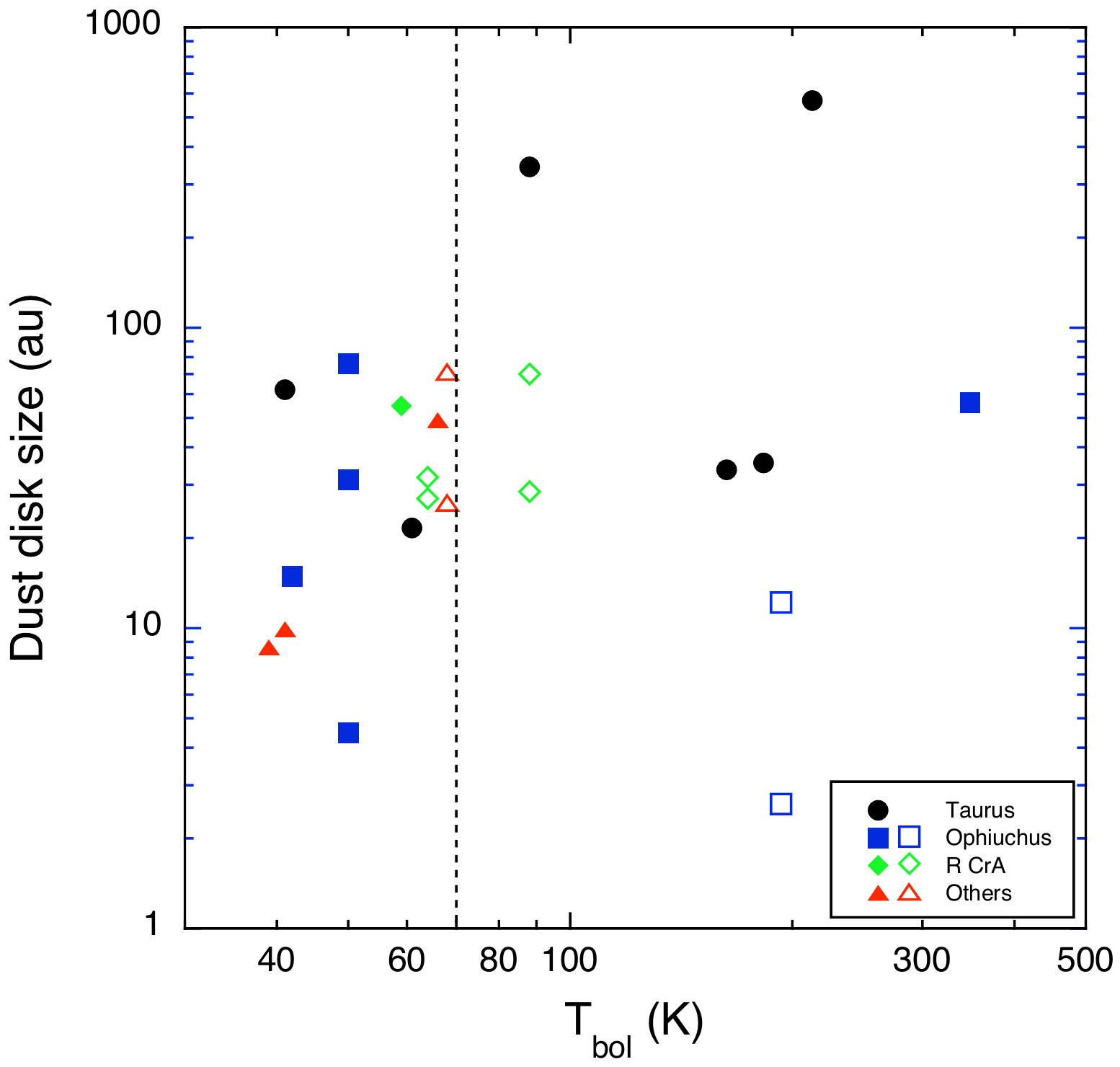}
\caption{Dust disk size in diameter is plotted as a function of bolometric temperature of their central protostellar system. Sources located in different regions are shown in different symbols. Open symbols show the 4 binary systems (Ced110~IRS4A/B, Oph IRS43 VLA1/VLA2, R CrA IRS7B-a/b, and R CrA IRAS 32A/B). The vertical dashed line shows the boundary between Class 0 and I sources.  \label{fig:Tbol-size}}
\end{figure}

All the continuum emission images
are spatially resolved, exhibiting elongated structures. These elongated structures strongly suggest that the dust continuum traces disks around the targeted sources. 
Note that for some sources envelope contamination may still play a role for both dust and gas emission: to assess that more detailed radiative transfer modeling will be needed on a source-by-source basis.
The extent of the continuum emission, measured through 2-D Gaussian fittings and listed in Table~\ref{tab:cont-tab}, varies significantly from source to source; L1489 IRS has the largest emission with an angular extension of $\sim3\farcs{9}\times1\farcs{3}$, while the companion of Oph IRS43 (VLA~2) is the most compact, marginally resolved ($2\sigma$) with an angular extent $\sim0\farcs{019}\times0\farcs{016}$.
We note that some of the continuum emissions were not well fitted with 2-D Gaussian, suggesting that the extents of the emissions provided in Table~\ref{tab:cont-tab} for those sources may not reflect their actual extents. In such a sense, the extents of the continuum emissions provided here should be considered as a quick reference of their extents measured by a simple procedure. More accurate measurements of the extents of the dust emissions through more sophisticated methods will be presented in future papers.
A comparison between the extents of the continuum emission from the different sources and their bolometric temperatures is shown in Figure \ref{fig:Tbol-size}. In agreement with the systematic studies of protostars in Orion with the VLA \citep{2020Tobin}, no clear correlation between size and evolutionary stage is seen - although the two largest disks in our sample are both associated with Class I protostars (L1489 IRS and IRAS 04302+2247). Most of the disks in the eDisk sample are smaller than about 100~au but we caution that these are estimates of the sizes of the dust disks and that the gas disks may be larger as it is often inferred for more evolved disks \citep[see, e.g.,][]{2022arXiv220309818M}.

\begin{figure}[ht!]
\epsscale{1.15}
\plotone{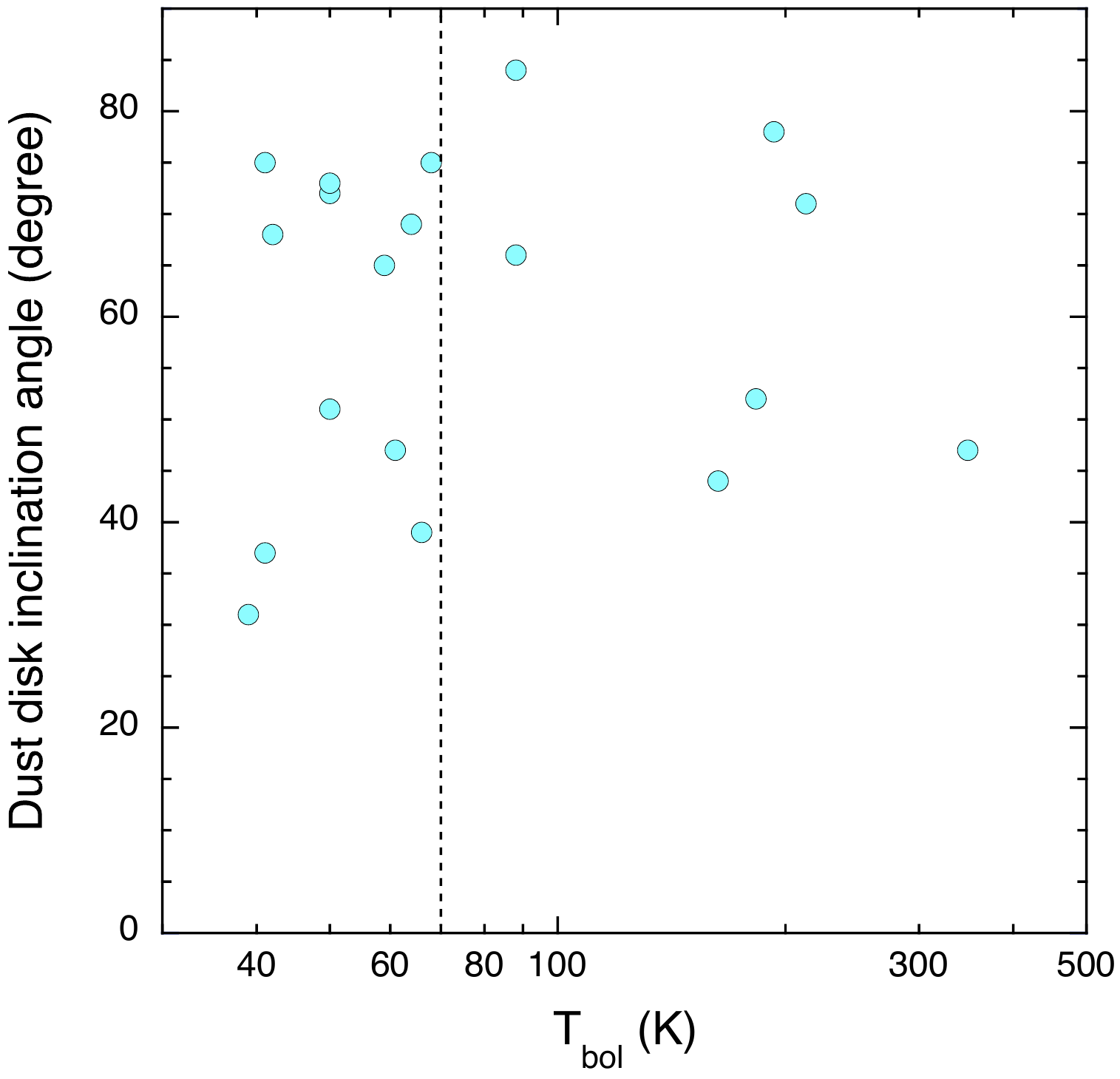}
\caption{Dust disk inclination angle is plotted as a function of bolometric temperature of their systems. Four binary companions (Ced110 IRS4B, Oph IRS43 VLA2, R~CrA~IRS7B-b, and R~CrA~IRAS~32B) are excluded. The vertical dashed line shows the boundary between Class 0 and I sources.  \label{fig:Tbol-inclination}}
\end{figure}

From a comparison between the major and minor axes of the continuum emission, the inclination of the system
can be also estimated on the assumption that the continuum emission is a vertically thin disk. We find that 7 out of 19 sources, excluding companions, have inclination angles of more than 70 degrees (see Table \ref{tab:cont-tab}), roughly consistent with the expectation from a random distribution. We caution that the estimate of the inclination is a lower limit to the true inclination because the disks have most likely some vertical thickness. Still, these are different from most Class II disks that have been observed at high angular resolution to date which tend to be more face-on \citep{2018ApJ...869L..41A,2021MNRAS.501.2934C}. Figure \ref{fig:Tbol-inclination} shows a comparison between the inclination angle and the bolometric temperature, demonstrating no correlation between them. This suggests that the disk inclination has less impact on our classification of the 19 protostellar systems based on the bolometric temperature even though it was also suggested that higher inclination angles of disks could misclassify protostars as Class 0 sources \citep{1996AJ....112.2076T,2003ApJ...583..322N,2008ApJ...679.1364T,2017ApJ...840...69F}. We also note that some of the continuum structures show relatively high brightness temperatures of more than 100~K. Such high brightness temperature may suggest that the detected continuum emission is at least partially optically thick.
Basic physical parameters of the continuum emission are summarized in Table \ref{tab:cont-tab}. We note that statistical errors are not given in this table because the actual uncertainties of these measurements are more likely determined by systematic errors, such as the flux calibration error and the gain calibration error.

\begin{deluxetable*}{lllcllllll}
\tablecaption{Physical parameters of continuum emission} \label{tab:cont-tab}
\tablewidth{0pt}
\tablehead{
\colhead{Source name} & \colhead{ICRS R.A.} & \colhead{ICRS Dec.} & 
\colhead{Robust} & \colhead{$\theta_{\rm b}$, $PA_{\rm b}$} &
\colhead{Extent, PA} &
\colhead{Peak $I_{\nu}$, $T_{\rm b}$} & \colhead{$F_{\nu}$} & \colhead{inc.} & \colhead{Ref.} \\
\colhead{} & \colhead{(h m s)} & \colhead{(d m s)} & \colhead{} & \colhead{(mas,\arcdeg)} &
\colhead{(mas,\arcdeg)} & \colhead{(mJy beam$^{-1}$, K)} & \colhead{(mJy)} & \colhead{(\arcdeg)} & \colhead{}
}
\decimalcolnumbers
\startdata
L1489 IRS & 04:04:43.080 & +26:18:56.12 & 1 & 105$\times$78, 12 & 3913$\times$1283, 66 & 5.4, 21 & 91 & 71 & 1 \\
IRAS 04166+2706 & 04:19:42.505 & +27:13:35.83 & 0 & 49$\times$37, 22 & 138$\times$94, 122 & 9.6, 133 & 71 & 47 & 2 \\
IRAS 04169+2702 & 04:19:58.477 & +27:09:56.82 & 0 & 48$\times$37, 21 & 216$\times$156, 139 & 4.3, 64 & 101 & 44 & 3 \\
IRAS 04302+2247 & 04:33:16.499 & +22:53:20.23 & 0.5 & 55$\times$50,3 & 2149$\times$239,175 & 1.1, 14 & 183 & 84 & 4\\
L1527 IRS & 04:39:53.878 & +26:03:09.43 & --0.5 & 56$\times$29, 9 & 444$\times$114, 2 & 4.3, 69 & 139 & 75 & 5\\
Ced110 IRS4A & 11:06:46.369 & --77:22:32.88 & 0 & 54$\times$35, --13 & 376$\times$99, 104 & 5.8, 79 & 80 & 75 & 6 \\
Ced110 IRS4B & 11:06:46.772 & --77:22:32.76 & 0 & 54$\times$35, --13 & 138$\times$41, 85 & 0.38, 9 & 2.0 & 73 & 6 \\
BHR71 IRS2 & 12:01:34.008 & --65:08:48.08 & 0.5 & 70$\times$50, 23 & 49$\times$42, 68 & 8.7, 65 & 14 & 31 & 7 \\
BHR71 IRS1 & 12:01:36.476 & --65:08:49.37 & 0.5 & 72$\times$53, 22 & 279$\times$215, 98 & 26, 170 & 384 & 39 & 7 \\
IRAS 15398-3359 & 15:43:02.232 & --34:09:06.96 & -0.5 & 43$\times$36, --56 & 29$\times$18, 117 & 5.9, 93 & 7.9 & 51 & 8 \\
GSS30 IRS3 & 16:26:21.715 & --24:22:51.09 & 0 & 68$\times$45, 77 & 550$\times$169, 109 & 6.4, 56 & 124 & 72 & 9 \\
Oph IRS43 VLA1 & 16:27:26.906 & --24:40:50.81 & --1 & 46$\times$30, 61 & 89$\times$18, 136 & 3.3, 62 & 11 & 78 & 10 \\
Oph IRS43 VLA2 & 16:27:26.911 & --24:40:51.40 & --1 & 46$\times$30, 61 & 19$\times$16, 131 & 0.88, 20 & 1.1 & 32 & 10 \\
IRAS 16253-2429 & 16:28:21.615 & --24:36:24.33 & 0 & 73$\times$48, 78 & 107$\times$40,113 & 5.1, 40 & 12 & 68 & 11 \\
Oph IRS63 & 16:31:35.654 & --24:01:30.08 & 0 & 34$\times$25, 57 & 426$\times$292, 149 & 5.8, 170 & 280 & 47 & 12 \\
IRAS 16544--1604 & 16:57:19.643 & --16:09:24.02 & 0 & 36$\times$27, 69 & 207$\times$60, 45 & 4.1, 103 & 52 & 73 & 13 \\
R CrA IRS5N & 19:01:48.480 & --36:57:15.39 & 0.5 & 50$\times$30,75 & 374$\times$157, 81 & 5.5, 94 & 99 & 65 & 14 \\
R CrA IRS7B-a & 19:01:56.420 & --36:57:28.66 & 0 & 54$\times$42, --73 & 462$\times$174, 115 & 18, 195 & 339 & 68 & 15 \\
R CrA IRS7B-b & 19:01:56.385 & --36:57:28.11 & 0 & 54$\times$42, --73 & 187$\times$75, 115 & 5.4, 63 &  36 & 66 & 15 \\
R CrA IRAS 32A & 19:02:58.722 & --37:07:37.39 & --0.5 & 35$\times$23, 67 & 212$\times$78, 135 & 4.9, 153 & 80 & 69 & 16 \\
R CrA IRAS 32B & 19:02:58.642 & --37:07:36.39 & --0.5 & 35$\times$23, 67 & 180$\times$58, 132 & 3.2, 102 & 44 & 72 & 16 \\
\hline
TMC-1A$^{\dagger}$ & 04:39:35.202 & +25:41:44.22 & 0 & 32$\times$21, --1 & 259$\times$159, 76 & 5.0, 187 & 184 & 52 & 17 \\
B335$^{\dagger}$ & 19:37:00.900 & +07:34:09.81 & 0 & 27$\times$19, --54 & 60$\times$48, 163 & 5.0, 243 & 29 & 37 & 17 \\
\enddata
\tablecomments{$^{\dagger}$For these sources, we used only ALMA archival data without new observations. Column 1: target names. Column 2: Right ascension of the continuum emission measured with single 2-D Gaussian fittings. For L1489 IRS 2-D Gaussian fittings with two-component were performed because single 2-D Gaussian fittings did not provide a solution. Column 3: Declination of the continuum emission measured with single 2-D Gaussian fittings. For L1489 IRS 2-D Gaussian fittings with two-component were performed because single 2-D Gaussian fittings did not provide a solution. Column 4: Robust parameter used to create the continuum image shown in Figures \ref{fig:continuum-gallery} and \ref{fig:zoom-cont-maps}. Column 5: synthesized beam FWHM and position angle of the representative map shown in Figure \ref{fig:continuum-gallery}. Column 6: Beam deconvolved size and position angle of the continuum emission measured with single 2-D Gaussian fittings. For L1489 IRS 2-D Gaussian fittings with two-component were performed because single 2-D Gaussian fittings did not provide a solution. Column 7: Peak intensity of the continuum emission and its brightness temperature estimated with the full Plank function. Column 8: Integrated flux density of the continuum emission measured with single 2-D Gaussian fittings. For L1489 IRS 2-D Gaussian fittings with two-component were performed because single 2-D Gaussian fittings did not provide a solution. Column 9: Inclination angle of the continuum emission estimated from the ratio of the major and minor axes of the continuum emission listed in the column 5. Column 10: References from the eDisk program. 1. \citet{2022yamato}; 2. \citet{2022phuong}; 3. \citet{2022han}; 4. \citet{2022Lin}; 5. \citet{2022vant}; 6. \citet{2022sai}; 7. \citet{2022gavino}; 8. \citet{2022thieme}; 9. \citet{2022santamaria}; 10. \citet{2022narayanan}; 11. \citet{2022aso}; 12. \citet{2022flores}; 13. \citet{2022Kido}; 14. \citet{2022sharma}; 15. \citet{2022ohashi}; 16. \citet{2022encalada}; 17. This work}
\end{deluxetable*}

When we carefully inspect the continuum maps, it is found that some of them show ring-like structures that can be visually identified by naked eye. The most obvious source is L1489 IRS. Oph IRS63 also weakly shows ring-like structures. More details of the continuum maps obtained in the eDisk program will be presented in subsequent papers. Ring-like structures in continuum emission around these two sources have been suggested by previous studies as well \citep{2022ApJ...933...23O,2020Natur.586..228S} although our map of L1489 IRS seems to show ring-like structures more clearly than in the previous work \citep{2022ApJ...933...23O}. The continuum emission around IRAS~04169+2702 seems to show a bean-like structure with brighter emission on its southwest side, which could also be a ring-like structure. We should stress that a more detailed and systematic analysis of continuum emission is definitely required to identify their substructures, including those that cannot be visually identified by the naked eye, as will be discussed in a forthcoming paper. Nevertheless, the first analysis on the observations suggests that continuum emission around our Class 0/I sample does not show very sharp ring or spiral structures that are often seen in Class II continuum disks (e.g., DSHARP, ODISEA). This striking difference will be discussed in more detail in Sec. \ref{sec:dis-cont}.  
On the other hand, several disk continuum images show brightness asymmetries, particularly along their minor axes. IRAS 04302+2247, GSS 30 IRS3, IRAS 16253-2429, IRAS 16544--1604, R~CrA IRS7B-a, R~CrA IRAS~32A, and R~CrA IRAS~32B show clear asymmetry that can be identified by eye, while asymmetry can be also identified with more careful inspection of continuum intensity distributions for some additional sources, such as L1527 IRS. For these sources showing brightness asymmetries, the geometric centers of the continuum emission derived from single 2-D Gaussian fittings listed in Table \ref{tab:cont-tab} are shifted from their actual continuum peak positions. The possible origin of the brightness asymmetry is discussed in Section \ref{sec:dis-cont}. 

\subsection{C$^{18}$O 2--1} \label{sec:C18O-results}
The C$^{18}$O 2--1 emission is also detected toward all 19 sources in our sample. Here, we show the C$^{18}$O emission detected in one of our sources, R~CrA IRS7B, as an example of what can be seen toward the eDisk targets. 

\begin{figure*}[ht!]
\epsscale{1.1}
\plotone{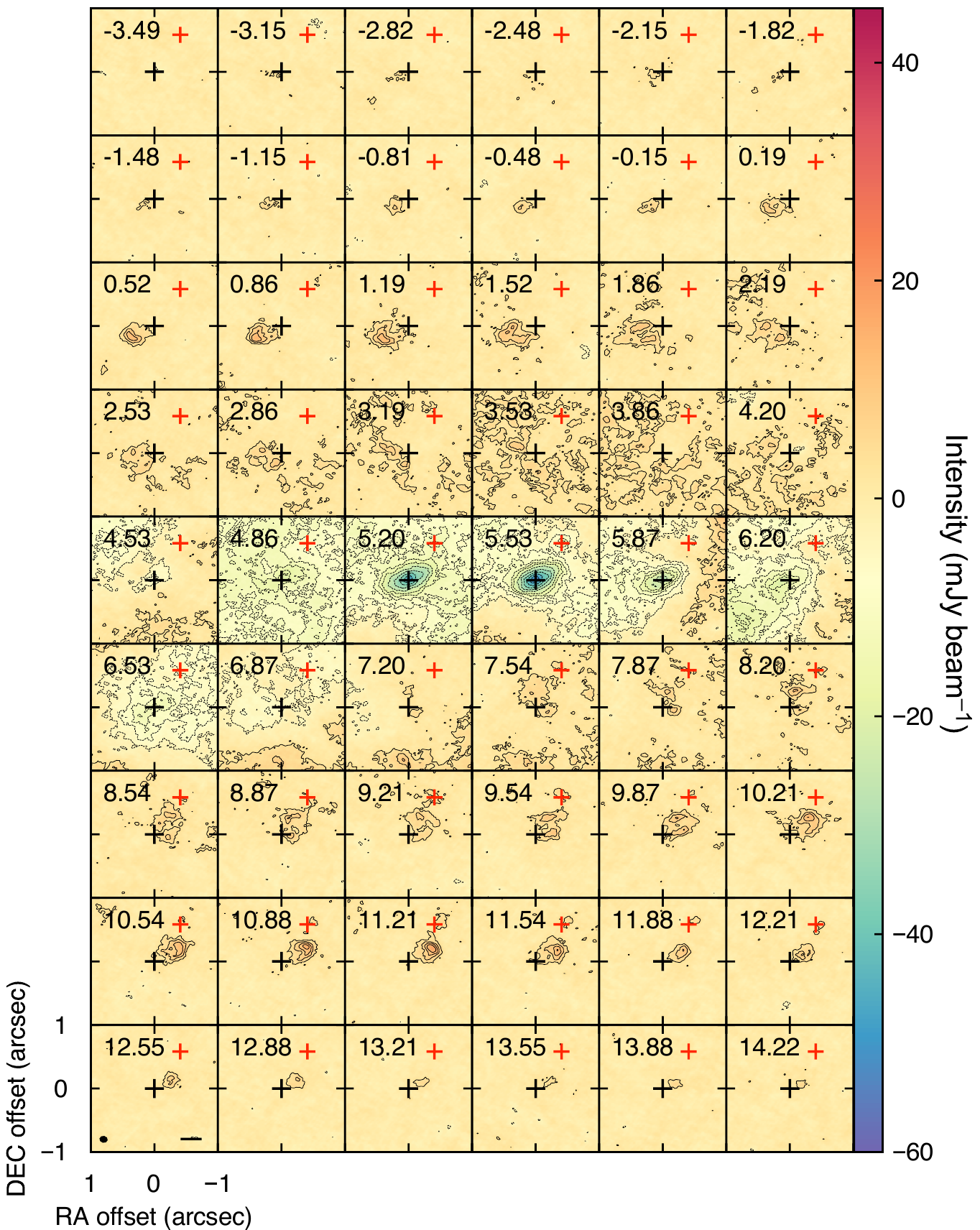}
\caption{C$^{18}$O channel maps with a velocity resolution of 0.334~km~s$^{-1}$ (two channels combined) in R CrA IRS7B. Solid contours are drawn from 3$\sigma$ to 15$\sigma$ every 3$\sigma$ 
for positive emissions, while dashed contours are drawn from -3$\sigma$ to -15$\sigma$ every -3$\sigma$ and less every -5$\sigma$. 1$\sigma$ is 1.2~mJy~beam$^{-1}$. The center of each channel map is the position of R~CrA IRS7B-a shown in Table \ref{tab:cont-tab}. The black and red crosses in each panel indicate the positions of IRS7B-a and IRS7B-b, respectively. The beam size and a scale bar of 50~au are shown at the bottom-left and the bottom-right corners in the bottom-left panel.  \label{fig:c18o-channel}}
\end{figure*}

\citet{2014A&A...566A..74L} have carried out ALMA observations of IRS7B in Band 7 at an angular resolution of $\sim0\farcs{4}$ and detected continuum emission and several molecular lines, including C$^{17}$O 3--2. Both continuum and C$^{17}$O emission have extended structures elongated in the southeast to northwest direction. The C$^{17}$O emission also shows a clear velocity gradient along the elongation, which was interpreted as Keplerian rotation \citep{2014A&A...566A..74L}. Although a systemic velocity of 6.2~km~s$^{-1}$ was adopted for the analysis of the velocity structures of C$^{17}$O in \citet{2014A&A...566A..74L}, we adopt $\sim$6.0~km~s$^{-1}$ as the systemic velocity of IRS7B in this paper (see more details below).

C$^{18}$O 2-1 was detected at more than 3$\sigma$ in the LSR velocity range between $-$3.15~km~s$^{-1}$ and 14.22~km~s$^{-1}$ at an angular resolution of $\sim0\farcs{13}\times0\farcs{11}$ as shown in Figure \ref{fig:c18o-channel}. Note that the velocity resolution of the C$^{18}$O data shown here is $\sim$0.33~km~s$^{-1}$, which is double the original velocity resolution to achieve a higher sensitivity. Since the systemic velocity of this system is $\sim$6~km~s$^{-1}$ (see Sec. \ref{sec:dis-C18O}), LSR velocities smaller and larger than 6~km~s$^{-1}$ correspond to blue- and red-shifted, respectively. At higher blue-shifted velocities ($V_{\rm LSR}=-3.15$ to 2.86~km~s$^{-1}$) and red-shifted velocities ($V_{\rm LSR}=8.20$ to 14.22~km~s$^{-1}$), C$^{18}$O is mainly detected on the southeast side and the northwest side of IRS7B-a within $\sim$0\farcs{5}, respectively. At lower blue-shifted ($V_{\rm LSR}=3.19$ to 4.20~km~s$^{-1}$) and red-shifted velocities ($V_{\rm LSR}=7.20$ to 7.87~km~s$^{-1}$), more extended C$^{18}$O emission is detected within the field of $2\arcsec\times2\arcsec$, demonstrating that our observations can also trace extended emission as well. At velocities close to the systemic velocity, absorption of C$^{18}$O is detected. In particular, very deep absorption is seen at the position of IRS7B-a at $V_{\rm LSR}$ between 5.20 and 5.87~km~s$^{-1}$. The extent of the deep absorption is almost the same as that of the continuum emission associated with IRS7B-a, suggesting that the deep absorption is due to cold, foreground C$^{18}$O emission against the bright continuum emission. Compact and weak C$^{18}$O emission is detected at the position of IRS7B-b at several channels of red-shifted velocities, such as $V_{\rm LSR}=9.21$ and 11.21~km~s$^{-1}$, and also absorption at $V_{\rm LSR}=5.20$ and 5.53~km~s$^{-1}$. The absorption at the position of IRS7B-b may be also due to cold foreground C$^{18}$O against the continuum emission associated with IRS7B-b. 

\begin{figure*}[ht!]
\epsscale{1.15}
\plotone{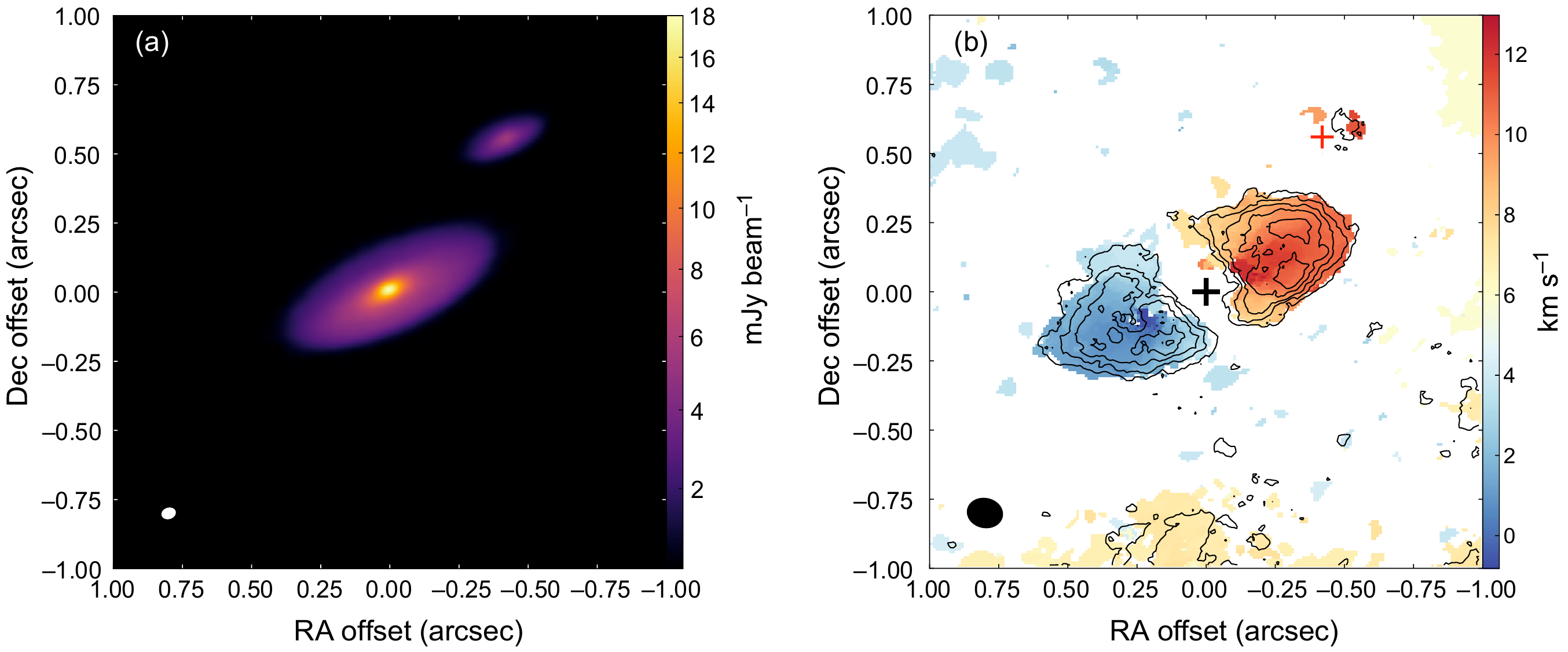}
\caption{(a) Continuum emission map in IRS7B. IRS7B-a is located at the center of the map, while IRS7B-b is located to the northwest side of IRS7B-a. The beam size is shown in the bottom-left corner as the white ellipse. (b) C$^{18}$O moment 0 map shown in contour and moment 1 map shown in color. The moment 0 map was created by integrating emission having intensity more than 3$\sigma$ (1 $\sigma$ = 1.2~mJy~beam$^{-1}$) in the velocity range of -3.15~km~s$^{-1}$ to 14.22~km~s$^{-1}$ to avoid the influence of the deep absorption features. The moment 1 map was created over the same velocity range as the moment 0 map with the emission having an intensity more than 5$\sigma$. The contours of the moment 0 map are drawn every 2$\sigma$ from 3$\sigma$ (1$\sigma$ = 2.9~mJy~beam$^{-1}$km~s$^{-1}$). The black and red crosses indicate the positions of IRS7B-a and IRS7B-b, respectively. The beam size is indicated in the bottom-left corner with the black ellipse. \label{fig:c18o-mom-maps}}
\end{figure*}

To investigate the overall distribution and velocity structures of C$^{18}$O around IRS7B-a, moment 0 and 1 maps of IRS7B are shown in Fig. \ref{fig:c18o-mom-maps}b. For comparison, the continuum map of IRS7B is also shown in Fig. \ref{fig:c18o-mom-maps}a. The moment 0 map shows two emission components with semi-elliptic shapes, one located to the southeastern side of IRS7B-a and the other located on the northwestern side of IRS7B-a, forming a disk-like structure divided into two along its minor axis as a whole. The C$^{18}$O disk-like structure is basically similar to that seen in the continuum associated with IRS7B-a, although it extends slightly larger than the continuum emission, suggesting that C$^{18}$O probably traces gas associated with the protoplanetary disk around IRS7B-a, but further investigation of the kinematics of C$^{18}$O is required for us to have a firm conclusion (see below). Although both continuum and C$^{18}$O show similar disk-like structures, their intensity distributions are quite different, e.g., C$^{18}$O is stronger in outer parts of the disk-like structure and it gets weaker toward the position of IRS7B-a, where the continuum emission shows a peak. This discrepancy would be most likely due to absorption of the warm continuum emission by colder C$^{18}$O gas.

The moment 1 map presented in Figure \ref{fig:c18o-mom-maps}b shows a velocity structure of the C$^{18}$O emission having blue-shifted velocities with respect to the systemic velocity of $\sim$6~km~s$^{-1}$ on the southeastern side and red-shifted velocities on the northwestern side. This clear velocity gradient along the major axis of the disk-like structure is naturally interpreted as rotation. The nature of the velocity structures can be investigated further with a position-velocity (PV) diagram cutting along the major axis, as discussed in Sec. \ref{sec:dis-C18O}. The diagram shows that the velocity of C$^{18}$O increases toward the position of IRS7B-a, a clear sign of differential rotation. Further discussions on the nature of the rotation are presented in Section \ref{sec:dis-C18O}.



\section{Discussion} \label{sec:dis}

\subsection{Substructures and asymmetry of continuum emission} \label{sec:dis-cont}
When we compare our continuum maps from Fig.~\ref{fig:continuum-gallery} with those toward Class II sources, from, e.g., the DSHARP and ODISEA programs, the absence of clear ring or spiralstructures identifiable by the naked eye with few exceptions is remarkable. 
 The noticeable difference between protostars and Class II sources could suggest that substructures are not well developed in the embedded phase yet, and they could be quickly formed when Class 0/I protostars evolve into Class II sources. Interestingly, the two sources, L1489~IRS and Oph~IRS~63 showing visually identified ring-like structures and another source, IRAS~04169+2702, possibly showing a ring-like structure in the eDisk sample are all Class I protostars. In addition, two protostars previously found to have ring-like structures in their continuum maps are also Class I protostars (WL~17 and GY~91). Note that substructures found in dust continuum around GY~91 are less sharpe than those seen around Class~II sources \citep[see][]{2018Sheehan,2021MNRAS.501.2934C} while WL~17 does show a sharp ring-like structure \citep{2017Sheehan}. Although seven sources from the Orion VANDAM survey exhibit relatively clear ring structures, these ring structures could be due to close binary formation \citep[][]{2020ApJ...902..141S}. 

We should note, however, that a reason behind the lack of obvious sharp substructures in continuum emission around the Class 0/I protostars could also be that the continuum emission at 225 GHz may be optically thick. In fact, some of the continuum emission observed in the eDisk program may be optically thick as indicated by their higher brightness temperatures (see Table \ref{tab:cont-tab}). To address this possibility, it will be necessary to observe continuum emission at lower frequencies, where dust continuum emission is expected to be optically thinner. One of the sources in the eDisk sample, L1527 IRS, has been observed in 7~mm continuum with VLA, finding no clear gaps \citep{2022Sheehan}. 

We should also note that some of the disk structures seem to be more inclined (see Table~\ref{tab:cont-tab}), particularly as compared with Class II disks observed at high angular resolution to date, which could make substructures more difficult to be observed in disks around our sample. In addition, some of the continuum emissions are also more compact, and our angular resolution may not be high enough to detect substructures. More detailed analyses and discussions of substructures in the continuum emission, including the visibility analysis, will be presented in forthcoming papers.

In contrast to substructures, several of the target sources show brightness asymmetries, particularly along their minor axes (see Figure \ref{fig:zoom-cont-maps}). 
A possible explanation for these  brightness asymmetries along the minor axes might come from a geometrical effect, namely, the far-side of a highly inclined disk with a finite vertical thickness is brighter than the near-side. Such a near-far side brightness asymmetry is an indication that the dust has yet to completely settle onto the disk midplane. In more edge-on systems such as HH~212, the non-settled dust produces a dark lane near the mid-plane sandwiched between two brighter features \citep{2017Lee,2021Lin}. In contrast to embedded disks, Class II disks rarely show any asymmetry along their minor axes \citep[e.g.,][]{2018ApJ...869L..41A,Long_2019}. The difference could suggest that dust is not quite settled in the embedded phase but quickly settles as embedded disks evolve into Class II disks.

\subsection{Identifying Keplerian rotation} \label{sec:dis-C18O}


\begin{deluxetable*}{lclcc}
\tablecaption{Results of the best fitting to the C$^{18}$O PV diagram\label{tab:c18o_pv_fit}}
\tablewidth{0pt}
\tablehead{
\colhead{Parameter} & \colhead{Unit}& \colhead{Description} & \multicolumn2c{Best fit value} \\
\cline{4-5}
\colhead{} & \colhead{} & \colhead{} & \colhead{Edge} &\colhead{Ridge}
}
\decimals
\startdata
$p_{\rm in}$ & & Power of the fitting function & 0.61$\pm$0.04 & 0.51$\pm$0.01 \\
$V_{\rm sys}$ & km~s$^{-1}$ & systemic velocity & 6.04$\pm$0.04 & 6.00$\pm$0.02 \\ 
Mass & $M_{\sun}$ & Dynamical mass of the central star & 3.21$\pm$0.04 & 2.06$\pm$0.02 \\
\enddata

\end{deluxetable*}

In addition to investigating substructures in continuum emission, another important purpose of the eDisk program is to identify Keplerian rotation around our target protostars. Here we discuss the case of R~CrA~IRS7B-a as an example showing how we identify Keplerian rotation.


One of the methods to identify Keplerian rotation, which is also adopted by the eDisk program, is fitting to position-velocity (PV) diagrams. Our fitting process includes two independent methods. One method uses the ridge in 1D intensity profiles along the positional axis (the velocity axis) at a given velocity (position). The other method uses the outer edge in the 1D intensity profiles. We call the former the ridge method, while we call the latter the edge method.
In each method, PV diagrams are fitted with a power-law function. The function is a single- or double-power function, which is defined with a power-law index, $p_{\rm in}$. The systemic velocity, $v_{\rm sys}$, is also included as a free parameter in the fitting process. More details of the fitting method and process are explained in Appendix~\ref{sec:app-PV-fit}.

\begin{figure}[ht!]
\epsscale{1.15}
\plotone{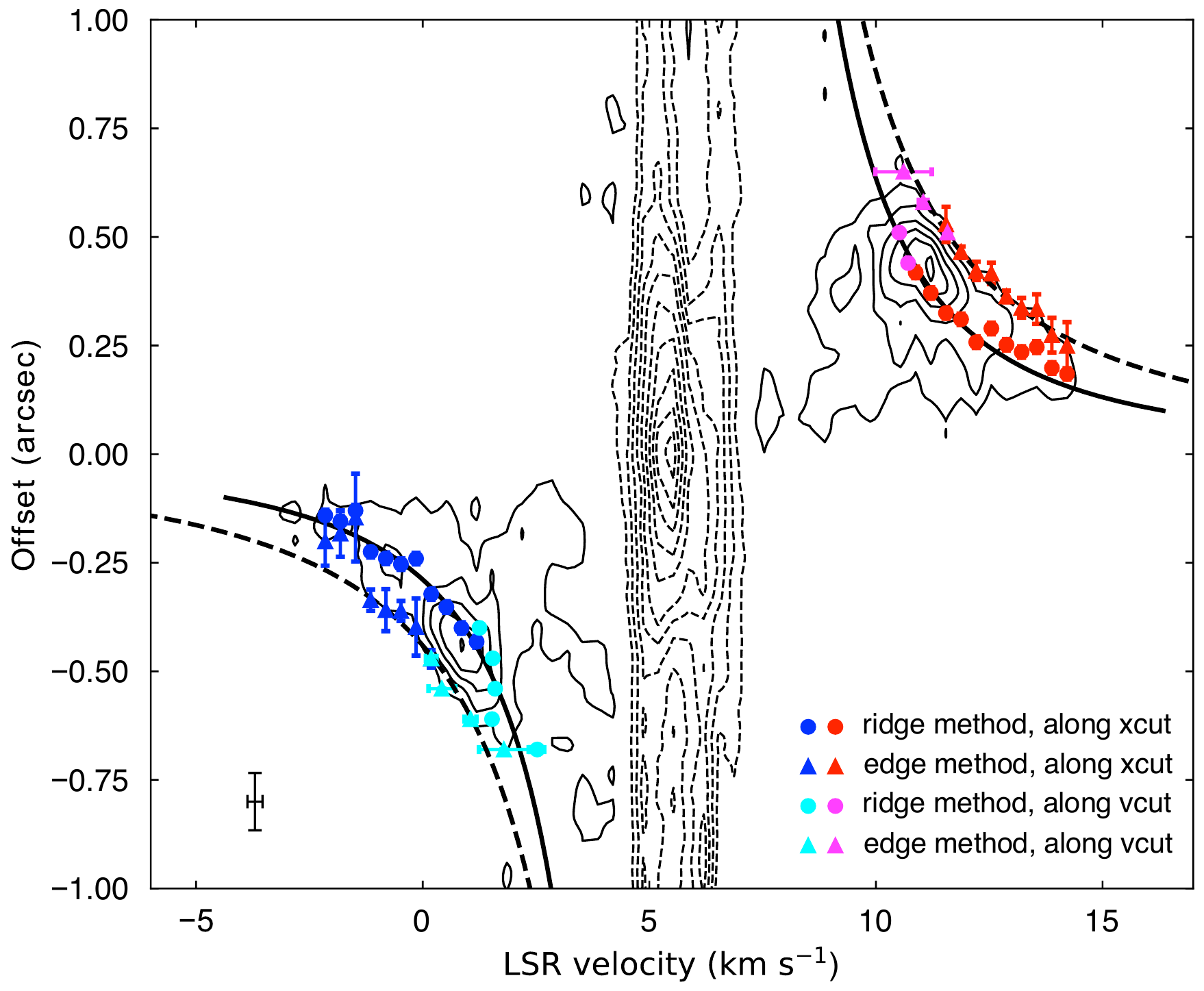}
\caption{C$^{18}$O position-velocity diagram of R~CrA~IRS7B-a with a velocity resolution of 0.334~km~s$^{-1}$. Contours are drawn from 3$\sigma$ to 15$\sigma$ in every 2$\sigma$ 
for positive emission, while dashed contours are drawn from -3$\sigma$ to -15$\sigma$ in every -3$\sigma$ and less every -5$\sigma$. 1$\sigma$ is 1.2~mJy~beam$^{-1}$. Cicles and triangles show data points obtained through the ridge and edge methods, respectively, while the solid and dashed curves represent fitting results obtained through the ridge and edge methods, respectively. Different colors show data points obtained along different direction: blue and red show data points obtained along the position axis (xcut), while cyan and magenta represent data points obtained along the velocity axis (vcut).   \label{fig:c18o-pv}}
\end{figure}

As was shown in Section \ref{sec:C18O-results}, the C$^{18}$O emission in R~CrA~IRS7B-a shows a velocity gradient along its disk-like elongated structure centered at the position of R~CrA~IRS7B-a, suggestive of rotation. The C$^{18}$O PV diagram cutting along the major axis of the dust emission around R~CrA~IRS7B-a presented in Figure \ref{fig:c18o-pv} shows that the rotation can be described as differential rotation, and it is fitted in both the edge and ridge methods described above.
The fitting results are presented in Table \ref{tab:c18o_pv_fit}. For demonstration purposes, only fittings with a single-power law function are presented here. Both edge and ridge fittings with a single power-law show that $p_{\rm in}$ is close to 0.5, suggesting that the rotation can be explained as Keplerian rotation. The dynamical mass of the central protostar was estimated to be $\sim3.2~M_{\sun}$ with the edge method, while it was estimated to be $\sim2.1~M_{\sun}$ with the ridge method. As explained in Appendix~\ref{sec:app-PV-fit}, the actual dynamical mass of the central star is considered to be between $\sim2.1~M_{\sun}$ and $\sim3.2~M_{\sun}$. Discussions on further details of the Keplerian disk around R~CrA~IRS7B-a are beyond the scope of this overview paper, and they will be presented in the subsequent paper.


\section{Data Release} \label{sec:data release}
As one of ALMA large programs, the eDisk program will release a suite of data products that have legacy values for the community. The released data sets, which will be available at https://almascience.org/alma-data/lp/edisk, include (1) CASA scripts and associated python modules used to calibrate and image the data; (2) calibrated visibility data from each execution block, including those from the ALMA archive; (3) continuum images created from combined visibility data with a robust value of 0.5; (4) line cube images created from combined visibility data with a robust value of 0.5; (5) fits file of the primary beam for each image mentioned above. These released data sets enable the community to reproduce all the images presented in this paper and forth coming papers.

\section{Summary} \label{sec:summary}
This article has provided an introduction to the ALMA large program, eDisk, including the background of the program, the sample selection of the 19 Class 0/I protostars, the ALMA observations at Band~6, and the data reduction processes. The first look results of the 1.3~mm continuum emission detected around all the 19 targets for the eDisk program and C$^{18}$O 2--1 emission detected around one of the 19 targets, R~CrA~IRS7B, are presented to demonstrate what kind of results we would expect to see in the eDisk program. 

All the continuum emission is spatially resolved at angular resolutions of $\sim0\farcs{04}$, revealing elongated disk-like structures. The detected continuum emission is considered to trace embedded disks around the sampled Class 0/I protostars although they could also partially arise from inner parts of the envelopes surrounding the protostars. More importantly, the continuum emission does not show any sharp ring-like or spiral-like structures with high contrast, which are often seen in Class II disks. This could suggest that substructures are formed in disks quickly when Class 0/I protostars evolve to Class II sources or otherwise that high optical depth of dust emission may obscure such features. Note that some of the continuum emission show more compact or inclined disk-like structures, which could prevent us from finding substructures there. Asymmetries in the brightness distributions along minor axes seen by eye for some sources may also reflect the dust optical thickness and orientations of the disks.

C$^{18}$O 2--1 emission detected in R~CrA~IRS7B-a also shows an elongated disk-like structure similar to that seen in its dust continuum emission. There is a clear velocity gradient along the elongation. The analysis of the velocity structures in the PV-diagram cutting along the elongation suggests that the velocity gradient seen in the C$^{18}$O elongation can be explained as Keplerian rotation with a central dynamical mass of 2.1-3.2~$M_{\sun}$.

The first results here already demonstrate the great potential of the data from the eDisk program. A set of accompanying papers go into more detail about the first results for individual sources setting the stage for future more synergistic studies. These analyses will provide new, important constraints on these pivotal stages in the formation and early evolution of protostars and their disks.



\acknowledgments

We would like to thank Elise Furlan, who kindly provided us with the Spitzer IRS spectra data for IRAS 04166+2702 and IRAS 04169+2709. We would like to also thank Neal Evans and Yao-Lun Yang, who kindly provided us with the Spitzer IRS spectra and the Herschel spectra for B335, respectively.
This paper makes use of the following ALMA data: ADS/JAO.ALMA\#2019.1.00261.L, ADS/JAO.ALMA\#2019.A.00034.S, \seqsplit{ADS/JAO.ALMA\#2013.1.00879.S}, \seqsplit{ADS/JAO.ALMA\#2013.1.01086.S}, ADS/JAO.ALMA\#2015.1.01415.S,
ADS/JAO.ALMA\#2015.1.01512.S, and
\seqsplit{ADS/JAO.ALMA\#2017.1.00288.S.}
ALMA is a partnership of ESO (representing its member states), NSF (USA) and NINS (Japan), together with NRC (Canada), NSTC and ASIAA (Taiwan), 
and KASI (Republic of Korea), in cooperation with the Republic of Chile. The Joint ALMA Observatory is operated by ESO, AUI/NRAO and NAOJ.
The National Radio Astronomy Observatory is a facility of the National Science Foundation operated under cooperative agreement by Associated Universities, Inc.
N.O. and  C.F. acknowledge support from National Science and Technology Council (NSTC) in Taiwan through the
grants NSTC 109-2112-M-001-051, 110-2112-M-001-031, 110-2124-M-001-007, and 111-2124-M-001-005.
J.K.J., R.S. and S.G. acknowledge support from the Independent Research Fund Denmark (grant No. 0135-00123B).
PDS acknowledges support from NSF AST-2001830 and NSF AST-2107784.
ZYL is supported in part by NASA 80NSSC20K0533 and NSF AST-1910106. 
LWL and FJE acknowledge support from NSF AST-2108794. 
C.W.L. is supported by the Basic Science Research Program through the National Research Foundation of Korea (NRF) funded by the Ministry of Education, Science and Technology (NRF-2019R1A2C1010851), and by the Korea Astronomy and Space Science Institute grant funded by the Korea government (MSIT; Project No. 2022-1-840-05).
JEL is supported by the National Research Foundation of Korea (NRF) grant funded by the Korean government (MSIT) (grant number 2021R1A2C1011718). M.L.R.H. acknowledges support from the Michigan Society of Fellows.
ZYDL acknowledges support from NASA 80NSSC18K1095, the Jefferson Scholars Foundation, the NRAO ALMA Student Observing Support (SOS) SOSPA8-003, the Achievements Rewards for College Scientists (ARCS) Foundation Washington Chapter, the Virginia Space Grant Consortium (VSGC), and UVA research computing (RIVANNA). J.J.T. acknowledges support from NASA 21-XRP21-0064, The National Radio Astronomy Observatory is a facility of the National Science Foundation operated under cooperative agreement by Associated Universities, Inc. Y.Y. is supported by the International Graduate Program for Excellence in Earth-Space Science (IGPEES), World-leading Innovative Graduate Study (WINGS) Program of the University of Tokyo. 

JPW acknowledges support from NSF AST-2107841. S.N. acknowledges support from the National Science Foundation through the Graduate Research Fellowship Program under Grant No. 2236415. 
SPL and TJT acknowledge grants from the National Science and Technology Council of Taiwan 
106-2119-M-007-021-MY3 and 109-2112-M-007-010-MY3.
IdG acknowledges support from grant \seqsplit{PID2020-114461GB-I00}, funded by \seqsplit{MCIN/AEI/10.13039/501100011033}. H.-W.Y. acknowledges support from the National Science and Technology Council (NSTC) in Taiwan through the grant NSTC \seqsplit{110-2628-M-001-003-MY3} and from the Academia Sinica Career Development Award (AS-CDA-111-M03). W.K. was supported by the National Research Foundation of Korea (NRF) grant funded by the Korea government (MSIT) \seqsplit{(NRF-2021R1F1A1061794)}.
PMK acknowledges support from NSTC 108-2112-M-001-012, NSTC 109-2112-M-001-022 and NSTC 110-2112-M-001-057.
This work was supported by JSPS KAKENHI Grant Numbers JP16H05998, JP18H05440, JP21H04487(KT), JP21H00048(ST), JP18H05222, JP20H05844, JP20H05847 (YA), JP21H04495(ST,KT,YA,KS), and by NAOJ ALMA Scientific Research grant No.2019-13B (YA) and No.2022-20A (ST).
J.J.T. acknowledges support from NASA XRP 80NSSC22K1159.

%
\facilities{ALMA}
\software{astropy \citep{2013A&A...558A..33A},
matplotlib \citep{Hunter2007},
bettermoments \citep{2018Teague,2019Teague},
PVextractor \citep{2016Ginsburg},
APLpy \citep{aplpy2012,aplpy2019},
SLAM \citep{Aso_2023_slamv1},
CASA \citep{Mcmullin2007}}


\clearpage
\appendix

\section{Spectral energy distributions}\label{sec:app-SEDs}

\begin{figure*}
\centering
  \includegraphics[width=0.24\linewidth]{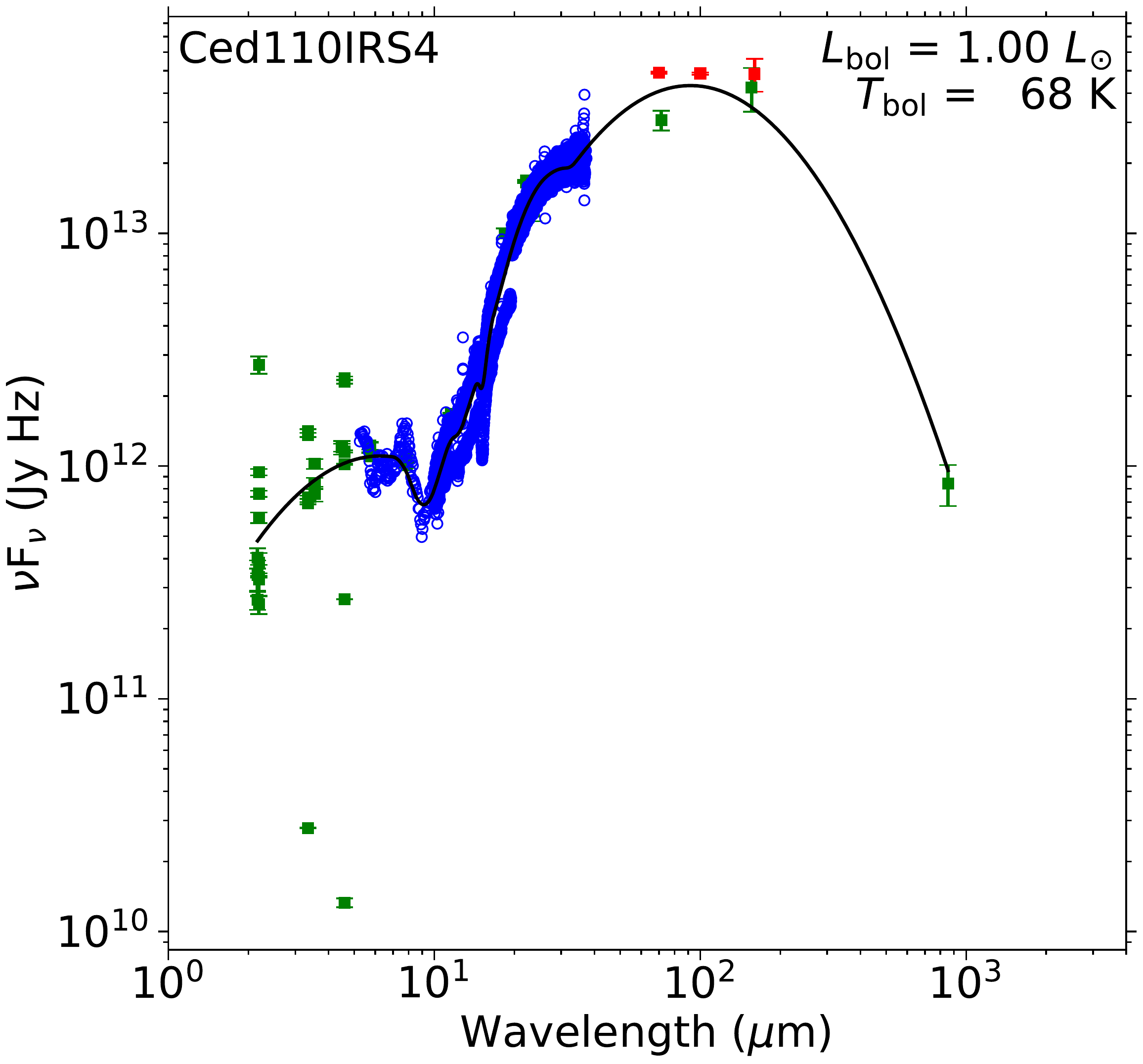}
  \includegraphics[width=0.24\linewidth]{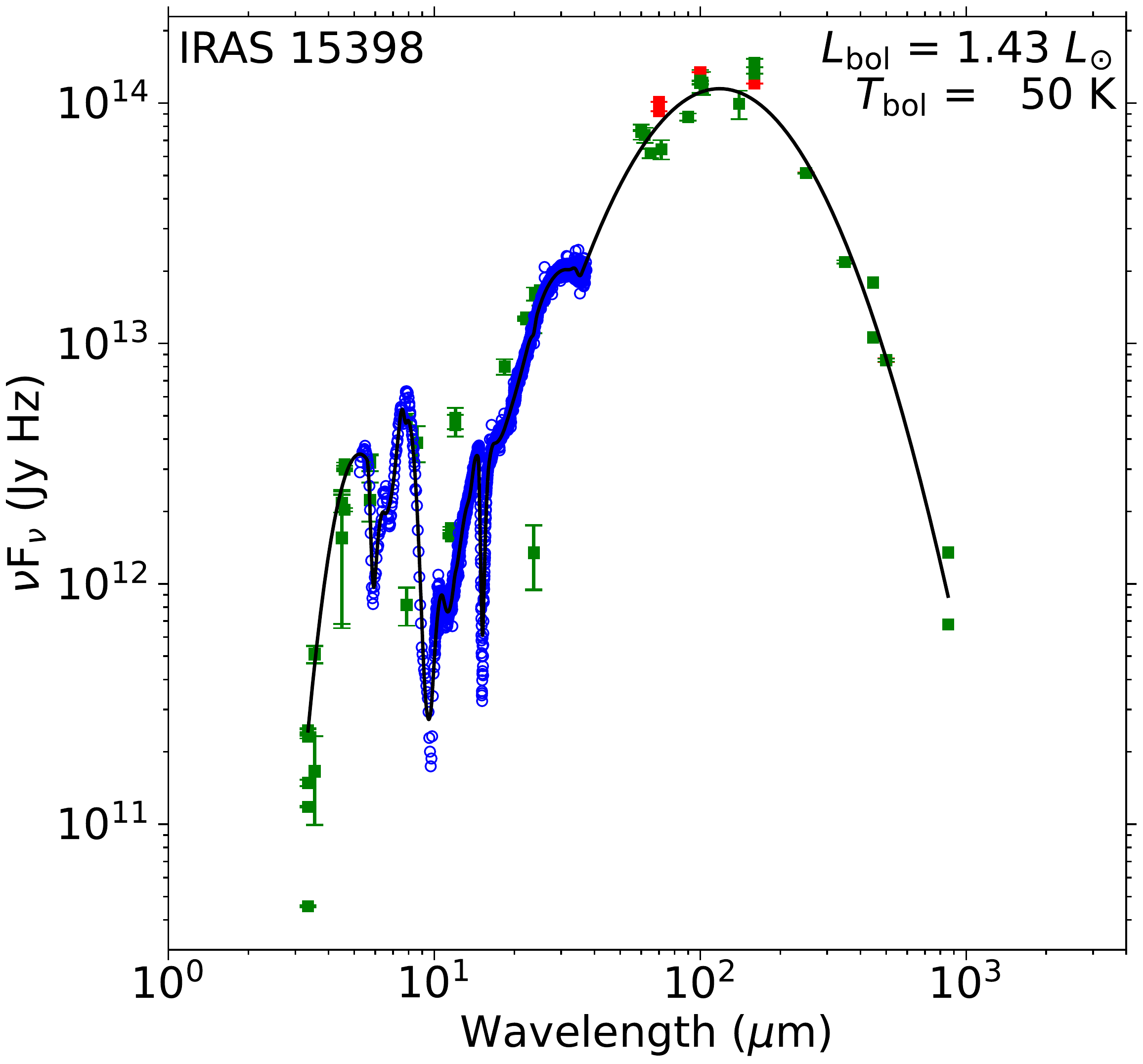}
  \includegraphics[width=0.24\linewidth]{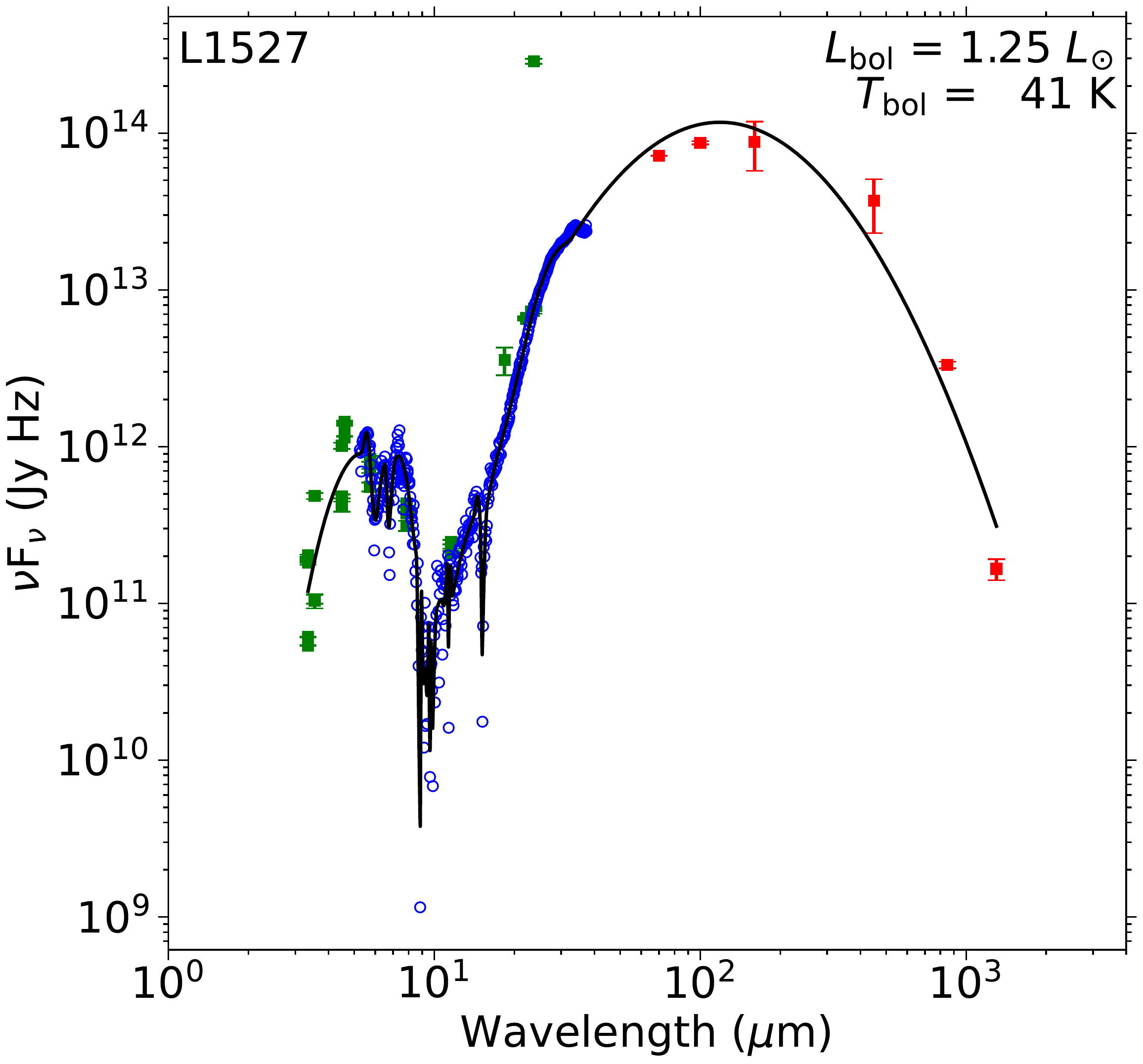}
  \includegraphics[width=0.24\linewidth]{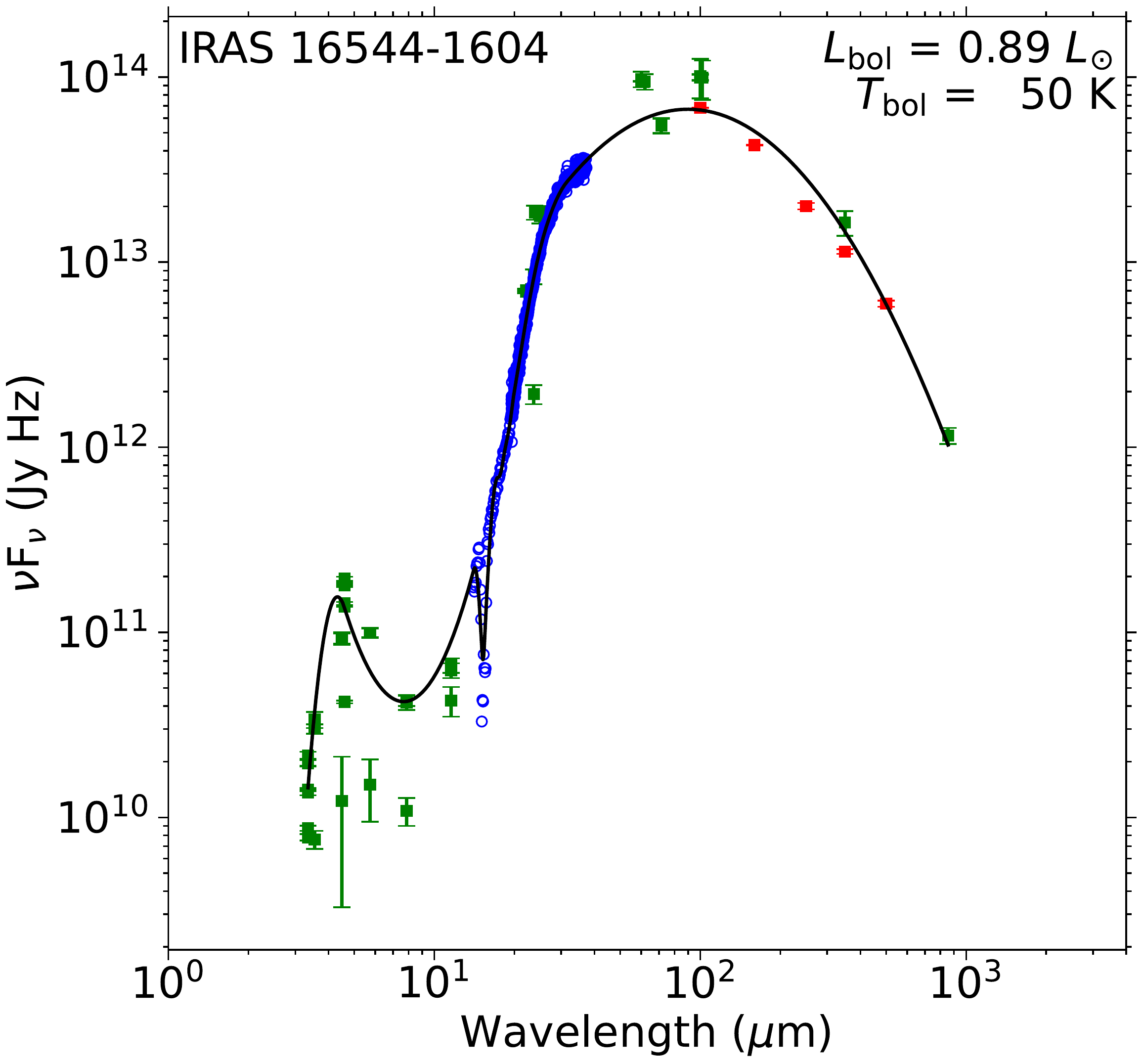}
  \includegraphics[width=0.24\linewidth]{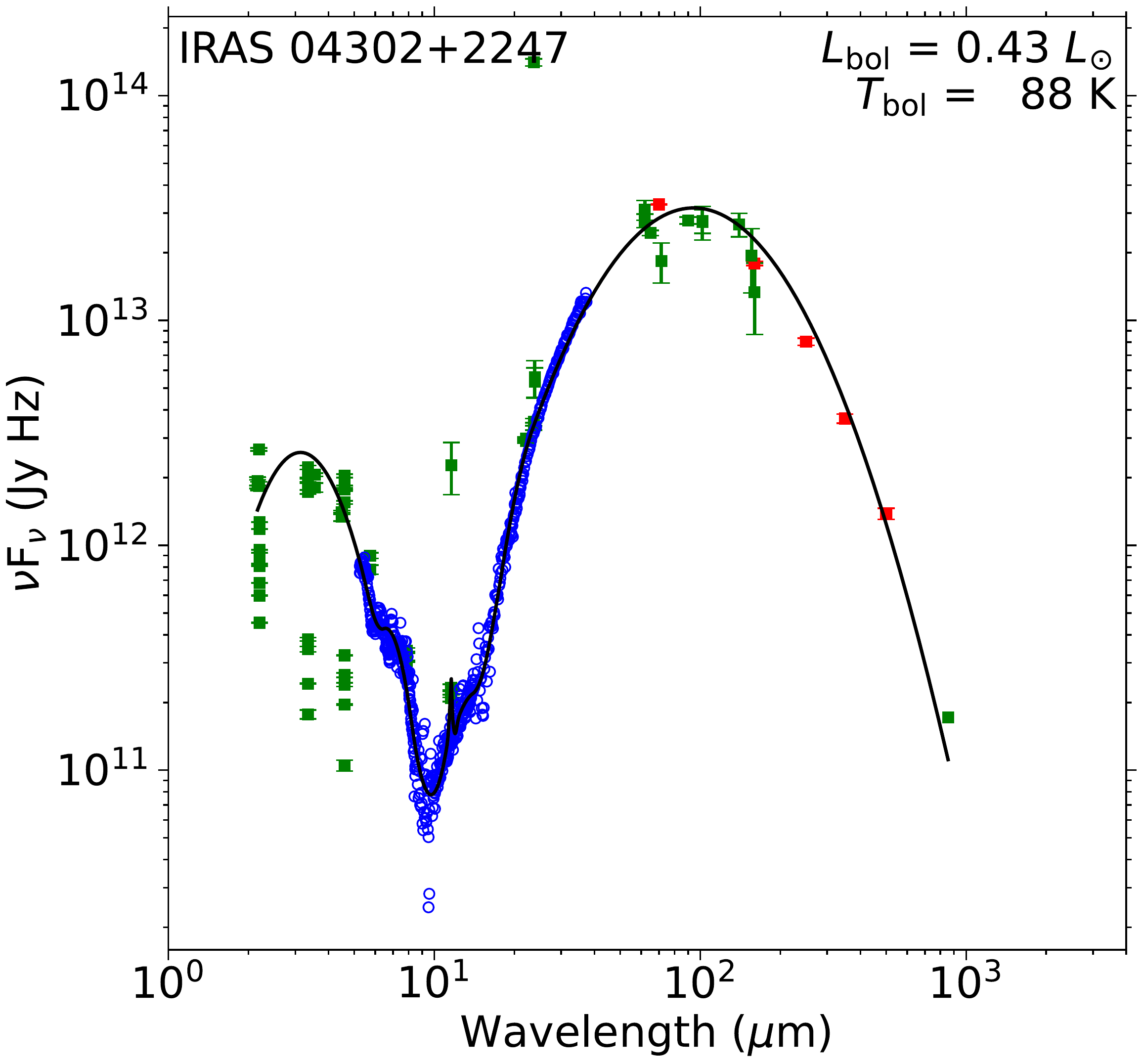}
  \includegraphics[width=0.24\linewidth]{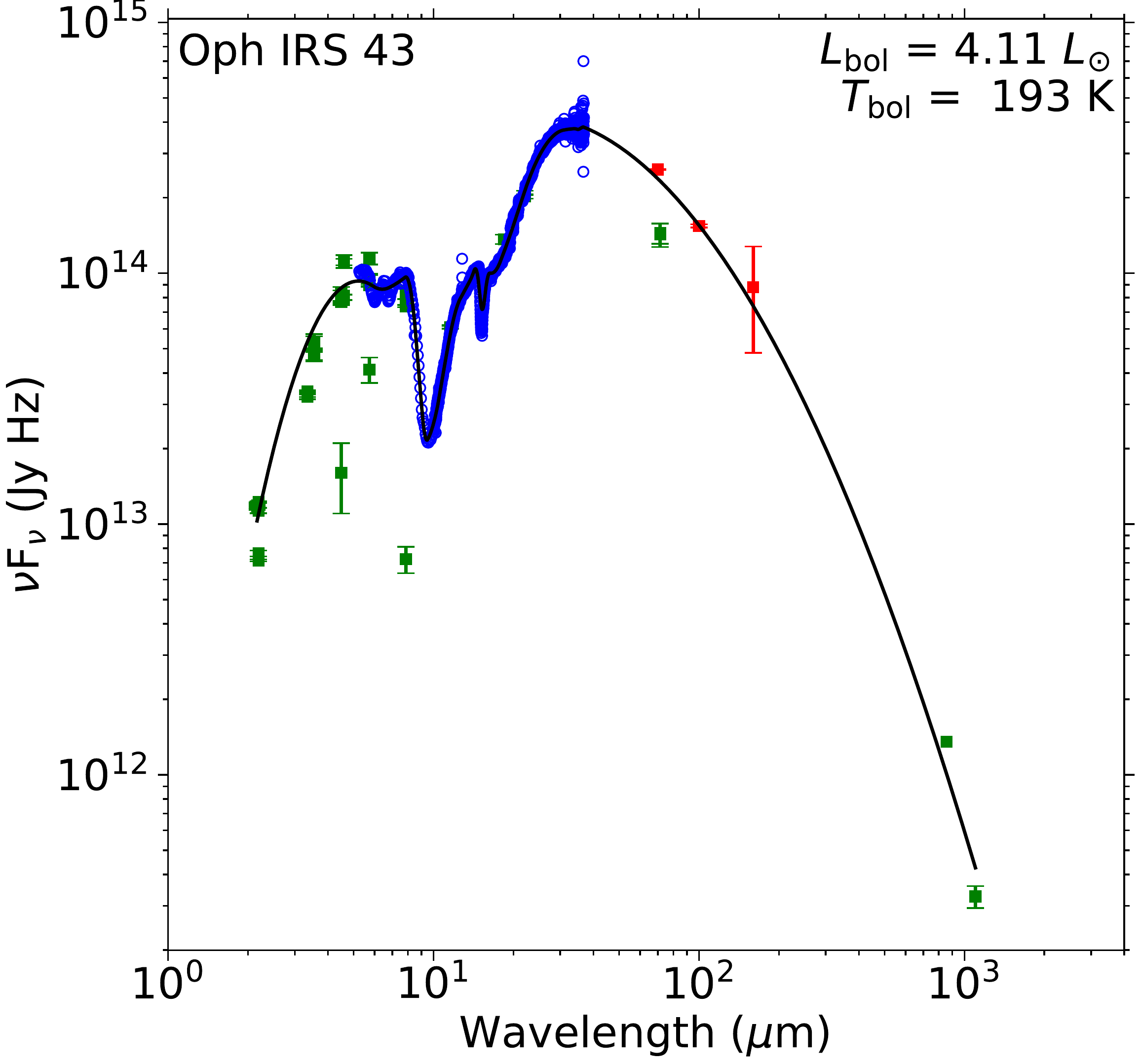}
  \includegraphics[width=0.24\linewidth]{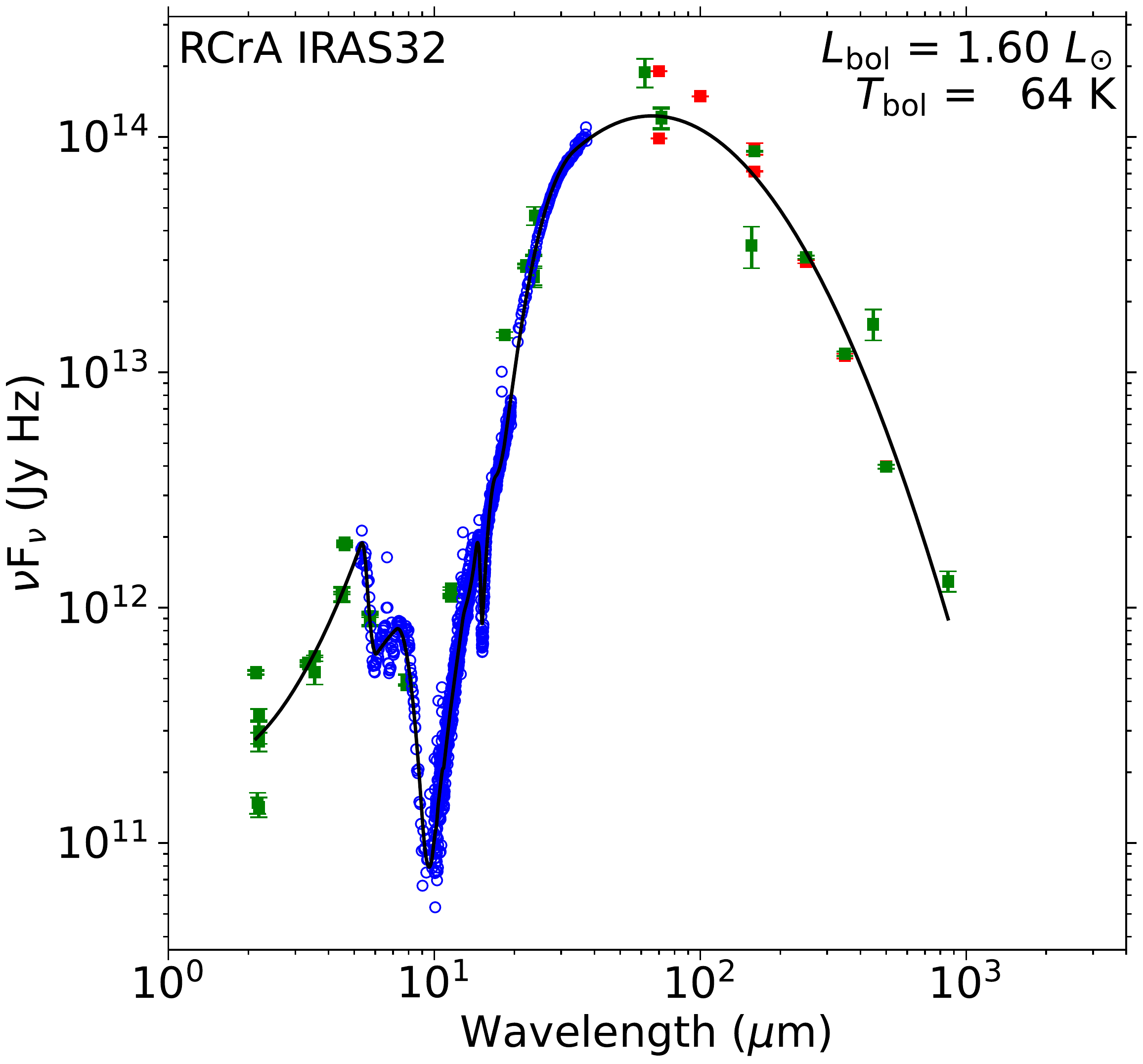}
  \includegraphics[width=0.24\linewidth]{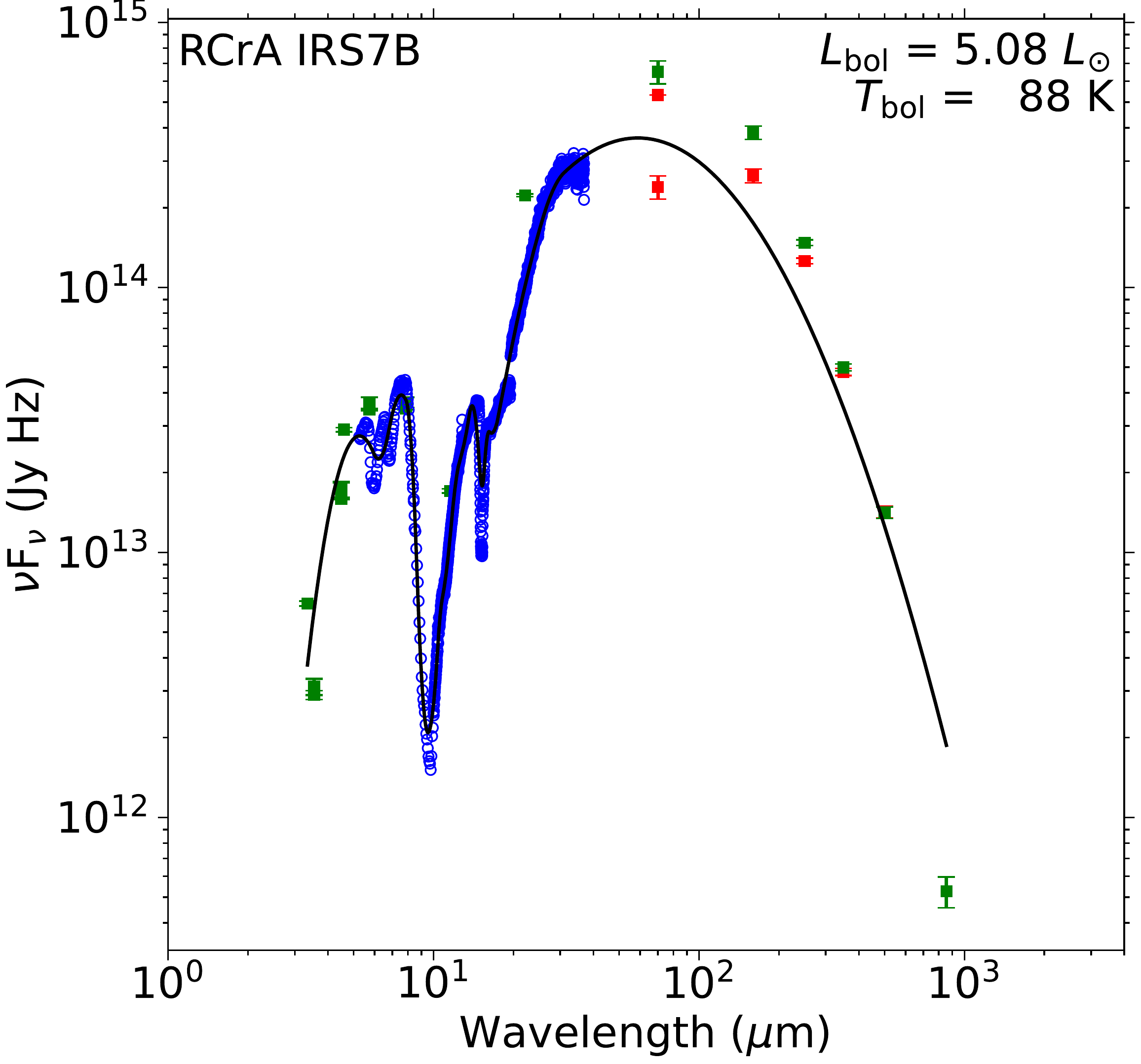}
  \includegraphics[width=0.24\linewidth]{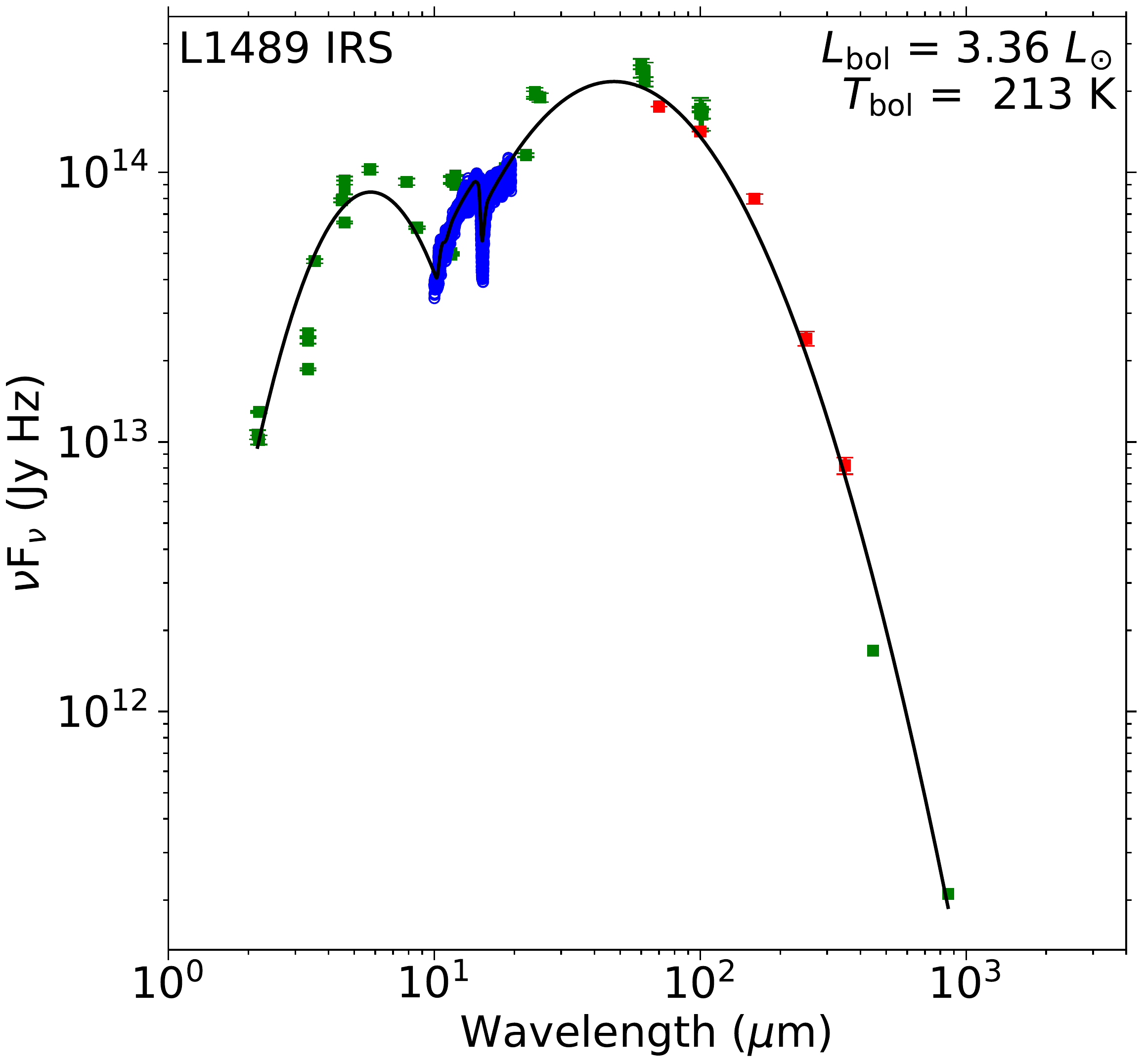}
  \includegraphics[width=0.24\linewidth]{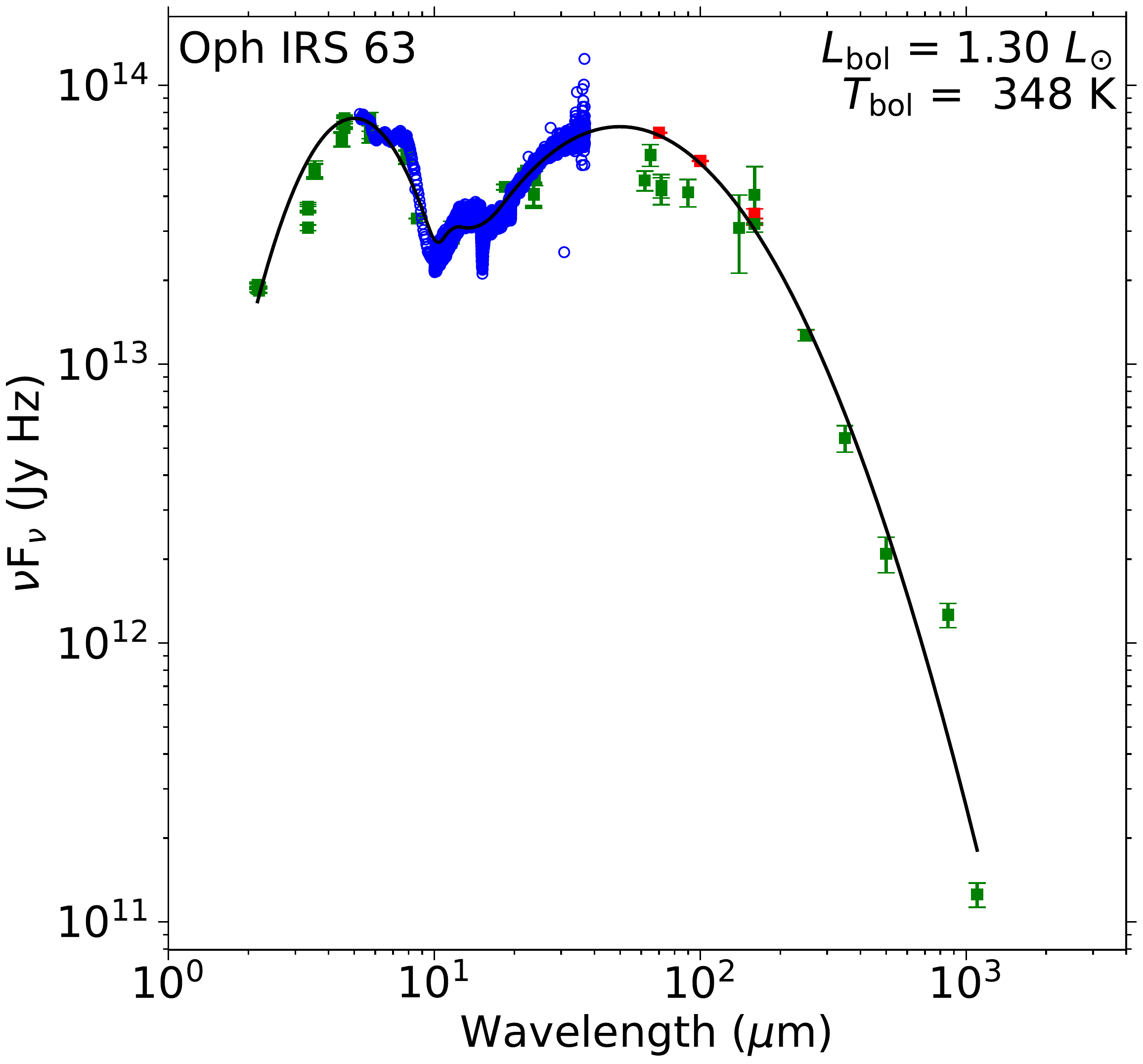}
  \includegraphics[width=0.24\linewidth]{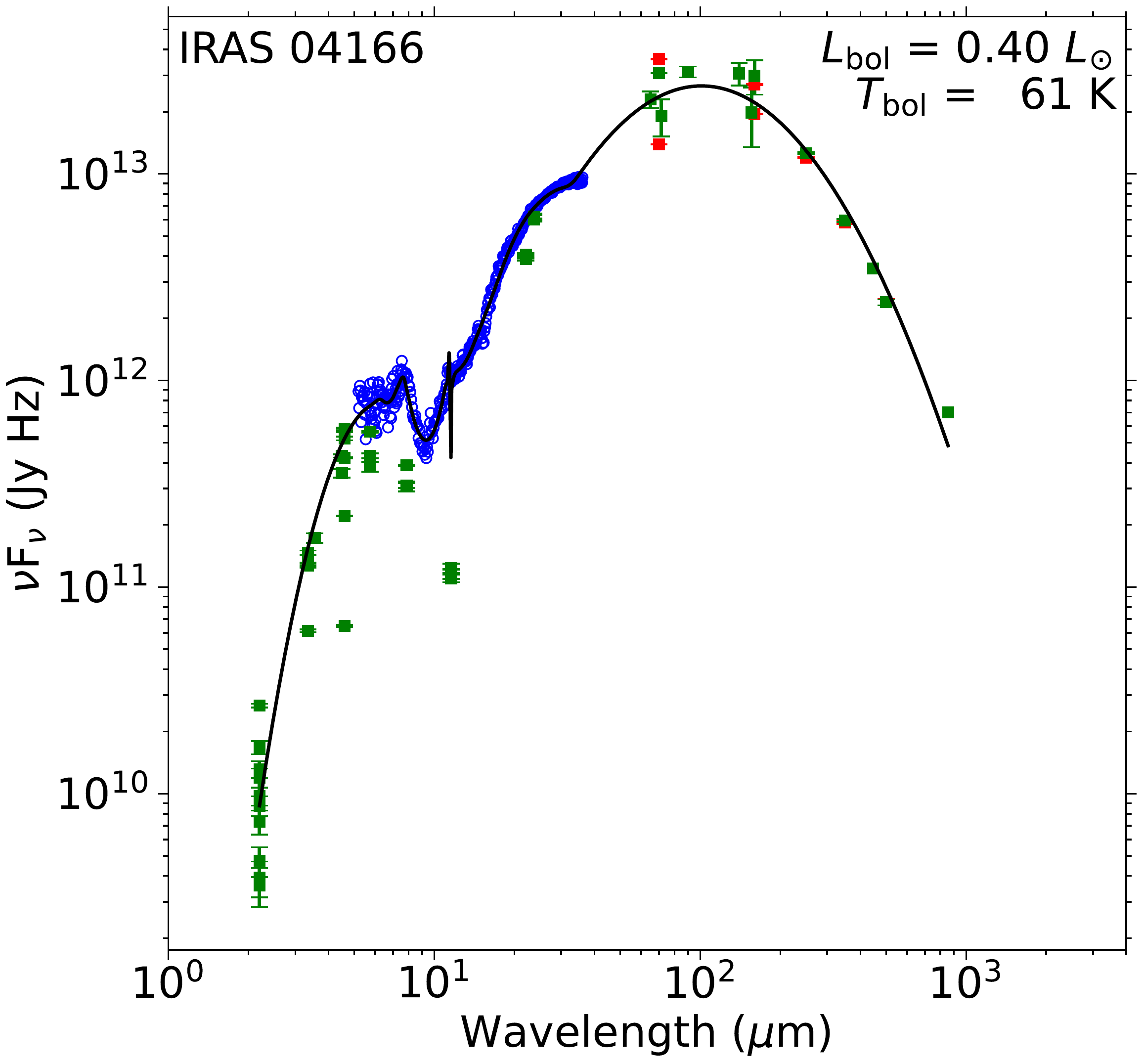}
  \includegraphics[width=0.24\linewidth]{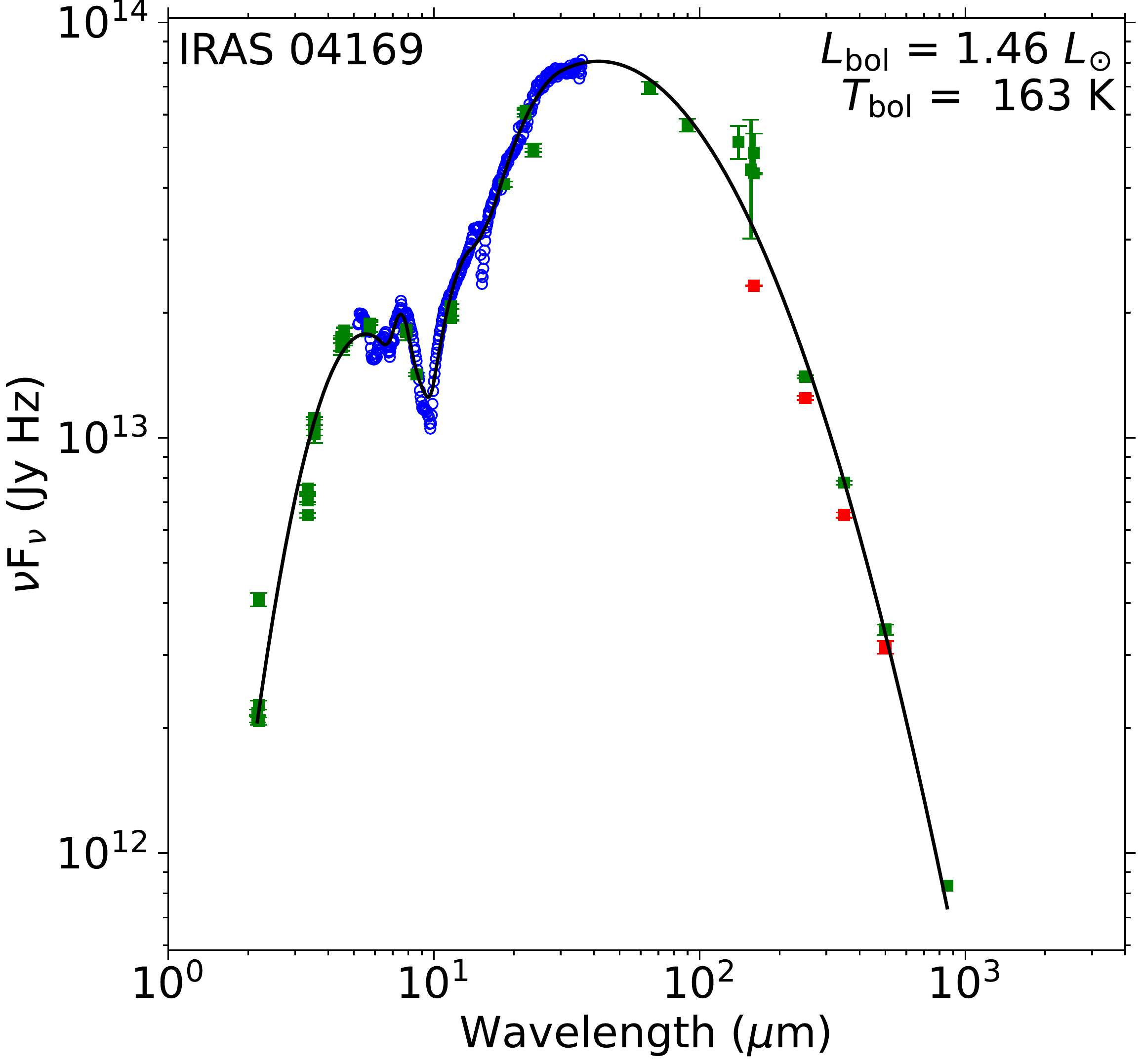}
  \includegraphics[width=0.24\linewidth]{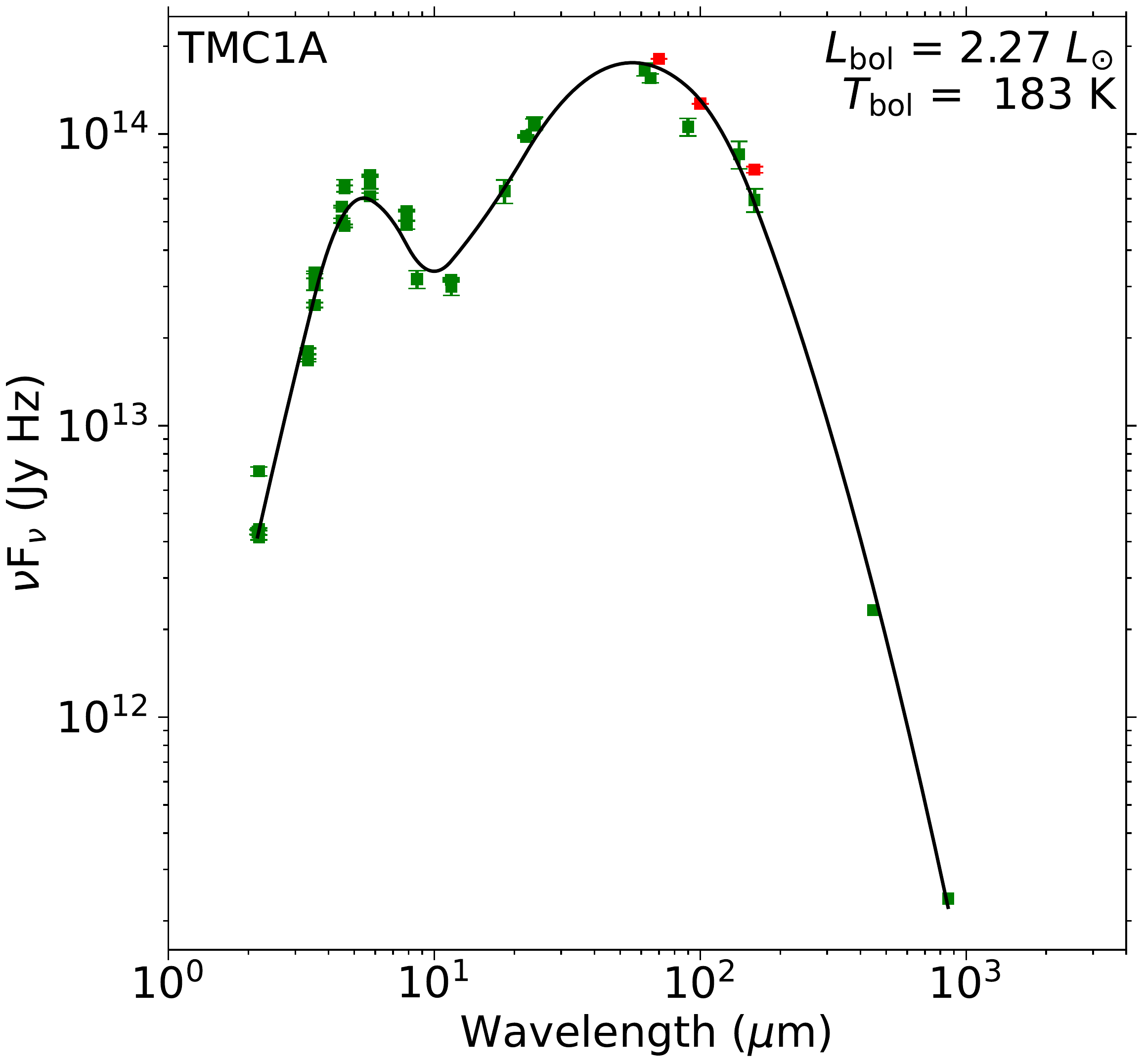}
  \includegraphics[width=0.24\linewidth]{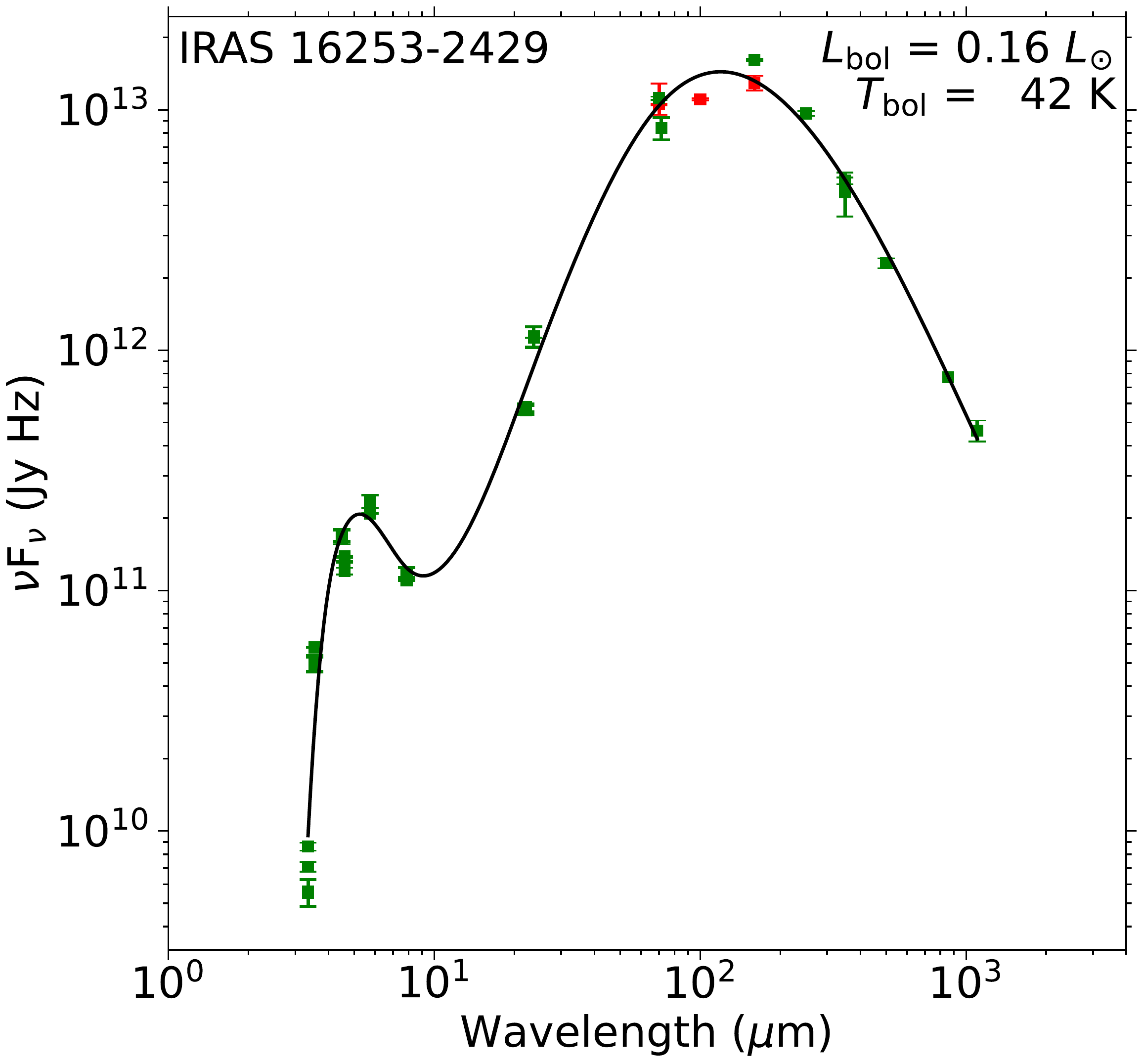}
  \includegraphics[width=0.24\linewidth]{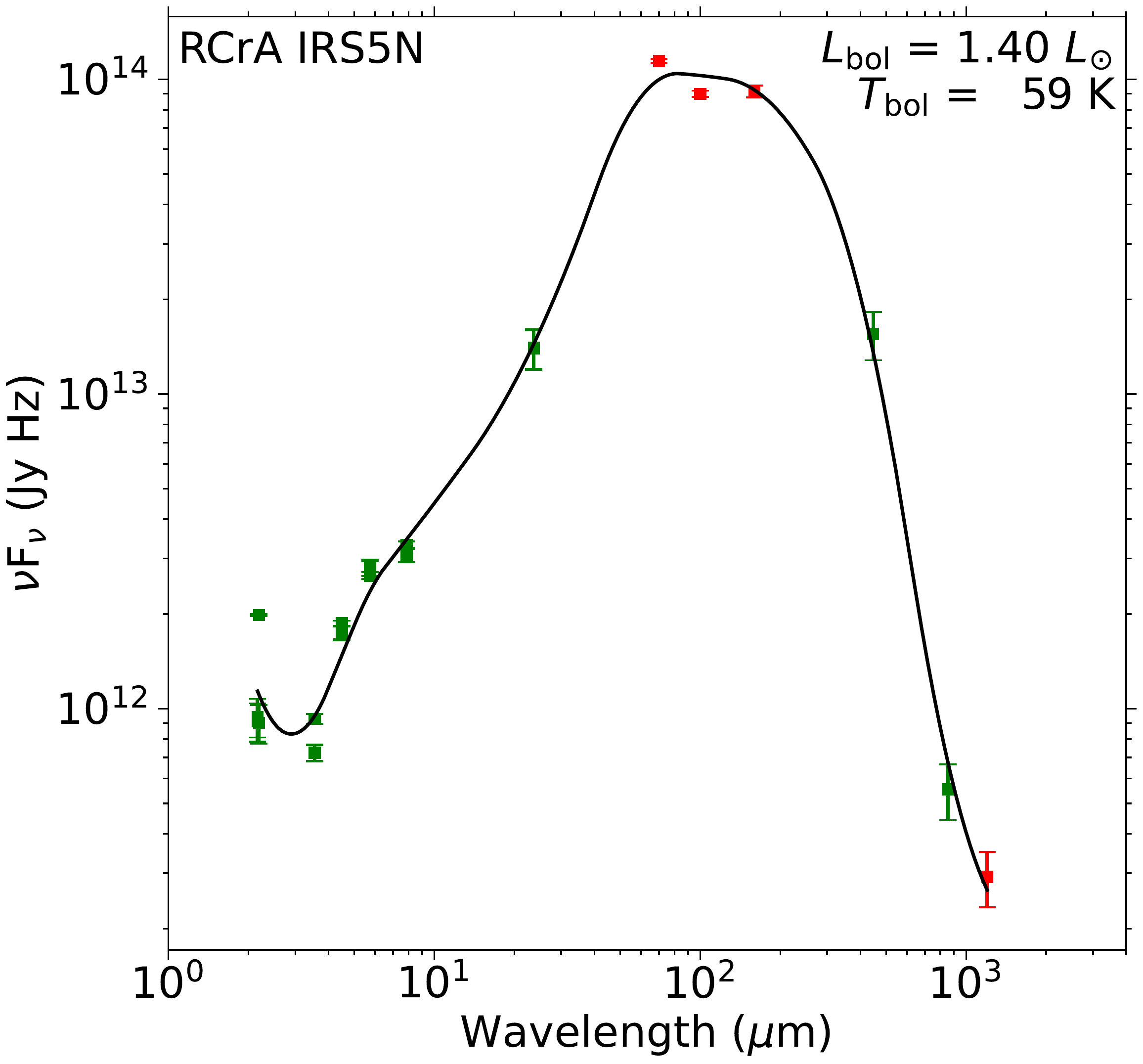}
  \includegraphics[width=0.24\linewidth]{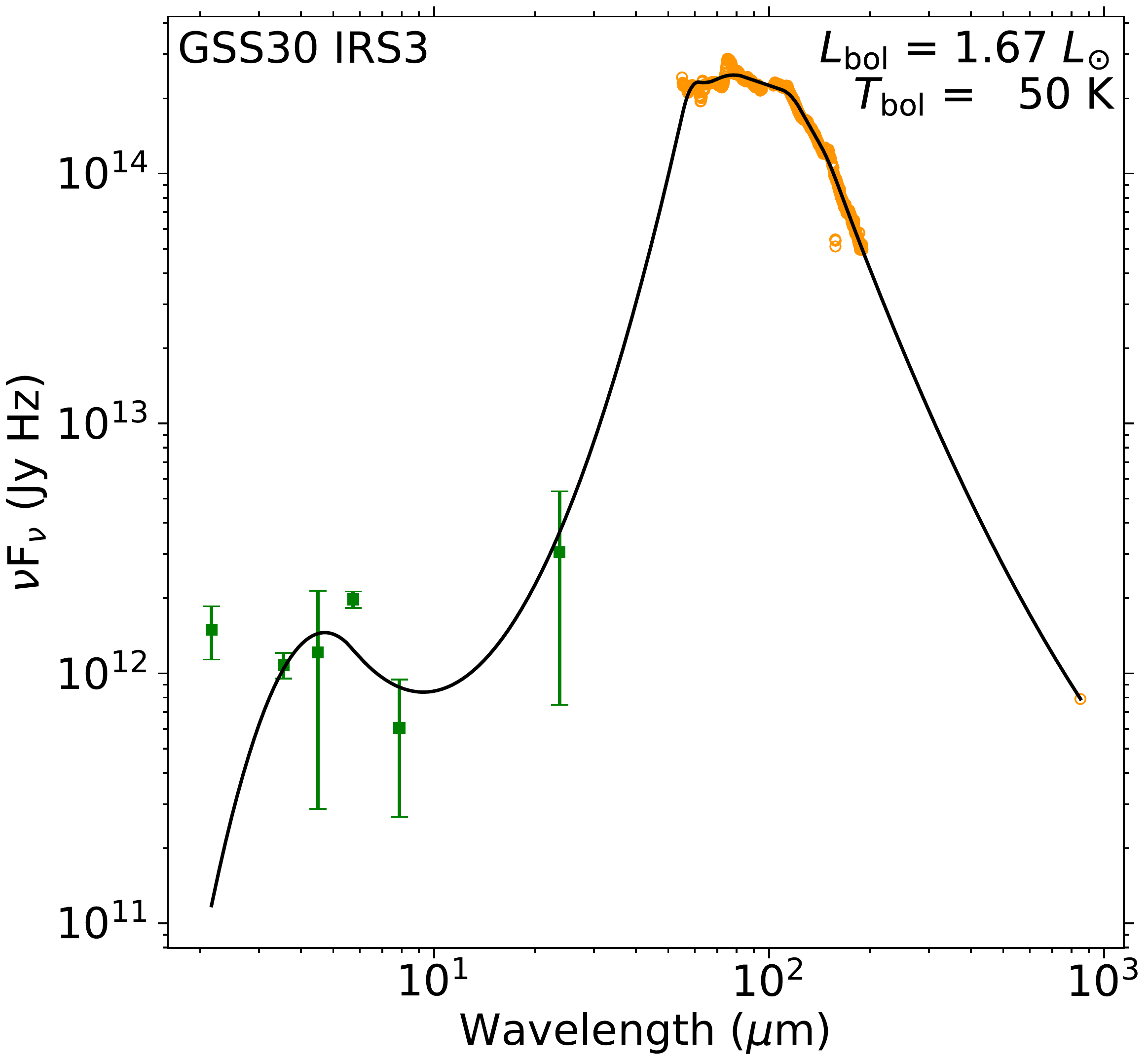}
  \includegraphics[width=0.24\linewidth]{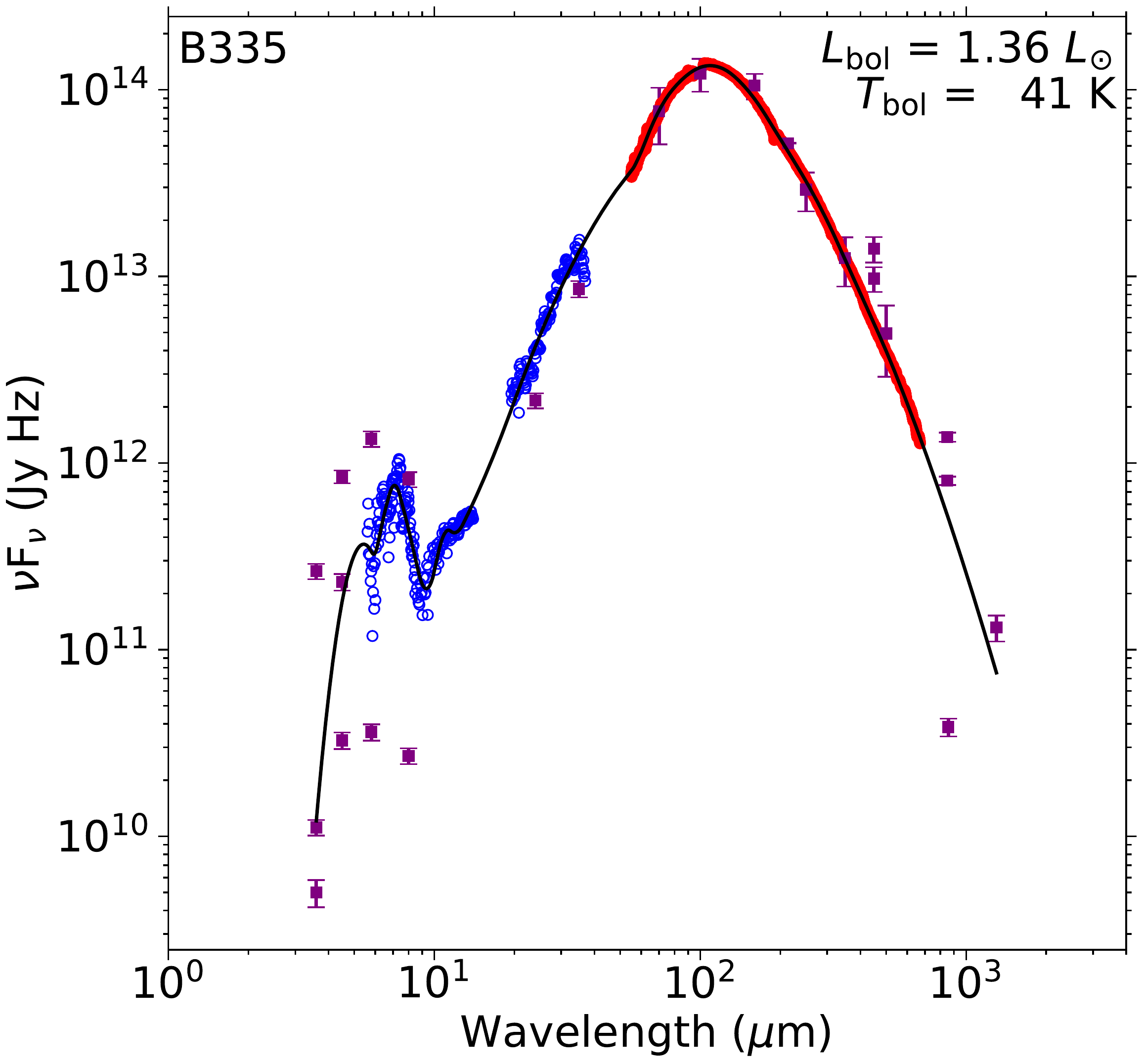}
  \includegraphics[width=0.24\linewidth]{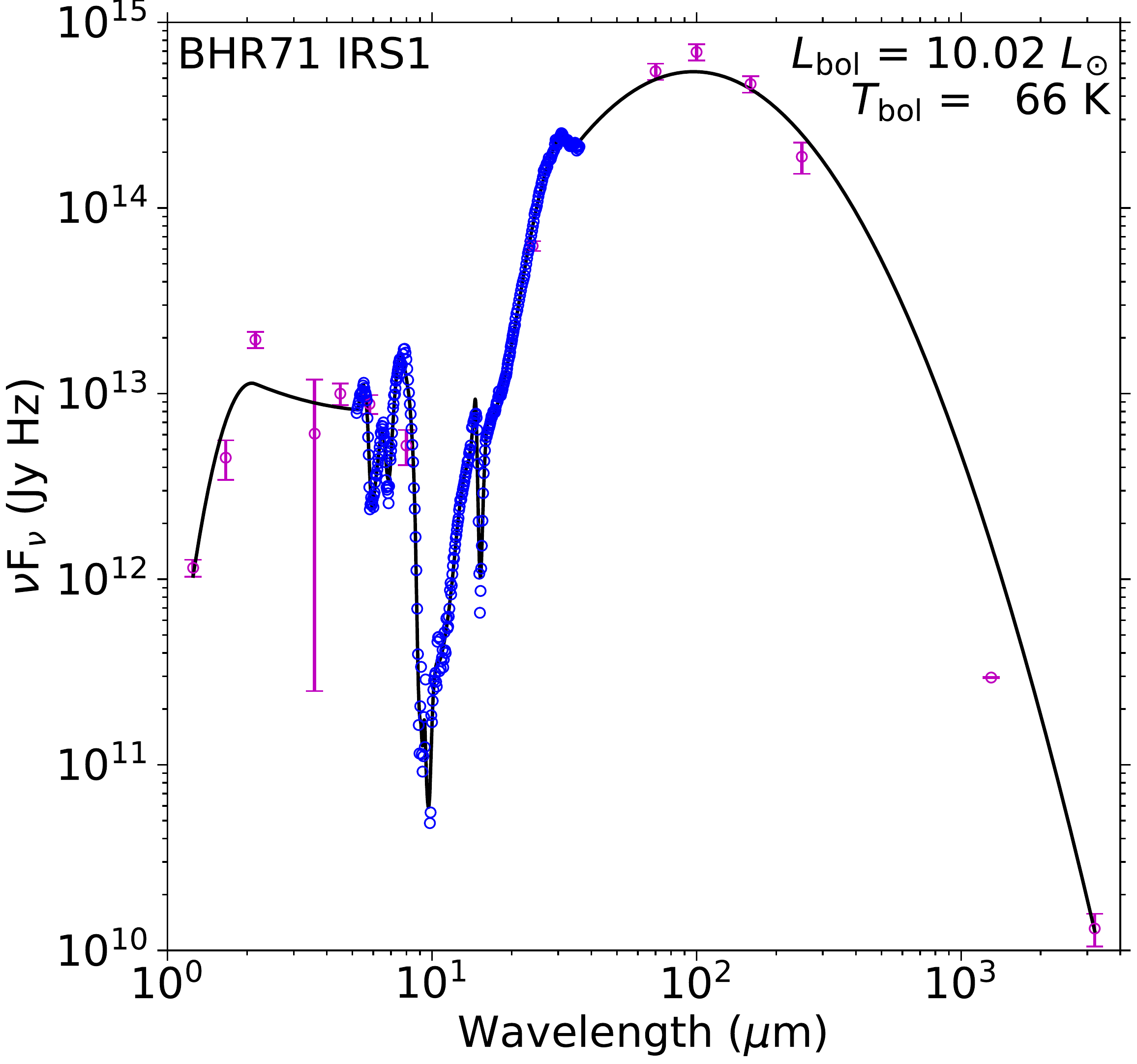}
  \includegraphics[width=0.24\linewidth]{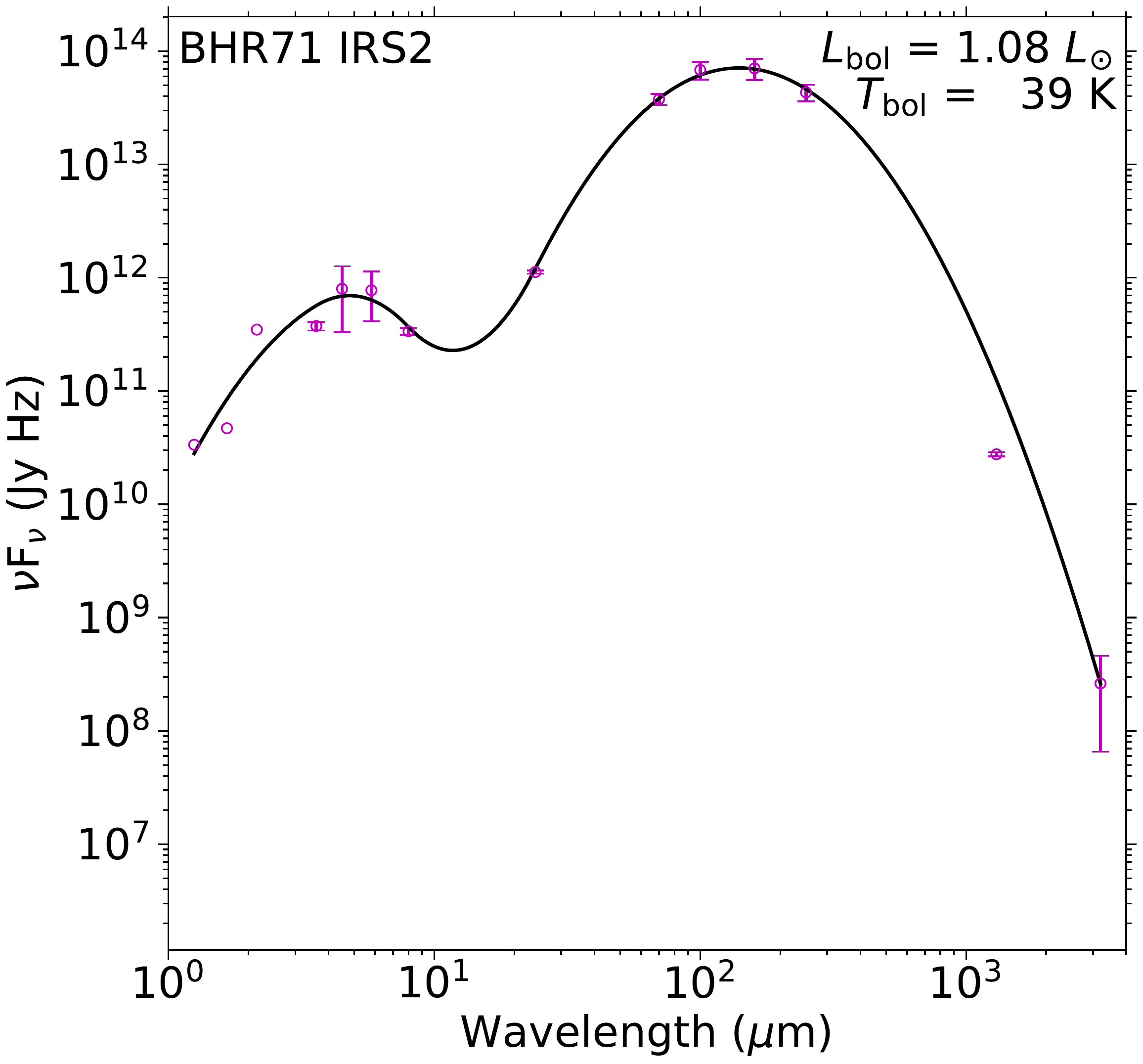}

  \caption{Spectral energy distributions from the near-infrared to millimeter ranges of the eDisk sources. Photometric data and spectroscopic data are shown in filled squares and empty circles, respectively. Flux data taken from the IRSA database are shown in green and \textit{Spitzer} IRS spectra are shown in blue. \textit{Herschel} PACS/SPIRE data are shown in red. Deconvolved GSS30 IRS3 data taken from \citet{Je_2015} are shown in orange. B335 flux data shown in purple are taken from \citet{2023ApJ...943...90E} and \textit{Herschel} PACS/SPIRE spectral data shown in red are taken from \citet{Green_2016}. Photometry data for BHR71 IRS1 and IRS2 data taken from \citet{Tobin_2019} are shown in magenta.  The black line represents the spline fit of the SED values used to calculate each source's  $L_{\textrm{bol}}$ and $T_{\textrm{bol}}$. \label{fig:seds}}
\end{figure*}

The spectral energy distributions (SEDs) for the sources in the eDisk sample are shown in Fig.~\ref{fig:seds}. The majority of the spectral data for the sources were compiled from the NASA/IPAC Infrared Science Archive (IRSA) database\footnote{https://irsa.ipac.caltech.edu/}. We searched toward the position of each source and include flux values and errors from catalog positions within 5$\arcsec$ of the source coordinates. These include data from \textit{Herschel} (PACS and SPIRE), \textit{Spitzer} (IRAC and MIPS), the SCUBA bolometer on the JCMT, WISE/NeoWISE, and \textit{Akari}. In addition, \textit{Spitzer} IRS spectra were available from the CASSIS database\footnote{https://cassis.sirtf.com/atlas/query.shtml} for some sources, which have also been included in the plots. In the eDisk sample, BHR71 IRS1 and IRS2 constitute a binary system and GSS30 IRS3 is part of a triple system each separated by of order 10$\arcsec$ from their companions. Hence, in particular, the long wavelength flux points include unresolved emission from all companions and the flux values from IRSA require deconvolution before they are used for SED fits. For BHR71 IRS1 and IRS2, the deconvolved values are taken from \citet{Tobin_2019}, whereas for GSS30 IRS3, the deconvolved values over 50 $\mu$m are taken from \citet{Je_2015}. For the closer binaries/multiples in the sample, the bolometric temperature and luminosities refer to the overall values for the systems.

For each source, we perform spline-fits to the SEDs and use these to calculate the $L_{\textrm{bol}}$ and $T_{\textrm{bol}}$ of the sources. The $L_{\textrm{bol}}$ is computed by integrating all the points on the spline fits with the \textit{trapezoid} function of the \textit{scipy} package. The $T_{\textrm{bol}}$ for each of the sources are calculated using the average frequency of the SED fits following the methods in \citet{Myers_1993}.

\section{Observation logs}\label{sec:app-obslog}
Here we present a table of the observation logs for the eDisk program (see Table~\ref{tab:logs}). In addition, we present a table of observation logs for the ALMA archival data we used in the eDisk Program (see Table~\ref{tab:archive-logs}).

\section{Scale factors}\label{sec:app-scale-factor}
Here we present a scale factor we applied to each EB in Table~\ref{tab:scale-factor}. See more details in Sec. \ref{sec:reduction}.

\section{Fittings to position-velocity diagrams}\label{sec:app-PV-fit}

One of the methods to identify Keplerian rotation, which is also adopted by the eDisk program, is fitting to position-velocity (PV) diagrams. Such an analytic approach has advantages of the lower computational cost and a simpler process than model fitting to visibility data including radiative transfer with ten or more free parameters \citep{2019ApJ...874..136S}, although the robustness of fitting the PV diagram would be less than that of fittings to visibility data due to fitting in the image vs. uv-plane and the reduced spatial dimension.
The analytic method also enables us to focus on a specific quantity, such as the central stellar mass, without complicated parameter degeneracy. Our fitting process includes two independent methods. One method uses the ridge in 1D intensity profiles along the positional axis (the velocity axis) at a given velocity (position). This ridge method is demonstrated in previous observational studies \citep{2014ApJ...796..131O, 2017ApJ...849...56A, 2017ApJ...834..178Y, 2020ApJ...893...51S} with a discussion about its potential problems \citep{2013ApJ...772...22Y, 2015ApJ...812...27A}, as well as in a study using synthetic observations of a magnetohydrostatic (MHD) simulation of protostellar evolution \citep{2020ApJ...905..174A}. The other method uses the outer edge in the 1D intensity profiles. This edge method was developed using an MHD simulation of protostellar evolution \citep{2016MNRAS.459.1892S} and is used in previous observational studies as well \citep[e.g.,][]{2017A&A...603L...3A,2021ApJ...907L..10R}. The advantage and disadvantage of each method are debatable. For example, \citet{2020A&A...635A..15M} reported that, depending on the spatial resolution, the ridge method can underestimate the central stellar mass by $\sim 30\%$, while the edge method can overestimate it by a factor of $\sim 2$. For this reason, we regard the radius, or central stellar mass, derived from the ridge method as the lower limit and the one derived from the edge method as the upper limit.

In the eDisk program, the fitting process with the edge and ridge methods is implemented through the public Python package \texttt{pvanalysis} in  \texttt{SLAM} \citep{Aso_2023_slamv1}\footnote{https://github.com/jinshisai/SLAM}. The process is composed of two main steps: obtaining the edge and ridge points in a PV diagram and fitting the obtained points with a power-law function. First, the edge and ridge radii are obtained in the 1D intensity profile along the positional axis at the given velocity (xcut). The edge radius is defined as the outermost position with a given threshold level ($3\sigma$ in this paper) of emission. The uncertainty of the edge radius is calculated from the emission gradient and the observational noise level. The ridge radius is defined as the intensity-weighted mean position in the 1D profile or the center derived from the Gaussian fitting. This fitting also calculates an uncertainty of the ridge radius from the observational noise level. Similarly, the edge and ridge velocities are obtained in the 1D profile along the velocity axis at a given position (vcut). Then, in each of edge and ridge methods, the xcut points at velocities higher than a ``middle" velocity and the vcut points at radii larger than a ``middle" radius are combined as the final pairs of $(r, v)$. The middle velocity and radius are determined from the closest pair of points between the xcut and vcut, as explained in \citet{2020ApJ...905..174A} in more detail. Note that 1D intensity profiles along only one direction (position or velocity) would be enough to obtain the edge and ridge points if a PV-diagram is mainly elongated along the velocity or positional axis, respectively.

Second, the obtained pairs of $(r, v)$ are fitted with a power-law function, in the edge and ridge methods separately. The function is a single- or double-power function as follows:
\begin{eqnarray}
v = {\rm sign}(r)\ v_b \left(\frac{|r|}{r_b}\right) ^{-p}+v_{\rm sys}, \\
\label{eq:power}
p = p_{\rm in}\ \ {\rm if}\ \ |r|<r_b\ \ {\rm else}\ \ p_{\rm in} + dp,
\end{eqnarray}
where $v_b$ is the velocity at the radius of $r_b$, $p_{\rm in}$ is the power-law index, $dp$ is the difference of the index between the inner and outer parts, and $v_{\rm sys}$ is the systemic velocity. $dp$ is limited to be $\geq 0$ to produce a steeper or the same index outside $r_b$, such as the inner Keplerian rotation versus the outer rotation where the angular momentum is conserved. For the single-power fitting, $v_b$ is fixed at a middle value and $dp=0$. Equation \ref{eq:power} provides a function of radius, defined here as $V_{\rm fit}(r)$. When $dp \geq 0$, the inverse function of velocity $R_{\rm fit}(v)$ can be defined. Using these functions, $\chi ^2$ to be minimized is defined as 
\begin{eqnarray}
\chi ^2 &=& \sum _{i~\rm for~xcut} \left(r_{i, {\rm obs}}- R_{\rm fit}(v_{i, {\rm obs}})\right) / \sigma^{2}_{i, {\rm obs}} \nonumber \\
&&+ \sum _{j~\rm for~vcut} \left(v_{j, {\rm obs}} - V_{\rm fit}(r_{j, {\rm obs}})\right)^2 / \sigma^{2}_{j, {\rm obs}},
\end{eqnarray}
where $\sigma$ is the uncertainty of each radius/velocity obtained above.

The power-law fitting provides the pair of $(r_b, v_b)$. From this pair, the central stellar mass can be calculated as $M_* = v_b ^2 r_b / G / \sin ^2 i$, where $G$ is the gravitational constant and $i$ is an inclination angle. The inclination angle is often estimated from the aspect ratio of the associated continuum emission or independent modeling of the associated outflow, which is beyond the scope of this fitting with PV diagrams. If the power-law index $p_{\rm in}$ is significantly different from the Keplerian index 0.5, the estimated $M_*$ does not have the meaning of the central stellar mass.

\movetabledown=15mm
\begin{longrotatetable}
\begin{deluxetable*}{lllcclclc}
\tablecaption{eDisk observing logs (2019.1.00261.L and 2019.A.00034.S)} \label{tab:logs}
\tablewidth{0pt}
\tablehead{
\colhead{Source name} & \colhead{UTC Date} & \colhead{Config.} & \colhead{Baselines} &
\colhead{N$_{ant}$} & \colhead{Elev.} & \colhead{PWV} & \colhead{Calibrators} & \colhead{Check source}\\
\colhead{} & \colhead{} & \colhead{} & \colhead{(m)} &
\colhead{} & \colhead{(deg.)} & \colhead{(mm)} &\colhead{} &\colhead{}
}
\decimalcolnumbers
\startdata
L1489 IRS & 2021 Aug 20 07:43 & C43-8 & 59--9934 & 43 & 31.7 & 0.7 & J0423--0120, J0423--0120, J0357+2319 & J0401+2110 \\
 & 2021 Aug 20 09:29 & C43-8 & 59--9934 & 46 & 40.0 & 0.5 &
J0510+1800, J0510+1800, J0357+2319 & J0401+2110 \\
L1489 IRS, L1527 IRS & 2021 Dec 03 03:02 & C43-5 & 15--2617 & 46 & 40.4 & 1.8 & J0510+1800, J0510+1800, J0438+3004 & J0435+2532 \\
 & 2021 Dec 16 03:29 & C43-5 & 15--1917 & 42 & 37.2 & 2.3 & J0510+1800, J0510+1800, J0438+3004 & J0435+2532 \\
 & 2022 Jul 03 12:58 & C43-5 & 15--1301 & 42 & 36.3 & 1.3 & J0237+2848, J0237+2848, J0438+3004 & J0418+3801 \\
IRAS 04166+2706, IRAS 04169+2702 & 2021 Sep 30 05:20$^{\dagger}$ & C43-8 & 70--11886 & 44 & $\cdots$ & 0.73 &
J0238+1636, J0238+1636, J0433+2905 & J0429+2724 \\
 & 2021 Oct 01 05:22$^{\dagger}$ & C43-8 & 70--10803 & 45 & $\cdots$ & 1.77 &
 J0238+1636, J0238+1636, J0438+3004 & J0440+2728 \\
 & 2021 Oct 18 04:32$^{\dagger}$ & C43-8 & 46--8983 & 43 & $\cdots$ & 0.70 & 
 J0238+1636, J0238+1636, J0438+3004 & J0435+2532 \\
 & 2021 Oct 24 07:20 & C43-8 & 46--8983 & 45 & 35.2 & 0.26 &
 J0510+1800, J0510+1800, J0438+3004 & J0440+2728 \\
  & 2022 Jul 3 13:15 & C43-5 & 15-1301 & 41 & 39.5 & 1.40 &
 J0510+1800, J0510+1800, J0438+3004 & J0435+2532 \\
 & 2022 Jul 3 14:26 & C43-5 & 15-1301 & 41 & 37.4 & 1.49 &
 J0510+1800, J0510+1800, J0438+3004 & J0435+2532 \\
IRAS 04302+2247 & 2021 Sep 30 08:48 & C43-8 & 70-11886 & 45 & $\cdots$ & 0.9 &
J0510+1800, J0510+1800, J0426+2327 & J0425+2235 \\
 & 2021 Oct 01 06:54 & C43-8 & 70--10803 & 43 & $\cdots$ & 2.1 &
J0510+1800, J0510+1800, J0426+3227 & J0425+2235 \\
 & 2021 Oct 01 08:31 & C43-8 & 70-10803 & 45 & 42.1 & 1.8 & 
 J0510+1800, J0510+1800, J0426+2327 & J0425+2235 \\
 & 2021 Dec 01 03:40 & C43-5 & 15-3638 & 48 & 43.8 & 1.0 &
 J0510+1800, J0510+1800, J0426+2327 & J0425+2235 \\
L1527 IRS & 2021 Oct 14 08:26 & C43-8 & 91-11469 & 42 & 35.5 & 0.5 &
J0510+1800, J0510+1800, J0438+3004 & J0433+2905 \\
 & 2021 Oct 15 06:01 & C43-8 & 91-8983 & 41 & 39.3 & 0.8 &
 J0510+1800, J0510+1800, J0438+3004 & J0433+2905 \\
 & 2021 Oct 15 07:51 & C43-8 & 91-8983 & 41 & 38.2 & 0.7 &
 J0510+1800, J0510+1800, J0438+3004 & J0433+2905 \\
Ced 110 IRS4 & 2021 Apr 24 01:30 & C43-5 & 15-1397 & 45 & 35.3 & 1.4 &
J1107-4449, J1107-4449, J1058-8003 & $\cdots$ \\
 & 2021 Oct 04 13:05 & C43-8 & 70-10803 & 42 & 35.2 & 0.3 &
 J0519-4546, J0519-4546, J1058-8003 & J1224-8313 \\
 & 2021 Oct 20 10:58 & C43-8 & 46-8983 & 47 & 33.7 & 0.7 &
 J0519-4546, J0519-4546, J1058-8003 & J0942-7731 \\
 & 2021 Oct 20 12:24 & C43-8 & 46-8983 & 44 & 35.4 & 0.7 &
 J1427-4206, J1427-4206, J1058-8003 & J0942-7731 \\
BHR 71 IRS1, BHR 71 IRS2 & 2021 May 04 00:26 & C43-5 & 15-2517 & 44 & 47.3 & 1.6 & J1107-4449, J1107-4449, J1147-6753 & J1206-6138 \\
 & 2021 May 09 01:26 & C43-5 & 15-2517 & 44 & 47.3 & 1.4 & J1107-4449, J1107-4449, J1147-6753 & J1206-6138 \\
 & 2021 Oct 14 14:48$^{\dagger}$ & C43-8 & 91-8983 & 41 & $\cdots$ & 0.88 & J1427-4206, J1427-4206, J1147-6753 & J1145-6954 \\
 & 2021 Oct 15 13:13 & C43-8 & 91-8983 & 43 & 46.7 & 0.5 & J1427-4206, J1427-4206, J1147-6753 & J1145-6954 \\
 & 2021 Oct 16 12:18$^{\dagger}$ & C43-8 & 46-8983 & 42 & $\cdots$ & 0.66 & J0635-7516, J0635-7516, J1147-6753 & J1145-6954 \\
 & 2021 Oct 19 14:34 & C43-8 & 91-8547 & 41 & 47.3 & 0.7 & J1427-4206, J1427-4206, J1147-6753 & J1145-6954 \\
 & 2021 Oct 20 13:57$^{\dagger}$ & C43-8 & 46-8983 & 43 & $\cdots$ & 0.76 & J1427-4206, J1427-4206, J1147-6753 & J1145-6954 \\
IRAS 15398-3359 & 2021 May 06 03:56 & C43-5 & 15-2517 & 45 & 74.9 & 1.2 & J1337-1257, J1337-1257, J1610-3958 & J1523-3338 \\
 & 2021 Aug 09 22:34 & C43-8 & 70-8282 & 39 & 77.9 & 0.7 & J1337-1257, J1337-1257, J1534-3526 & J1523-3338 \\
 & 2021 Aug 13 23:41 & C43-8 & 70-8282 & 42 & 66.3 & 1.7 & J1517-2422, J1517-2422, J1534-3526 & J1523-3338 \\
 & 2021 Oct 20 15:28 & C43-8 & 46-8983 & 43 & 59 & 0.8 & J1427-4206, J1427-4206, J1534-3526 & J1536-3151 \\
GSS 30 IRS3, IRAS 16253-2429 & 2021 Oct 05 22:14 & C43-8 & 70-11614 & 46 & 49.7 & 3.5 & J1517-2422, J1517-2422, J1633-2557 & J1650-2943 \\
 & 2021 Oct 26 19:22 & C43-8 & 63-8547 & 45 & 71.1 & 3.3 & J1427-4206, J1427-4206, J1633-2557 & J1642-2849 \\
 & 2021 Oct 27 20:35 & C43-8 & 63-8547 & 46 & 52.2 & 1.3 & J1427-4206, J1427-4206, J1633-2557 & J1642-2849 \\
 & 2021 Oct 28 17:02 & C43-8 & 63-8282 & 43 & 79.7 & 1.0 & J1427-4206, J1427-4206, J1633-2557 & J1642-2849 \\
GSS 30 IRS3, IRAS 16253-2429, Oph IRS63 & 2022 Jun 14 04:13 & C43-5 & 15-1301 & 42 & 70.3 & 0.3 & J1517-2422, J1517-2422, J1700-2610 & J1650-2943 \\
 & 2022 Jun 15 00:45 & C43-5 & 15-1301 & 43 & 62.8 & 0.5 & J1517-2422, J1517-2422, J1700-2610 & J1650-2943 \\
Oph IRS 43 & 2021 May 26 07:32 & C43-5 & 15-2280 & 42 &  $\cdots$ & 3.6 & J1924-2914, J1924-2914, J1700-2610 & J1642-2849 \\
 & 2021 Jun 08 05:01 & C43-5 & 15-2386 & 45 & $\cdots$ & 0.6 & J1517-2422, J1517-2422, J1700-2610 & J1642-2849 \\
 & 2021 Oct 10 22:11 & C43-8 & 70-11469 & 44 & 47.1 & 1.8 & J1517-2422, J1517-2422, J1633-2557 & J1650-2943 \\
 & 2021 Oct 19 21:37$^{\dagger}$ & C43-8 & 46-8547 & 41 & $\cdots$ & 1.59 & J1924-2914, J1924-2914, J1633-2557 & J1642-2849 \\
 & 2021 Oct 25 19:14 & C43-8 & 46-8547 & 46 & 73.5 & 1.5 & J1427-4206, J1427-4206, J1633-2557 & J1642-2849 \\
IRAS 16544--1604 & 2021 Aug 24 01:21 & C43-8 & 47-12594.5 & 50 & 56 & 0.4 & J1924-2914, J1924-2914, J1653-1551 & J1657-2004 \\
 & 2021 Oct 04 22:48 & C43-8 & 70-11614.8 & 45 & 48.7 & 0.6 & J1924-2914, J1924-2914, J1653-1551 & J1657-2004 \\
 & 2021 Oct 05 20:59 & C43-8 & 70--11614 & 46 & 72.3 & 3.1 & J1517-2422, J1517-2422, J1653-1551 & J1657-2004 \\
 & 2022 Jun 14 03:11 & C43-5 & 15-1301 & 42 & 81.6 & 0.3 & J1517-2422, J1517-2422, J1733-1304 & J1745-0753 \\
R~CrA IRS5N, R~CrA IRAS 32 & 2021 May 04 10:17 & C43-5 & 15-2517 & 42 & $\cdots$ & 1.7 & J1924-2914, J1924-2914, J1937-3958 & J1839-3453 \\
 & 2021 May 15 08:52 & C43-5 & 15-2517 & 45 & $\cdots$ & 1.0 & J1924-2914, J1924-2914, J1937-3958 & J1839-3453 \\
 & 2021 Aug 18 04:20 & C43-8 & 92-8282 & 40 & 45.8 & 0.3 & J1924-2914, J1924-2914, J1839-3453 & J1823-3454 \\
 & 2021 Aug 22 04:13 & C43-8 & 47-11614 & 49 & 44.1 & 0.3 & J1924-2914, J1924-2914, J1839-3453 & J1823-3454 \\
 & 2021 Oct 01 01:30 & C43-8 & 70-11614 & 45 & 45.1 & 1.5 & J1924-2914, J1924-2914, J1839-3453 & J1823-3454 \\
 & 2021 Oct 02 23:25 & C43-8 & 70-11614 & 43 & 67.6 & 0.7 & J1924-2914, J1924-2914, J1839-3453 & J1826-3650 \\
R~CrA IRS7B & 2021 May 03 09:05 & C43-5 & 15-1997 & 43 & 72.2 & 0.5 & J1924-2914, J1924-2914, J1937-3958 & J1839-3453 \\
 & 2021 Aug 10 01:36 & C43-8 & 59-8282 & 46 & 75.5 & 0.5 & J1924-2914, J1924-2914, J1839-3453 & J1823-3454 \\
 & 2021 Aug 10 04:21 & C43-8 & 59-8282 & 46 & 52.8 & 0.4 & J1924-2914, J1924-2914, J1839-3453 & J1823-3454 \\
\enddata
\tablecomments{$^{\dagger}$ QA0 was Semi-passed. Column 1: Observed source name included in each execution block. Column 2: Observation start time in UTC. Column 3: Antenna configulation. Column 4: Range of Baseline length. Column 5: Number of antennas used for observation. Column 6: Averaged elevation during observation. An entry of ``$\cdots$'' indicates no information was available for the averaged elevation. Column 7: Averaged precipitable water vapor (PWV) during observation. Column 8: from left to right, the quasars observed for calibrating the bandpass, amplitude scale and phase variations. Column 9: Checking the phase transfer. An entry of ``$\cdots$'' indicates no check source was observed for checking the phase transfer.}
\end{deluxetable*}
\end{longrotatetable}


\movetabledown=15mm
\begin{longrotatetable}
\begin{deluxetable*}{lllcclclcl}
\tablecaption{Observing logs of archival data used in the eDisk program} \label{tab:archive-logs}
\tablewidth{0pt}
\tablehead{
\colhead{Source name} & \colhead{UTC Date} & \colhead{Config.} & \colhead{Baselines} &
\colhead{N$_{ant}$} & \colhead{Elev.} & \colhead{PWV} & \colhead{Calibrators} & \colhead{Check source} & \colhead{Program ID} \\
\colhead{} & \colhead{} & \colhead{} & \colhead{(m)} &
\colhead{} & \colhead{(deg.)} & \colhead{(mm)} &\colhead{} & \colhead{} & \colhead{}
}
\decimalcolnumbers
\startdata
Oph IRS 63 & 2015 Nov 01 18:49 & C36-8/7 & 84--14969 & 41 & 71.2 & 1.4 & J1517-2422, J1517-2422, J1625-2527 & J1633-2557 & 2015.1.01512.S \\
 & 2015 Nov 08 14:09 & C36-8/7 & 84--16196 & 42 & 50.9 & 0.57 & J1517-2422, J1517-2422, J1625-2527 & J1633-2557 & 2015.1.01512.S \\
 & 2017 Sep 18 23:42 & C40-8/9 & 41--12145 & 43 & 43.5 & 1.5 & J1517-2422, J1733-1304, J1625-2527 & J1633-2557 & 2015.1.01512.S \\
TMC-1A & 2015 Sep 20 06:47 & C34-7/(6) & 41--2270 & 35 & 37.2 & 1.76 & J0510+1800, J0423-013, J0429+2724 & $\cdots$ & 2013.1.01086.S \\
 & 2015 Oct 23 06:10 & C36-8 & 84--16196 & 40 & 41 & 0.58 & J0510+1800, J0510+1800, J0440+2728 & J0433+2905 & 2015.1.01415.S$^{\sharp}$ \\
 & 2015 Oct 30 06:10 & C36-8/7 & 84--16196 & 43 & 41 & 0.37 & J0510+1800, J0510+1800, J0440+2728 & J0433+2905 & 2015.1.01415.S$^{\sharp}$ \\
B335 & 2014 Sep 02 01:30 & C34-6 & 33--1091 & 34 & $\cdots$ & 1.1 & J1751+0939, J1751+096, J1955+1358 & $\cdots$ & 2013.1.00879.S \\
 & 2017 Oct 08 00:45 & C43-10 & 41--16196 & 51 & 41.9 & 0.5 & J2148+0657, J2148+0657, J1938+0448 & J1929+0507 & 2017.1.00288.S$^{\S}$ \\
 & 2017 Oct 21 00:21 & C43-10 & 41--16196 & 49 & 36.7 & 1.2 & J2134-0153, J2134-0153, J1938+0448 & J1929+0507 & 2017.1.00288.S$^{\S}$ \\
 & 2017 Oct 22 00:29 & C43-10 & 41--16196 & 46 & 34.5 & 1.1 & J2134-0153, J2134-0153, J1938+0448 & J1929+0507 & 2017.1.00288.S$^{\S}$ \\
 & 2017 Oct 26 23:59 & C43-9 & 135--14851 & 47 & 36.3 & 0.7 & J2000-1748, J2000-1748, J1938+0448 & J1929+0507 & 2017.1.00288.S$^{\S}$ \\
 & 2017 Oct 29 23:10 & C43-9 & 113--13894 & 45 & 43.5 & 1.3 & J2000-1748, J2000-1748, J1938+0448 & J1929+0507 & 2017.1.00288.S$^{\S}$ \\
\enddata
\tablecomments{$^{\sharp}$The declination of the phase center for TMC-1A was 25:41:44.27 in this program. $^{\S}$The phase center for B335 was (19:37:00.894, 07:34:09.59) in this program. Column 1: Observed source name included in each execution block. Column 2: Observation start time in UTC. Column 3: Antenna configuration. Column 4: Range of Baseline length. Column 5: Number of antennas used for observation. Column 6: Averaged elevation during observation. An entry of ``$\cdots$'' indicates no information was available for the averaged elevation. Column 7: Averaged precipitable water vapor (PWV) during observation. Column 8: from left to right, the quasars observed for calibrating the bandpass, amplitude scale, and phase variations. Column 9: Checking the phase transfer. An entry of ``$\cdots$'' indicates no calibrator was observed for checking the phase transfer. Column 10: Program ID.}
\end{deluxetable*}
\end{longrotatetable}

\startlongtable
\begin{deluxetable*}{lll}
\tablecaption{Scale factors} \label{tab:scale-factor}
\tablewidth{0pt}
\tablehead{
\colhead{Source name} & \colhead{UTC Date} & \colhead{Scale factor}
}
\decimalcolnumbers
\startdata
L1489 IRS & 2021 Aug 20 07:43 & 0.86 \\
 & 2021 Aug 20 09:29 & 0.95 \\
 & 2021 Dec 03 03:02 & 1.000 \\
 & 2021 Dec 16 03:29 & 0.91 \\
 & 2022 Jul 03 12:58 & 0.87 \\
IRAS 04166+2706 & 2021 Sep 30 05:20$^{\dagger}$ & 1.000 \\
 & 2021 Oct 01 05:22$^{\dagger}$ & 0.91 \\
 & 2021 Oct 18 04:32$^{\dagger}$ & 0.87 \\
 & 2021 Oct 24 07:20 & 0.86 \\
 & 2022 Jul 3 13:15 & 0.93 \\
 & 2022 Jul 3 14:26 & 0.94 \\
IRAS 04169+2702 & 2021 Sep 30 05:20$^{\dagger}$ & 0.96 \\
 & 2021 Oct 01 05:22$^{\dagger}$ & 0.86 \\
 & 2021 Oct 18 04:32$^{\dagger}$ & 0.90 \\
 & 2021 Oct 24 07:20 & 0.93 \\
 & 2022 Jul 3 13:15 & 1.000 \\
 & 2022 Jul 3 14:26 & 0.99 \\
IRAS 04302+2247 & 2021 Sep 30 08:48 & 0.9 \\
 & 2021 Oct 01 06:54 & 1.000 \\
 & 2021 Oct 01 08:31 & 0.74 \\
 & 2021 Dec 01 03:40 & 0.97 \\
L1527 IRS & 2021 Dec 03 03:02 & 1.18 \\
 & 2021 Dec 16 03:29 & 1.13 \\
 & 2022 Jul 03 12:58 & 0.92 \\
 & 2021 Oct 14 08:26 & 0.95 \\
 & 2021 Oct 15 06:01 & 0.96 \\
 & 2021 Oct 15 07:51 & 1.000 \\
Ced110 IRS4 & 2021 Apr 24 01:30 & 0.91 \\
 & 2021 Oct 04 13:05 & 1.02 \\
 & 2021 Oct 20 10:58 & 1.01 \\
 & 2021 Oct 20 12:24 & 1.000 \\
BHR71 IRS2 & 2021 May 04 00:26 & 0.82 \\
 & 2021 May 09 01:26 & 1.000 \\
 & 2021 Oct 14 14:48$^{\dagger}$ & 1.01 \\
 & 2021 Oct 15 13:13 & 1.10 \\
 & 2021 Oct 16 12:18$^{\dagger}$ & 1.14 \\
 & 2021 Oct 19 14:34 & 1.14 \\
 & 2021 Oct 20 13:57$^{\dagger}$ & 1.34 \\
BHR71 IRS1 & 2021 May 04 00:26 & 1.04 \\
 & 2021 May 09 01:26 & 1.000 \\
 & 2021 Oct 14 14:48$^{\dagger}$ & 0.97 \\
 & 2021 Oct 15 13:13 & 1.11 \\
 & 2021 Oct 16 12:18$^{\dagger}$ & 1.13 \\
 & 2021 Oct 19 14:34 & 1.20 \\
 & 2021 Oct 20 13:57$^{\dagger}$ & 1.17 \\
IRAS 15398-3359 & 2021 May 06 03:56 & 1.0 \\
 & 2021 Aug 09 22:34 & 1.04 \\
 & 2021 Aug 13 23:41 & 1.04 \\
 & 2021 Oct 20 15:28 & 0.62 \\
GSS30 IRS3 & 2021 Oct 05 22:14 & 1.0 \\
 & 2021 Oct 26 19:22 & 0.92 \\
 & 2021 Oct 27 20:35 & 1.000 \\
 & 2021 Oct 28 17:02 & 1.07 \\
 & 2022 Jun 14 04:13 & 0.87 \\
 & 2022 Jun 15 00:45 & 0.86 \\
Oph IRS43 & 2021 May 26 07:32 & 1.20 \\
 & 2021 Jun 08 05:01 & 1.50 \\
 & 2021 Oct 10 22:11 & 0.93 \\
 & 2021 Oct 19 21:37$^{\dagger}$ & 0.95 \\
 & 2021 Oct 25 19:14 & 1.000 \\
IRAS 16253-2429 & 2021 Oct 05 22:14 & 0.71 \\
 & 2021 Oct 26 19:22 & 0.68 \\
 & 2021 Oct 27 20:35 & 0.88 \\
 & 2021 Oct 28 17:02 & 0.76 \\
 & 2022 Jun 14 04:13 & 0.99 \\
 & 2022 Jun 15 00:45 & 1.000 \\
Oph~IRS63 & 2015 Nov 01 18:49$^{\ddagger}$ & 0.98\\
 & 2015 Nov 08 14:09$^{\ddagger}$ & 1.05\\
 & 2017 Sep 18 23:42$^{\ddagger}$ & 1.05\\
 & 2022 Jun 14 04:13 & 1.01 \\
 & 2022 Jun 15 00:45 & 1.000 \\
IRAS 16544--1604 & 2021 Aug 24 01:21 & 1.02 \\
 & 2021 Oct 04 22:48 & 1.11 \\
 & 2021 Oct 05 20:59 & 1.000 \\
 & 2022 Jun 14 03:11 & 0.82 \\
R CrA IRS5N & 2021 May 04 10:17 & 1.03 \\
 & 2021 May 15 08:52 & 1.00 \\
 & 2021 Aug 18 04:20 & 0.99 \\
 & 2021 Aug 22 04:13 & 0.96 \\
 & 2021 Oct 01 01:30 & 1.000 \\
 & 2021 Oct 02 23:25 & 0.96 \\
R CrA IRS7B & 2021 May 03 09:05 & 0.95 \\
 & 2021 Aug 10 01:36 & 1.000 \\
 & 2021 Aug 10 04:21 & 0.99 \\
R CrA IRAS 32 & 2021 May 04 10:17 & 1.01 \\
 & 2021 May 15 08:52 & 1.04 \\
 & 2021 Aug 18 04:20 & 1.000 \\
 & 2021 Aug 22 04:13 & 0.93 \\
 & 2021 Oct 01 01:30 & 0.98  \\
 & 2021 Oct 02 23:25 & 0.93 \\
TMC-1A & 2015 Sep 20 06:47$^{\ddagger}$ & 1.000\\
 & 2015 Oct 23 06:10$^{\ddagger}$ & 0.74\\
 & 2015 Oct 30 06:10$^{\ddagger}$ & 0.63\\
B335 & 2014 Sep 02 01:30$^{\ddagger}$ & 0.75\\
& 2017 Oct 08 00:45$^{\ddagger}$ & 1.00\\
& 2017 Oct 21 00:21$^{\ddagger}$ & 1.000\\
& 2017 Oct 22 00:29$^{\ddagger}$ & 1.04\\
& 2017 Oct 26 23:59$^{\ddagger}$ & 0.99 \\
& 2017 Oct 29 23:10$^{\ddagger}$ & 1.01\\
\enddata
\tablecomments{$^{\dagger}$ QA0 was Semi-passed. $^{\ddagger}$ALMA archival data. Column 1: target names. Column 2: Observation start time in UTC. Column 3: Scale factor applied to each EB. The scale factor of the reference EB is shown as  1.000.}
\end{deluxetable*}

\bibliography{ref}{}
\bibliographystyle{aasjournal}

\end{document}